\documentclass[12pt]{article}

\usepackage[utf8]{inputenc}
\usepackage{amsmath,amssymb,amsfonts,amsbsy,bm,bbm,mathabx,latexsym}
\usepackage{enumitem}
\usepackage{natbib,verbatim}
\usepackage{algorithm,algorithmic}
\usepackage{color,xcolor}
\usepackage{hyperref}
\usepackage{graphicx}
\usepackage{booktabs}
\newtheorem{lemma}{\underline{\bf Lemma}}
\newtheorem{pro}{\underline{\bf Proposition}}
\newtheorem{theorem}{\underline{\bf Theorem}}

\newtheorem{cor}{\underline{\bf Corollary}}

\def\bse{\begin{eqnarray*}}
\def\ese{\end{eqnarray*}}
\def\be{\begin{eqnarray}}
\def\ee{\end{eqnarray}}
\def\n{\nonumber}

\def\I{\mathrm{I}}
\def\E{\mathrm{E}}

\def\var{\mathrm{var}}
\def\indep{\perp \!\!\! \perp}
\DeclareMathOperator*{\argmax}{argmax}

\def\Normal{\hbox{Normal}}

\def\wh{\widehat}
\def\wt{\widetilde}
\def\wc{\widecheck}
\def\trans{^{\rm T}}
\def\eff{_{\rm eff}}

\def\<{{\langle}}
\def\>{{\rangle}}

\def\sumi{\sum_{i=1}^n}

\def\a{{\mathbf a}}
\def\A{{\mathbf A}}
\def\b{{\mathbf b}}
\def\B{{\mathbf B}}

\def\g{{\mathbf g}}
\def\h{{\mathbf h}}

\def\S{{\mathbf S}}

\def\u{{\mathbf u}}

\def\V{{\mathbf V}}

\def\z{{\mathbf z}}
\def\Z{{\mathbf Z}}
\def\0{{\mathbf 0}}

\def\H{{\cal H}}
\def\calL{{\cal L}}
\def\mR{{\mathbb R}}

\def\ba{{\boldsymbol\alpha}}
\def\bb{{\boldsymbol\beta}}
\def\bB{\mathbf{B}}

\def\bSig{{\boldsymbol\Sigma}}
\def\bphi{{\boldsymbol\phi}}

\def\fyxz{f_{Y|X,\Z}}
\def\fcz{f_{C|\Z}}
\def\fxz{f_{X|\Z}}
\def\fz{f_{\Z}}

\def\boxit#1{\vbox{\hrule\hbox{\vrule\kern6pt\vbox{\kern6pt#1\kern6pt}\kern6pt\vrule}\hrule}}

\newcommand{\blind}{1}

\title{\bf Robust and efficient estimation in the presence of a randomly %right
censored covariate}

\if1\blind
{
  \author{Seong-ho Lee$^{1*}$, Brian D. Richardson$^{2*}$, Yanyuan Ma$^3$,\\
  Karen S. Marder$^4$, and Tanya P. Garcia$^2$\medskip\\
  $^1$Department of Statistics, University of Seoul, South Korea\\
  $^2$Department of Biostatistics,\\
  University of North Carolina at Chapel Hill, USA\\
  $^3$Department of Statistics, Pennsylvania State University, USA\\
  $^4$Department of Neurology, Columbia University Medical Center, USA}
} \fi

\if0\blind
{
  \author{}
} \fi

\date{}

\begin{document}
\def\spacingset#1{\renewcommand{\baselinestretch}{#1}\small\normalsize}
\spacingset{1}

\clearpage
\pagenumbering{arabic}
\maketitle
\def\thefootnote{*}\footnotetext{These authors contributed equally to this work and share first authorship}\def\thefootnote{\arabic{footnote}}

% Current word count: 200/200
\begin{abstract}
\noindent
In Huntington’s disease research, a current goal is to understand how
symptoms change prior to a clinical diagnosis. Statistically, this
entails modeling symptom severity as a function of the covariate
`time until diagnosis', which is often heavily right-censored in
observational studies. Existing estimators that handle right-censored
covariates have varying statistical efficiency and robustness to misspecified models for nuisance distributions (those of the censored covariate and
censoring variable). On one extreme, complete case estimation, which
utilizes uncensored data only, is free of nuisance distribution models
but discards informative censored observations. On the other extreme,
maximum likelihood estimation is maximally efficient but inconsistent
when the covariate's distribution is misspecified. We propose a
semiparametric estimator that is robust and efficient. When the
nuisance distributions are modeled parametrically, the estimator is
doubly robust, i.e., consistent if at least one distribution is
correctly specified, and semiparametric efficient if both models are
correctly specified. When the nuisance distributions are estimated via nonparametric or machine learning methods,
the estimator is consistent and semiparametric 
efficient. We show empirically that the proposed estimator, implemented in the R
package \texttt{sparcc}, has its claimed properties, and we apply it
to study Huntington's disease symptom trajectories using data from the
Enroll-HD study. 
\end{abstract}


%%%%%%%%%%%%%%%%%%%%%%%%%%%%%%%%%%%%%%%%%%%%%%%%%%

\clearpage
\spacingset{1.9}

%%%%%%%%%%%%%%%%%%%%%
\section{Introduction}
%%%%%%%%%%%%%%%%%%%%%

Huntington's disease is a genetic neurodegenerative disease with
progressive cognitive, motor, and functional symptoms
\citep{dutra_2020}. Unlike most neurodegenerative diseases, its exact
cause is known: a mutation in the Huntingtin gene that contains an
abnormally large number of CAG (cytosine, adenine, guanine) repeats. A
genetic blood test can detect this mutation, and individuals with 
sufficient CAG repeats are virtually guaranteed to exhibit symptoms
during their lifetime. This ability to identify, with near certainty, who will develop
Huntington's disease symptoms offers a scientific
opportunity: in observational studies, individuals with the
gene mutation can be tracked to see how symptom severity changes over time (i.e.,
the symptom trajectory) in the period \emph{before} a clinical
diagnosis. If the symptom trajectory can be accurately estimated
during this period, researchers can test interventions to slow or halt
disease progression before irreparable damage is done
\citep{scahill_biological_2020}.

However, estimating the symptom trajectory in
the period before a diagnosis is challenging because Huntington's
disease progresses slowly. Patient decline spans decades
\citep{dutra_2020}, so observational studies often end before all
at-risk individuals are diagnosed. In such cases, an individual's time
of diagnosis is not observed, but is known to be some time after they exit the
study, and thus their time until diagnosis is
right-censored, i.e., known to be greater than some observed
value. The challenge then is estimating the symptom trajectory as a
function of time until diagnosis, when that \emph{time until diagnosis
  is a right-censored covariate}. 

Currently, there are a range of estimators that address the right-censored
covariate problem \citep{lotspeich2024making}. These estimators target
parameters that define the symptom trajectory, and each estimator
depends to varying degrees on the distributions of the right-censored
covariate  (e.g., time until diagnosis) and censoring variable (e.g.,
time until study exit).  These distributions are not of direct
interest---they are nuisance distributions---but they impact the
estimators' consistency and statistical efficiency (i.e.,
variance). The complete case estimator only uses data from uncensored
individuals and the inverse probability weighting estimator assigns
weights to those individuals to make the sample representative of the
entire population--censored or not \citep{AshnerGarcia2023,
  MatsouakaAtem2020}. Both estimators are maximally robust in the
sense that their consistency does not depend on correctly specified models for
nuisance distributions \citep{vazquez2024establishingparallelsdifferencesrightcensored}, but relatively
inefficient because they discard data from censored individuals.  

%Estimators that do not discard data are:
Estimators that make more direct use of censored observations are: (i) the augmented complete
case and augmented inverse probability weighting estimators, which
modify their respective original estimators to increase efficiency \citep{ahn2018cox, vazquez2024establishingparallelsdifferencesrightcensored}; (ii)
the imputation estimator, which replaces censored covariate values
with imputed values \citep{Bernhardtetal2015, Ateme&Matsouaka2017};
and (iii) the maximum likelihood estimator (MLE), which maximizes the
likelihood of the observed data accounting for the right-censored
covariate \citep{kongetal2018, chen2022semiparametric}. By making more direct use of
available data, these estimators gain efficiency over the complete
case and inverse probability weighting estimators, with the MLE
achieving maximal efficiency. That efficiency gain, however, requires nuisance distributions to be correctly specified to ensure
consistency, making the estimators less robust. The two augmented
estimators require correct specification of the censoring
distribution, while the imputation estimator and MLE require correct
specification of the censored covariate distribution; under incorrect specifications, these estimators are inconsistent. \citep{Bernhardtetal2015, vazquez2024establishingparallelsdifferencesrightcensored}. 

Existing solutions that aim to compromise less efficiency and robustness include adaptations to these estimators, where the nuisance distributions are modeled flexibly, e.g., with a semiparametric model (as in \citet{kongetal2018,
  atem2019cox, chen2022semiparametric}) or a nonparametric model (as in
\citet{CalleGomez2005}). Yet, semiparametric and nonparametric models
can still be wrong. These models are not free of assumptions; e.g.,
the semiparametric Cox model imposes a proportional hazards
assumption, and nonparametric models require unknown functions to be
sufficiently smooth. When these assumptions do not hold, even flexible
semiparametric or nonparametric models will be wrong, and, if wrong,
will result in inconsistent estimators of the symptom trajectory.  

These points highlight the opportunity for an estimator that strikes a
better balance between robustness and efficiency. In this paper, we address
that need by proposing the ``\underline{S}emi\underline{pa}rametric
\underline{R}ight-\underline{C}ensored \underline{C}ovariate" (SPARCC) estimator. We develop this estimator by applying the general theory of semiparametric models to the specific problem of randomly right-censored covariates and using a geometric approach to construct influence functions \citep{bickel1993efficient,tsiatis2006semiparametric}. The SPARCC
estimator is doubly robust: it remains consistent if we
misspecify parametric models for the distribution of either the censored covariate
or the censoring variable. If we are not willing to specify parametric models for the
nuisance distributions, we can instead use standard nonparametric
or machine learning methods for estimating the nuisance
distributions, and the SPARCC estimator 
remains consistent. Under correct parametric or nonparametric models
for the nuisance distributions, the SPARCC estimator achieves the
semiparametric efficiency bound, making it maximally efficient among
all consistent estimators under the same semiparametric model assumptions. 

The remainder of the paper proceeds as follows. The SPARCC estimator
is derived in Section~\ref{sec:deriv_est} and its implementation is
shown in Section~\ref{sec:imp}. The empirical performance of the
SPARCC estimator, which is available in the R package
\texttt{sparcc}, is compared to that of existing estimators in
Section~\ref{sec:sim}. In
Section~\ref{sec:dat}, the estimator is then applied to data from Enroll-HD, an
observational study of Huntington's disease, to assess symptom
trajectories prior to diagnosis. Finally, a discussion is provided in
Section~\ref{sec:con}.

%%%%%%%%%%%%%%%%%%%%%%%%%%%%%%%%%%%%%%%%%%%%%%%%%%

%%%%%%%%%%%%%%%%%%%%%%%%
\section{Deriving the SPARCC estimator}
\label{sec:deriv_est}
%%%%%%%%%%%%%%%%%%%%%%%%

%%%%%%%%%%%%%%%%%%%%%%%%
\subsection{Notation and framework}
\label{sec:not}
%%%%%%%%%%%%%%%%%%%%%%%%

Consider a general regression model $\fyxz(y,x,\z,\bb)$ that describes
the relationship between an outcome $Y$ and covariates $X$ and $\Z$,
using the $p$-dimensional parameter $\bb$. Here and throughout, $f$
denotes (conditional) probability density (or mass) functions
described by the subindices. The main challenge when estimating $\bb$
is that, whereas $\Z$ is fully observed, $X$ is randomly
right-censored. Instead of observing $X$ directly, we observe $W =
\min(X, C)$ and $\Delta = I(X \leq C)$, where $C$ is a random
censoring variable and $\Delta$ indicates whether $X$ is observed. Thus,
the observed data are $(Y, W, \Delta, \Z)$. In the context of
Huntington's disease, for example, $Y$ is symptom severity, $X$ is
time until diagnosis, $\Z$ contains other patient variables, and $C$ is
time until study exit. 

We will derive the SPARCC estimator under a common assumption known as
noninformative covariate censoring, meaning $C$ is independent of
$(Y,X)$ given $\Z$, or, equivalently, that $C\indep Y\mid\Z$ and
$C\indep X\mid\Z$, where $\indep$ denotes independence. This
assumption may be valid in Huntington's disease studies where time
until diagnosis is censored by time until study exit, provided participants exit the study for administrative reasons not related to their health outcomes. Under the noninformative censoring assumption, the likelihood for the observed data
$(Y,W,\Delta,\Z)$ can be constructed by integrating over the support
of the full data $(Y,X,C,\Z)$, noting that when $\Delta=1$, we have
$X\le C$, and when $\Delta=0$, we have $X>C$. This likelihood is 
\be\label{eq:model}
f_{Y,W,\Delta,\Z}(y,w,\delta,\z,\bb)
&\equiv&
\left\{\fyxz(y,w,\z,\bb)\fxz(w,\z)\int_w^\infty\fcz(c,\z)dc\right\}^{\delta}\n\\
&&\times\left\{\int_w^\infty\fyxz(y,x,\z,\bb)\fxz(x,\z)dx\fcz(w,\z)\right\}^{1-\delta}\fz(\z).
\ee

\begin{lemma}\label{lem:identifiability}
All unknown components in \eqref{eq:model}, i.e., $\fxz$, $\fcz$, $\fz$, and $\bb$, are identifiable.
\end{lemma}

The proof of identifiability in Lemma \ref{lem:identifiability} is in
Section  \ref{sec:lem-identifiability-proof} (Supplementary
Material). Having identifiability ensures there is a well-defined and
unique estimator for $\bb$ even without specifying the distributions
$\fxz$, $\fcz$, or $\fz$ a priori. This result is especially
appealing, as these distributions are often unknown in practice. 

%%%%%%%%%%%%%%%%%%%%%%%%
\subsection{The semiparametric method}
%%%%%%%%%%%%%%%%%%%%%%%%

We treat the distributions $\fxz$, $\fcz$, and $\fz$ as
infinite-dimensional nuisance parameters that are not restricted to a
specific parametric form. To reflect this fact, we write them as
$\eta_1(x,\z)\equiv f_{X|\Z}(x,\z)$,  $\eta_2(c,\z)\equiv
f_{C|\Z}(c,\z)$, and $\eta_3(\z)\equiv  f_{\Z}(\z)$. We then aim
  to construct the robust and efficient
SPARCC estimator for $\bb$.

Under this framework, we consider estimators that are roots of $p$-dimensional and mean-zero estimating functions of $(Y, W, \Delta, \Z)$ and $\bb$.
Since any given estimating function $\g$ has mean zero, the estimator
of $\bb$ obtained by solving $\sumi
\g(y_i,w_i,\delta_i,\z_i,\bb,\eta_1,\eta_2,\eta_3)=\0$ is guaranteed
to be consistent under mild regularity conditions \citep{Foutz1977}.
The set of all estimating functions makes up a Hilbert space $\H$,
with the covariance inner product defined as
$\<\h_1,\h_2\>=\E(\h_1\trans\h_2)$. The corresponding notion of
orthogonality is then $\h_1\perp\h_2$ if and only if $\<\h_1,\h_2\>=0$. 
Using this definition of orthogonality, we can decompose $\H$ into two
key orthogonal subspaces that lead us to the SPARCC estimator. The first is the
nuisance tangent space, $\Lambda$, which is formed by considering all
possible parametric models for the nuisance distributions, deriving
their associated score vectors, then taking the span of those score
vectors and their limits. The second subspace is the orthogonal
complement, $\Lambda^{\perp}$, which is the space of estimating
functions orthogonal to $\Lambda$. We prove in Sections
\ref{sec:pro-lambda-proof} and \ref{sec:pro-lambdaperp-proof}
(Supplementary Material) that for our model in \eqref{eq:model}, the
nuisance tangent space and its orthogonal complement are as given
  in Proposition \ref{pro:lambda}.
\begin{pro}\label{pro:lambda}
The nuisance tangent space is $\Lambda\equiv\Lambda_1\oplus\Lambda_2\oplus\Lambda_3$, where   $\Lambda_1$, $\Lambda_2$,
    and $\Lambda_3$ are the nuisance tangent spaces for $\eta_1(x,\z)$,
    $\eta_2(c,\z)$, and $\eta_3(\z)$, respectively. More exactly,
\bse
\Lambda_1
&\equiv&\left[\delta\a_1(w,\z)+(1-\delta)
\frac{\E\{\I(X>w)\a_1(X,\z)\mid y,\z\}}{\E\{\I(X>w)\mid y,\z\}}
:\E\{\a_1(X,\z)\mid\z\}=\0\right],\\
\Lambda_2
&\equiv&\left[\delta\frac{\E\{\I(C\ge w)\a_2(C,\z)\mid\z\}}{\E\{\I(C\ge w)\mid\z\}}
+(1-\delta)\a_2(w,\z)
:\E\{\a_2(C,\z)\mid\z\}=\0\right],\\
\Lambda_3
&\equiv&[\a_3(\z):\E\{\a_3(\Z)\}=\0].
\ese
\end{pro}

\begin{pro}\label{pro:lambdaperp}
The orthogonal complement is the intersection of $\Lambda_1^{\perp}$, $\Lambda_2^{\perp}$, $\Lambda_3^{\perp}$, which can be explicitly written as
\bse
\Lambda^\perp
&\equiv&\left[
\g(y,w,\delta,\z)=\delta\g_1(y,w,\z)+(1-\delta)\g_0(y,w,\z):\right.\\
&&\E\{\I(X\le c)\g_1(Y,X,\z)+\I(X>c)\g_0(Y,c,\z)\mid c,\z\}=\0,\\
&&\left.\E\{\I(x\le C)\g_1(Y,x,\z)+\I(x>C)\g_0(Y,C,\z)\mid x,\z\}=\0\right].
\ese
\end{pro}

Intuitively, since $\Lambda$ includes estimating functions
correlated with nuisance distribution score vectors, estimators based
on these functions will generally be affected by misspecification of
nuisance distributions.  
In contrast, the orthogonality of $\Lambda^{\perp}$  to $\Lambda$
implies that nuisance distributions minimally impact the estimating  
functions in $\Lambda^{\perp}$ \citep{bickel1993efficient,
  tsiatis2006semiparametric}. We thus build our SPARCC estimator from
an estimating function in $\Lambda^{\perp}$, in particular, the
efficient score vector, $\S\eff\in\Lambda^{\perp}$, which is the orthogonal projection of the score vector
$\S_\bb(y,w,\delta,\z,\bb)=\partial\log
f_{Y,W,\Delta,\Z}(y,w,\delta,\z,\bb)/\partial\bb$ onto
$\Lambda^{\perp}$. The SPARCC estimator is then the solution to
$\sum_i \S\eff(y_i,w_i,\delta_i,\z_i,\bb) = \0$. While all
functions in $\Lambda^{\perp}$ are minimally impacted by
misspecification of nuisance distributions, $\S\eff$ results in an
estimator that is maximally efficient. The explicit form of $\S\eff$ is given in
Proposition \ref{pro:efficientscore}.
\begin{pro}\label{pro:efficientscore}
The efficient score vector for $\bb$ is
\be
\label{eq:seff}
\S\eff(y,w,\delta,\z,\bb)
&\equiv&\delta\{\S_\bb^{\rm F}(y,w,\z,\bb)-\a(w,z,\bb)\}\nonumber \\
&&+(1-\delta)\frac{\E_1[\I(X>w)\{\S_\bb^{\rm F}(y,X,\z,\bb)-\a(X,\z,\bb)\}\mid y,\z]}{\E_1\{\I(X>w)\mid y,\z\}},
\ee
where $\S_\bb^{\rm F}(y,x,\z,\bb)\equiv\partial\log\fyxz(y,x,\z,\bb)/\partial\bb$, and 
$\a(x,\z,\bb)$ satisfies
\be
&&\E_2\{\I(x\le C)\mid\z\}\a(x,\z,\bb)
+\E_2\left[\I(x>C)\frac{\E_1\{\I(X>C)\a(X,\z,\bb)\mid Y,C,\z\}}{\E_1\{\I(X>C)\mid Y,C,\z\}}\mid x,\z\right]\nonumber\\
&=&\E_2\left[\I(x>C)\frac{\E_1\{\I(X>C)\S_\bb^{\rm F}(Y,X,\z,\bb)\mid Y,C,\z\}}{\E_1\{\I(X>C)\mid Y,C,\z\}}\mid x,\z\right]. \label{eq:solve-a}
\ee
Here, $\E_1,\E_2$ denote expectations with respect to
$\eta_1(x,\z),\eta_2(c,\z)$, respectively. 
\end{pro}

The proof of Proposition \ref{pro:efficientscore} is in Section
\ref{sec:pro-efficientscore-proof} (Supplementary Material).  
There are two important observations we can make about $\S\eff$.  
The first is that computing $\S\eff$ requires solving the integral equation in
\eqref{eq:solve-a} for  $\a(x,\z,\bb)$ and evaluating multiple
integrals that may have no closed form:
\bse
\frac{\E_1\{\I(X>c)\S_\bb^{\rm F}(y,X,\z,\bb)\mid y, \z,\bb\}}{\E_1\{\I(X>c)\mid y, \z,\bb\}}
&=&\frac{\int\I(x>c)\S_\bb^{\rm F}(y,x,\z,\bb)\fyxz(y,x,\z,\bb)\eta_1(x,\z)dx}{\int\I(x>c)\fyxz(y,x,\z,\bb)\eta_1(x,\z)dx},\\
\frac{\E_1\{\I(X>c)\a(X,\z,\bb)\mid y, \z,\bb\}}{\E_1\{\I(X>c)\mid y, \z,\bb\}}
&=&\frac{\int\I(x>c)\a(x,\z,\bb)\fyxz(y,x,\z,\bb)\eta_1(x,\z)dx}{\int\I(x>c)\fyxz(y,x,\z,\bb)\eta_1(x,\z)dx},\\
\E_2\{\I(x\leq C)\mid \z\}
&=&\int\I(x\leq c)\eta_2(c,\z)dc,
\ese
and
\bse
&&\E_2\left\{\I(x>C)\frac{\E_1\{\I(X>C)\S_\bb^{\rm F}(Y,X,\z,\bb)\mid Y,C,\z,\bb\}}{\E_1\{\I(X>C)\mid Y,C,\z,\bb\}}\mid x,\z,\bb\right\}\\
&=&\int\I(x>c)\frac{\E_1\{\I(X>c)\S_\bb^{\rm F}(y,X,\z,\bb)\mid y,\z,\bb\}}{\E_1\{\I(X>c)\mid y,\z,\bb\}}\fyxz(y,x,\z,\bb)\eta_2(c,\z)dydc,\\
&&\E_2\left\{\I(x>C)\frac{\E_1\{\I(X>C)\a(X,\z,\bb)\mid Y,C,\z,\bb\}}{\E_1\{\I(X>C)\mid Y,C,\z,\bb\}}\mid x,\z,\bb\right\}\\
&=&\int\I(x>c)\frac{\E_1\{\I(X>c)\a(X,\z,\bb)\mid y,\z,\bb\}}{\E_1\{\I(X>c)\mid y,\z,\bb\}}\fyxz(y,x,\z,\bb)\eta_2(c,\z)dydc.
\ese
These challenges can be resolved with a numerical approach. Since we
ultimately solve for $\bb$ using $\sumi
\S\eff(y_i,w_i,\delta_i,\z_i,\bb)=\0$, we only need to construct the
function $\a(x,\z,\bb)$ at observed values
$\z=\z_i~(i=1,\cdots,n)$. For each fixed $\z_i$, the integral equation
in  \eqref{eq:solve-a} is a univariate Fredholm integral equation of
the second kind, which we solve using an adaptation of the algorithm
from \cite{atkinson1976automatic}. This adaptation involves solving a
linear system through discretizing \eqref{eq:solve-a}  at chosen
values for $x$ and numerically approximating the integrals through
quadrature rules, e.g., Gauss-Hermite quadrature or the trapezoidal
rule.
Details of this numerical procedure are given in Section
\ref{sec:num-soln-Seff} (Supplementary Material). All other integrals
in \eqref{eq:solve-a} can be similarly approximated using quadrature
rules.

The second observation is that $\S\eff$ involves $\eta_1(x,\z)$ and
$\eta_2(c,\z)$, but not $\eta_3(\z)$. Therefore, to estimate $\bb$, we
need to specify $\eta_1(x,\z)$ and $\eta_2(c,\z)$, but not
$\eta_3(\z)$.  In the next section, we demonstrate how $\S\eff$, and
the consistency and efficiency of the resulting SPARCC estimator, are
impacted by modeling choices for $\eta_1(x,\z)$ and $\eta_2(c,\z)$.

%%%%%%%%%%%%%%%%%%%%%%%%
\section{Implementing the SPARCC estimator}
\label{sec:imp}
%%%%%%%%%%%%%%%%%%%%%%%%

We propose and analyze two options for specifying the nuisance
distributions $\eta_1(x,\z)$ and $\eta_2(c,\z)$: parametric working
models and nonparametric/machine learning methods.

%%%%%%%%%%%%%%%%%%%%%%%%
\subsection{Parametric working models}
\label{sec:par}
%%%%%%%%%%%%%%%%%%%%%%%%

One way to specify $\eta_1(x,\z)$ and $\eta_2(c,\z)$ is with
parametric working models, $\eta_1^*(x,\z)$ and
$\eta_2^\dagger(c,\z)$, which depend on finite-dimensional parameters
$\ba_1,\ba_2$ that can be known or unknown. The name `working model'
reflects the fact that these models may not capture the true
data-generating process. We are free to choose any working models, so
long as they are valid conditional density functions given $\z$ (i.e.,
are nonnegative and integrate to one). With working models
$\eta_1^*(x,\z)$ and $\eta_2^\dagger(c,\z)$, the functions in
Proposition \ref{pro:efficientscore} are changed to $\E_1^*, \E_2^\dagger$,
$\a^{*\dagger}$,  and $\S\eff^{*\dagger}$ to reflect their dependency
on those working models. In short, using parametric working models to
obtain the SPARCC estimator involves the following steps in Algorithm \ref{alg:par}.

\begin{algorithm}[H]
    \caption{SPARCC estimator with parametric working models}
 	\begin{algorithmic}\label{alg:par}
        \STATE \textbf{Input}: data $\{(y_i,w_i,\delta_i,\z_i):i=1,\cdots,n\}$.
            \STATE \textbf{do}
            \STATE (a) adopt working models for $\eta_1(x,\z), \eta_2(c,\z)$, denoted by $\eta_1^*(x,\z,\ba_1), \eta_2^\dagger(c,\z,\ba_2)$ and, if needed, estimate the unknown parameters $\ba_1,\ba_2$,
            \STATE (b) compute $\E_1^*$ and $\E_2^\dagger$ with respect to $\eta_1^*(x,\z)$ and $\eta_2^\dagger(c,\z)$  in Proposition \ref{pro:efficientscore},            
            \STATE (c) obtain $\a^{*\dagger}(x,\z_i,\bb)$ by solving \eqref{eq:solve-a} at $\z=\z_i~(i=1,\cdots,n)$,
            \STATE (d) obtain $\wh\bb$ by solving $\sumi\S\eff^{*\dagger}(y_i,w_i,\delta_i,\z_i,\bb)=\0$ with respect to $\bb$.
 		\STATE \textbf{Output}: $\wh\bb$.
    \end{algorithmic}
\end{algorithm}

%%%%%%%%%%%%%%%%%%%%%%%%
\subsubsection{Constructing parametric working models}
\label{sec:par_est}
%%%%%%%%%%%%%%%%%%%%%%%%

There are many possible parametric working models we can use for
$\eta_1^*(x,\z)$ and $\eta_2^\dagger(c,\z)$. For example, because $X$
(or $C$) is right-censored, we can posit $\eta_1^*(x,\z,\ba_1)$ (or
$\eta_2^\dagger(c,\z,\ba_2)$) as a log-normal distribution
corresponding to an accelerated failure time model, where $X$ (or $C$)
is the response and $\Z$ is the predictor. Another option is a
generalized linear model where $X$ (or $C$) is the response and $\Z$
is the predictor. 

If the parametric working models contain unknown parameters
$\ba_1,\ba_2$, the next step is to estimate those parameters  and
substitute them with their estimators $\wh\ba_1$ and $\wh\ba_2$. In
general, these estimators may depend on $\bb$ as well. For example,
the MLEs for $\ba_1$ and $\ba_2$ based on the log-likelihood of the
full observed data $(Y,W,\Delta,\Z)$ are 
\be
\wh\ba_1(\bb)&\equiv&\argmax_{\ba_1}\sumi\left\{\delta_i\log\eta_1^*(w_i,\z_i,\ba_1)
+(1-\delta_i)\log\int_{w_i}^\infty\fyxz(y_i,x,\z_i,\bb)\eta_1^*(x,\z_i,\ba_1)dx\right\},\nonumber\\
\wh\ba_2
&\equiv&\argmax_{\ba_2}\sumi\left\{\delta_i\log\int_{w_i}^\infty\eta_2^\dagger(c,\z_i,\ba_2)dc
+(1-\delta_i)\log\eta_2^\dagger(w_i,\z_i,\ba_2)\right\}.\label{eq:mle-fulldata}
\ee
When we estimate $\ba_1$ and $\ba_2$ based on the log-likelihood of
just $(W,\Delta,\Z)$,  the MLE for $\ba_1$ no longer depends on $\bb$
and is simply
\be
\wh\ba_1
&\equiv&\argmax_{\ba_1}\sumi\left\{\delta_i\log\eta_1^*(w_i,\z_i,\ba_1)
+(1-\delta_i)\log\int_{w_i}^\infty\eta_1^*(x,\z_i,\ba_1)dx\right\},\label{eq:mle2}
\ee
and the MLE for $\ba_2$ remains as in
\eqref{eq:mle-fulldata}. Other options besides maximum
likelihood estimation exist to estimate parametric working
models; see Chapter 3 of \cite{kalbfleisch2002statistical} for a
detailed discussion.

Standard statistical results show that whether or not the parametric
working models are correctly specified, $\wh\ba_1(\bb)$ and $\wh\ba_2$
converge in probability to some $\ba_1^*(\bb)$ and $\ba_2^\dagger$
respectively at the root-$n$ rate
\citep{huber1967behavior,white1982maximum}. When the parametric
working models are correctly specified, $\wh\ba_1(\bb)$ and $\wh\ba_2$
converge to the true parameter values $\ba_1$ and $\ba_2$. This result
is important for establishing the asymptotic
properties of $\wh\bb$ when the working models are possibly
misspecified.

%%%%%%%%%%%%%%%%%%%%%%%%
\subsubsection{Theoretical properties with parametric working models}
\label{sec:par_th}
%%%%%%%%%%%%%%%%%%%%%%%%

Given the wide flexibility of parametric working models, one natural
question is: how does our choice of working models impact the
consistency of the SPARCC estimator $\wh\bb$? The answer is that the
SPARCC estimator is doubly robust: it remains consistent when either
$\eta_1(x,\z)$ or  $\eta_2(c,\z)$ is misspecified and in this sense
has two ``safety nets."

\begin{pro}\label{pro:consistency}
Let  $\wh\bb$ be the solution to
$\sumi\S\eff^{*\dagger}(y_i,w_i,\delta_i,\z_i,\bb)=\0$, where
$\S\eff^{*\dagger}$ is as in \eqref{eq:seff}. Let $\bb_0$ denote the
true value of $\bb$ and assume $\bb_0\in\Omega$,  where $\Omega$ is
compact. Assume also 
$\E\{\sup_{\bb\in\Omega}\|\S\eff^{*\dagger}(Y,W,\Delta,\Z,\bb)\|_2\}<\infty$
(where $\|\cdot\|_2$ is the vector $L_2$ norm), and
$\E\{\partial\S\eff^{*\dagger}(Y,W,\Delta,\Z,\bb_0)/\partial\bb\trans\}$ 
is invertible. Then $\wh\bb$ is a consistent estimator for $\bb_0$ if either
$\eta_1^*(x,\z)$ or $\eta_2^\dagger(c,\z)$ is correctly specified. 
\end{pro}

The proof of Proposition \ref{pro:consistency} is in Section
\ref{sec:pro-consistency-proof} (Supplementary Material). Having
established double robustness, another natural question is: how does
our choice of working models impact the efficiency of the SPARCC
estimator $\wh\bb$? We answer this question by establishing the
asymptotic distribution of $\wh\bb$ in three scenarios:  when only
$\eta_1^*(x,\z)$ is correctly specified, when only
$\eta_2^\dagger(c,\z)$ is correctly specified, and when both  are
correctly specified.  

To show the asymptotic distribution in these three scenarios, we  make
the following assumptions. We consider the general setting where
$\wh\ba_1$ may depend on $\bb$, e.g., as in \eqref{eq:mle-fulldata};
we assume that $\wh\ba_1(\bb), \wh\ba_2$ have arbitrary influence
functions $\bphi_1(y,w,\delta,\z,\bb), \bphi_2(w,\delta,\z)$, respectively; we
assume regularity conditions \ref{con:bb}-\ref{con:B'}, which are
stated and justified in Section~\ref{sec:reg-parprofile}
(Supplementary Material); and we define the following matrices:
\bse
\A_1(\bb)&\equiv&\E[\partial\S\eff^{*\dagger}\{Y,W,\Delta,\Z,\ba_1^*(\bb),\ba_2^\dagger,\bb\}/\partial\ba_1\trans],\\
\A_2(\bb)&\equiv&\E[\partial\S\eff^{*\dagger}\{Y,W,\Delta,\Z,\ba_1^*(\bb),\ba_2^\dagger,\bb\}/\partial\ba_2\trans].
\ese
The asymptotic distribution  of $\wh\bb$ is summarized in Theorem \ref{th:parprofile}.
\begin{theorem}\label{th:parprofile}
Assume conditions \ref{con:bb}-\ref{con:B'}, and
\bse
\wh\ba_1(\bb)-\ba_1^*(\bb)&=&n^{-1}\sumi\bphi_1\{y_i,w_i,\delta_i,\z_i,\ba_1^*(\bb),\bb\}+o_p(n^{-1/2}),\\
\wh\ba_2-\ba_2^\dagger&=&n^{-1}\sumi\bphi_2(w_i,\delta_i,\z_i)+o_p(n^{-1/2}),
\ese
for some $\bphi_1, \bphi_2$ such that $\E[\bphi_1\{Y,W,\Delta,\Z,\ba_1^*(\bb),\bb\}]=\0, \E\{\bphi_2(W,\Delta,\Z)\}=\0$.
\begin{enumerate}[label=(\roman*)]
    \item
    If $\eta_1^*\{x,\z,\ba_1^*(\bb_0)\}=\eta_1(x,\z)$ (i.e., truth), then $\sqrt{n}(\wh\bb-\bb_0)\to \Normal\{\0,\B^{-1}\bSig^{\dagger}(\B^{-1})\trans\}$ in distribution as $n\to\infty$, where $\V_1\equiv-(\E[\partial\bphi_1\{Y,W,\Delta,\Z,\ba_1^*(\bb_0),\bb_0\}/\partial\ba_1\trans])^{-1}$ and 
    \bse
    \B&=&\E\left[\partial\S\eff^\dagger\{Y,W,\Delta,\Z,\ba_1^*(\bb_0),\ba_2^\dagger,\bb_0\}/\partial\bb\trans\right.\\
    &&\left.+\A_1(\bb_0)\V_1\partial\bphi_1\{Y,W,\Delta,\Z,\ba_1^*(\bb_0),\bb_0\}/\partial\bb\trans\right],\\
    \bSig^{\dagger}&\equiv&\var[\S\eff^\dagger\{Y,W,\Delta,\Z,\ba_1^*(\bb_0),\ba_2^\dagger,\bb_0\}\\
    &&+\A_1(\bb_0)\bphi_1\{Y,W,\Delta,\Z,\ba_1^*(\bb_0),\bb_0\}].
    \ese
    \item
    If $\eta_2^\dagger(c,\z,\ba_2^\dagger)=\eta_2(c,\z)$ (i.e., truth), then
    $\sqrt{n}(\wh\bb-\bb_0)\to \Normal\{\0,\B^{-1}\bSig^*(\B^{-1})\trans\}$
    in distribution as $n\to\infty$, where
    \bse
    \B&=&\E[\partial\S\eff^*\{Y,W,\Delta,\Z,\ba_1^*(\bb_0),\ba_2^\dagger,\bb_0\}/\partial\bb\trans],\\
    \bSig^*&\equiv&\var[\S\eff^*\{Y,W,\Delta,\Z,\ba_1^*(\bb_0),\ba_2^\dagger,\bb_0\}+\A_2(\bb_0)\bphi_2(W,\Delta,\Z)].
    \ese
    \item
    If $\eta_1^*\{x,\z,\ba_1^*(\bb_0)\}=\eta_1(x,\z)$ and
    $\eta_2^\dagger(c,\z,\ba_2^\dagger)=\eta_2(c,\z)$, then $\wh\bb$ is the
    semiparametric efficient estimator of $\bb_0$, i.e.,
    $\sqrt{n}(\wh\bb-\bb_0)\to \Normal(\0,[\E\{\S\eff^{\otimes2}(Y,W,\Delta,\Z,\bb_0)\}]^{-1})$
    in distribution as $n\to\infty$.
\end{enumerate}
\end{theorem}
The proof of Theorem \ref{th:parprofile} is in Section
\ref{sec:th-parprofile-proof} (Supplementary Material). 
Theorem \ref{th:parprofile} implies that, provided at least one of the
nuisance distributions is correctly specified, the SPARCC estimator
$\wh\bb$ is consistent at the parametric root-$n$ convergence rate.
Further, from Theorem \ref{th:parprofile} (iii), $\wh\bb$ achieves the
semiparametric efficiency bound when both
nuisance distributions are correctly specified. That is, $\wh\bb$ is
semiparametric efficient, in that it achieves the smallest possible
asymptotic variance among all semiparametric estimators, including the
complete case and inverse probability weighting estimators.   

These efficiency results combined with double robustness show that the
SPARCC estimator lies between the complete case estimator and the MLE on a
spectrum of robustness versus efficiency. On one end of that spectrum,
the complete case estimator is maximally robust (i.e., does not require correct specification of
$\eta_1^*(x,\z)$ or $\eta_2^\dagger(c,\z)$) but inefficient. On the
other end, the MLE, the root of
$\sumi\S_\bb(y_i,w_i,\delta_i,\z_i,\bb)=\0$, is maximally efficient, (i.e., achieves the
smallest possible variance among parametric estimators---the
Cram\'er-Rao lower bound---when $\eta_1^*(x,\z)$ is correctly
specified), but not robust in that it is inconsistent when $\eta_1^*(x,\z)$ is
misspecified. The SPARCC estimator strikes a balance between robustness and
efficiency: it is doubly robust to misspecification of
$\eta_1^*(x,\z)$ or $\eta_2^\dagger(c,\z)$, meaning it is less robust
than the complete case estimator but more robust than the MLE, and
is less efficient than the MLE
but more efficient than the complete case estimator.

Up to now, we have allowed $\wh\ba_1$ to depend on $\bb$. When it does
not, e.g., as in \eqref{eq:mle2}, the asymptotic distribution of
$\wh\bb$ simplifies as in Corollary \ref{cor:par}. Let the limit of $\wh\ba_1$ be
$\ba_1^*$. 
\begin{cor}\label{cor:par}
Assume conditions \ref{con:bb}-\ref{con:B'},
$\wh\ba_1-\ba_1^*=n^{-1}\sumi\bphi_1(w_i,\delta_i,\z_i)+o_p(n^{-1/2})$,
and
$\wh\ba_2-\ba_2^\dagger=n^{-1}\sumi\bphi_2(w_i,\delta_i,\z_i)+o_p(n^{-1/2})$
for some $\bphi_1, \bphi_2$ such that $\E\{\bphi_1(W,\Delta,\Z)\}=\0,
\E\{\bphi_2(W,\Delta,\Z)\}=\0$. 
\begin{enumerate}[label=(\roman*)]
    \item
    If $\eta_1^*(x,\z,\ba_1^*)=\eta_1(x,\z)$ (i.e., truth), then
    $\sqrt{n}(\wh\bb-\bb_0)\to
    \Normal\{\0,\B^{-1}\bSig^\dagger(\B^{-1})\trans\}$ in distribution
    as $n\to\infty$, where  
    \bse
    \B&=&\E\{\partial\S\eff^\dagger(Y,W,\Delta,\Z,\ba_1^*,\ba_2^\dagger,\bb_0)/\partial\bb\trans\},\\
    \bSig^\dagger&\equiv&\var\{\S\eff^\dagger(Y,W,\Delta,\Z,\ba_1^*,\ba_2^\dagger,\bb_0)+\A_1(\bb_0)\bphi_1(W,\Delta,\Z)\}.
    \ese
    \item
    If $\eta_2^\dagger(c,\z,\ba_2^\dagger)=\eta_2(c,\z)$ (i.e., truth), then
    $\sqrt{n}(\wh\bb-\bb_0)\to \Normal\{\0,\B^{-1}\bSig^*(\B^{-1})\trans\}$
    in distribution as $n\to\infty$, where
    \bse
    \B&=&\E\{\partial\S\eff^*(Y,W,\Delta,\Z,\ba_1^*,\ba_2^\dagger,\bb_0)/\partial\bb\trans\},\\
    \bSig^*&\equiv&\var\{\S\eff^*(Y,W,\Delta,\Z,\ba_1^*,\ba_2^\dagger,\bb_0)+\A_2(\bb_0)\bphi_2(W,\Delta,\Z)\}.
    \ese
    \item
    If $\eta_1^*(x,\z,\ba_1^*)=\eta_1(x,\z)$ and
    $\eta_2^\dagger(c,\z,\ba_2^\dagger)=\eta_2(c,\z)$, then $\wh\bb$ is the
    semiparametric efficient estimator of $\bb_0$, i.e.,
    $\sqrt{n}(\wh\bb-\bb_0)\to \Normal(\0,[\E\{\S\eff^{\otimes2}(Y,W,\Delta,\Z,\bb_0)\}]^{-1})$
    in distribution as $n\to\infty$.
\end{enumerate}
\end{cor}

The proof of Corollary \ref{cor:par} is in Section \ref{sec:cor-par-proof} (Supplementary Material).
The asymptotic variances of $\wh\bb$ presented in Theorem
\ref{th:parprofile} and Corollary \ref{cor:par} do not guarantee that
more efficient estimation of  
$\ba_1,\ba_2$ (i.e., using $(Y,W,\Delta,\Z)$ instead of $(W,\Delta,\Z)$) leads to a more efficient $\wh\bb$ when only $\eta_1^*(x,\z)$, or $,\eta_2^\dagger(c,\z)$ is correctly specified.
When both $\eta_1^*(x,\z)$ and $\eta_2^\dagger(c,\z)$ are correctly
specified, however, the variance of $\wh\ba_1, \wh\ba_2$ does not
impact the variance of $\wh\bb$. That is, $\wh\bb$ is semiparametric
efficient if $\wh\ba_1,\wh\ba_2$ are consistent at the parametric
convergence rate, regardless of the asymptotic variance of
$\wh\ba_1,\wh\ba_2$. 

%%%%%%%%%%%%%%%%%%%%%%%%
\subsection{Nonparametric/machine learning methods}
\label{sec:non}
%%%%%%%%%%%%%%%%%%%%%%%%

While the double robustness of the SPARCC estimator provides two
safety nets,  we still require at least one of the parametric working
models $\eta_1^*(x,\z),\eta_2^\dagger(c,\z)$ to be correctly
specified. There may be scenarios where neither can be correctly
specified, and, for these scenarios, we propose an implementation that
uses nonparametric/machine learning methods, completely bypassing
the challenge of correct parametric specification.

The implementation generally proceeds in the same way as with
parametric working models (Section~\ref{sec:par_est}), except that now
there is dependency on some potentially data-driven
$\wh\eta_1(x,\z),\wh\eta_2(c,\z)$. To emphasize this dependency, we
write  $\E_1$, $\E_2$, $\a$, and $\S\eff$ as $\wh\E_1$, $\wh\E_2$,
$\wh\a(x,\z,\bb)$, and $\S\eff(y,w,\delta,\z,\wh\E_1,\wh\E_2,\bb)$,
respectively. The use of nonparametric/machine learning methods to obtain the SPARCC estimator is summarized in Algorithm \ref{alg:non}.

\begin{algorithm}[H]
    \caption{SPARCC estimator with nonparametric/machine learning methods}
    \label{alg:non}

 	\begin{algorithmic}
        \STATE \textbf{Input}: data $\{(y_i,w_i,\delta_i,\z_i):i=1,\cdots,n\}$.
            \STATE \textbf{do}
            \STATE (a)  compute $\wh\E_1,\wh\E_2$ with respect to
            $\eta_1(x,\z), \eta_2(c,\z)$ 
              via nonparametric/machine learning methods,
            \STATE (b) obtain $\wh\a(x,\z_i,\bb)$ by solving \eqref{eq:solve-a} at $\z=\z_i~(i=1,\cdots,n)$,
            \STATE (c) obtain $\wt\bb$ by solving $\sumi\S\eff(y_i,w_i,\delta_i,\z_i,\wh\E_1,\wh\E_2,\bb)=\0$ with respect to $\bb$.
 		\STATE \textbf{Output}: $\wt\bb$.
    \end{algorithmic}
\end{algorithm}

%%%%%%%%%%%%%%%%%%%%%%%%
\subsubsection{Constructing nonparametric/machine learning estimators}%{\red data-driven estimators}}
\label{sec:non_est}
%%%%%%%%%%%%%%%%%%%%%%%%

Since $X$ and $C$ are right-censored variables, candidate 
estimators should adjust for this censoring. Many such
estimators exist, among which we highlight a few. When
$\z$ is low-dimensional
(i.e., the number of variables is small), we can use nonparametric
kernel-based estimators like the extension of the Kaplan-Meier
estimator from \cite{beran1981nonparametric} and
\cite{dabrowska1987non}, a recursive kernel-based estimator from
\cite{khardani2014nonparametric}, and a presmoothed kernel density
estimator from  \citet{cao_presmoothed_2004}.  When $\z$ is
high-dimensional, possible methods include a
regression-tree estimator \citep{molinaro2004tree}, a neural network-based estimator  \citep{ren2019deep}, and a Laplace approximation for
nonparametric double additive location-scale models
\citep{lambert_fast_2021}, among others. Regardless of the estimator adopted,
  the estimators $\wh\eta_1(x,\z)$ and 
$\wh\eta_2(c,\z)$ need to be consistent at the rate of $o_p(n^{-1/4})$
to meet the assumptions of Theorem~\ref{th:non}
(Section~\ref{sec:non_th}). The required convergence rate can be
achieved by diverse nonparametric and machine learning estimators
\citep{chernozhukov2018double}, shown by, e.g.,
\cite{wager2015adaptive} for a class of regression trees and random
forests, and \cite{chen1999improved} for a class of neural network
models.

For the simulation study and real data application in this paper,
where $\z$ is low-dimensional, we model $\eta_1(x,\z)$ and
$\eta_2(c,\z)$ using a B-spline density estimator that accounts for
right-censored data. We model $\eta_1(x,\z)=\ba_1(\z)^{\trans}\bB(x)$,
where $\bB(\cdot) = \{B_1(\cdot),\dots,B_m(\cdot)\}^{\trans}$ are
positive, smooth, piece-wise polynomial functions and  $\ba_1(\z) =
\{\alpha_{11}(\z),\dots,\alpha_{1m}(\z)\}^{\trans}$ are unknown
coefficients. A similar model is used for $\eta_2(c,\z)$. To ensure
that these coefficients lead to a valid density, i.e., a non-negative
function that integrates to one, $\ba_1(\z)$ is constrained to have
positive components satisfying $\ba_1(\z)^{\trans}\int\bB(x)dx=1$.  
The coefficients $\ba_1(\z)$ are then chosen to maximize the
log-likelihood of $(W,\Delta)$ given $\Z=\z$. That is, 
\bse
\wh\ba_1(\z) = \argmax_{\ba_1(\z)} \sumi I(\z_i=\z) \left[\delta_i\log
  \{\ba_1(\z)^{\trans}\bB(w_i)\}
+(1-\delta_i)\log\int_{w_i}^\infty \ba_1(\z)^{\trans}\bB(x)dx\right].
\ese

%%%%%%%%%%%%%%%%%%%%%%%%%%%%%%%%%%%%%%%%%%%%%%%%%%

%%%%%%%%%%%%%%%%%%%%%%%%
\subsubsection{Theoretical properties with nonparametric/machine
  learning estimators}
\label{sec:non_th}
%%%%%%%%%%%%%%%%%%%%%%%%

Whereas parametric working models $\eta_1^*(x,\z)$ and
$\eta_2^\dagger(c,\z)$ will not converge to the truth if the models
are misspecified, 
nonparametric estimators $\wh\eta_1(x,\z)$ and $\wh\eta_2(c,\z)$ are
guaranteed to converge to the truth under mild conditions. Assuming
the convergence happens at the rate of $o_p(n^{-1/4})$, we have
the following asymptotic results for $\wt\bb$, the SPARCC estimator
from Algorithm \ref{alg:non}. We assume regularity conditions
\ref{con:bb-non}-\ref{con:B-non}, which are listed and justified in
Section~\ref{sec:reg-non} (Supplementary Material). 
\begin{theorem}\label{th:non}
Assume  conditions \ref{con:bb-non}-\ref{con:B-non},
$\sup_{x,\z}|(\wh\eta_1-\eta_1)(x,\z)|=o_p(n^{-1/4})$, and
$\sup_{c,\z}|(\wh\eta_2-\eta_2)(c,\z)|=o_p(n^{-1/4})$. Then, $\wt\bb$
is a semiparametric efficient estimator of $\bb_0$, i.e.,
$\sqrt{n}(\wt\bb-\bb_0)\to
\Normal(\0,[\E\{\S\eff^{\otimes2}(Y,W,\Delta,\Z,\bb_0)\}]^{-1})$ in
distribution as $n\to\infty$. 
\end{theorem}

The proof  of Theorem \ref{th:non} is  in Section
\ref{sec:th-non-proof} (Supplementary Material). The asymptotic
properties in Theorem \ref{th:non} establish the consistency and
efficiency of the SPARCC estimator $\wt\bb$ when using nonparametric
models $\wh\eta_1(x,\z)$ and $\wh\eta_2(c,\z)$. Even though we allow
$\wh\eta_1(x,\z),\wh\eta_2(c,\z)$ to converge at the slower
convergence rate of $o_p(n^{-1/4})$, the SPARCC estimator $\wt\bb$ is
consistent at the faster  root-$n$ convergence rate. 
Intuitively, the convergence rate follows because the SPARCC estimator is built
using the efficient score vector $\S\eff$, which diminishes the
leading order error from $\wh\eta_1(x,\z)$ and $\wh\eta_2(c,\z)$.

The SPARCC estimator $\wt\bb$ is also semiparametric efficient. Recall
that the SPARCC estimator $\wh\bb$ using parametric working models
(Section \ref{sec:par_th}) achieves semiparametric efficiency, but
only when both nuisance distribution models are correctly
specified. Theorem \ref{th:non} presents a stronger result: using
nonparametric models does not impact the asymptotic variance of
$\wt\bb$. In other words, the asymptotic variance of $\wt\bb$ is
equivalent to that of an estimator constructed with the true but
unknown nuisance distributions. This result eliminates the need to
choose a correct parametric working model for $\eta_1^*(x,\z)$ or
$\eta_2^\dagger(c,\z)$. As long as the nonparametric models converge
at a rate of $o_p(n^{-1/4})$, which many do (Section
\ref{sec:non_est}), the SPARCC estimator is both consistent and
semiparametric efficient.

%%%%%%%%%%%%%%%%%%%%%%%%%%%%%%%%%%%%%%%%%%%%%%%%%%

\section{Simulation study}
\label{sec:sim}

We assessed the finite sample robustness and efficiency of the SPARCC
estimator and compared the results against those of other estimators
through a simulation study. For this study, $n=8000$ independent
copies of the complete data $(Y, X, C, Z)$ were generated. Here, $Z
\sim \textrm{Bernoulli}(0.5)$ is a binary fully observed covariate,
$X$ is a censored covariate with $X|Z \sim \textrm{beta}(\alpha_{11} +
\alpha_{12}Z, \alpha_{13} + \alpha_{14}Z)$, $C$ is the censoring
variable with $C|Z \sim \textrm{beta}(\alpha_{21} + \alpha_{22}Z,
\alpha_{23} + \alpha_{24}Z)$, and $Y$ is the outcome with $Y|X,Z \sim
\Normal(\beta_0 + \beta_1X + \beta_2Z, \sigma^2)$. From the complete
data $(Y, X, C, Z)$, the observed data $(Y, W, \Delta, Z)$ were
constructed using $W = \min(X, C)$ and $\Delta = I(X \leq C)$. The
parameter $\ba_1 = (\alpha_{11}, \alpha_{12}, \alpha_{13},
\alpha_{14})\trans = (1.5, 1, 2.5, -1)\trans$ characterizing $f_{X|Z}$
was held constant across all simulations, whereas the parameter
$\ba_2 = (\alpha_{21}, \alpha_{22}, \alpha_{23}, \alpha_{24})\trans$
for $f_{C|Z}$ was varied to yield a desired censoring
proportion $q=\textrm{P}(X>C) \in (0,1)$. The parameter of interest
was fixed at $\bb = (\beta_0, \beta_1, \beta_2, \log\sigma^2)\trans =
(1, 10, 2, 0)\trans$.

Four estimators were considered in our study: (i) the SPARCC
estimator, (ii) the MLE, (iii) the complete case estimator, and (iv)
the oracle estimator (which makes use of the unobserved $X$ for each
subject). We did not consider the inverse probability weighting
estimator because though it is as robust as the complete case estimator, it
is less efficient \citep{vazquez2024establishingparallelsdifferencesrightcensored}. We also did not consider an
imputation estimator, as it would be no more robust than the MLE and
less efficient.

The SPARCC estimator uses both nuisance distributions $\eta_1$ and
$\eta_2$, which were specified in one of five ways: using parametric
working models with (i) both $\eta_1$ and $\eta_2$ correctly
specified, (ii) only $\eta_1$ correctly specified, (iii) only $\eta_2$
correctly specified, or (iv) neither $\eta_1$ nor $\eta_2$ correctly
specified; or (v) using nonparametric models. Scenarios (i), (ii),
(iii), and (v) appropriately model nuisance distributions in the sense
that they lead to consistent estimation of $\bb$. The MLE only
uses the nuisance distribution $\eta_1$, which was specified in
one of three ways: using (i) a correct parametric working model, (ii)
an incorrect parametric working model, or (iii) a nonparametric
model. The complete case and oracle estimators do not use nuisance
distributions. To misspecify $\eta_1$, we assumed $X|Z \sim
\textrm{beta}(\alpha_{11}^*, \alpha_{12}^*)$; the other nuisance
distribution $f_{C|Z}$ was similarly misspecified as
$\textrm{beta}(\alpha_{21}^\dagger, \alpha_{22}^\dagger)$. Here, the
superscripts $^*$ and $^\dagger$ indicate that these posited
distributions are incorrect. Nonparametric models for $\eta_1$ and
$\eta_2$ used the B-spline estimator described in Section
\ref{sec:non_est}. The MLE with the nonparametric model for $\eta_1$
can be viewed as a more robust version of the estimators by
\cite{KongNan2016} and \cite{chen2022semiparametric} that model
$\eta_1$ semiparametrically. The oracle estimator was implemented only
as a benchmark, as it utilizes the censored $X$, which is not available
in practice. 

The variance of each estimator was estimated using the empirical
sandwich technique \citep{stefanski2002}. For the MLE with the
parametric or B-spline $\eta_1$ model, its asymptotic distribution
depends on uncertainty in the estimation of $\ba$. To account for this
dependence, the sandwich variance estimator used the stacked
estimating functions $\partial\log
f_{W,\Delta|Z}(w,\delta,z,\ba)/\partial\ba$ for the $X|Z$ model and
$\S_\bb(y,w,\delta,z,\bb,\ba)$ for the outcome model.

\begin{figure}
    \centering
    \includegraphics[width=0.7\textwidth]{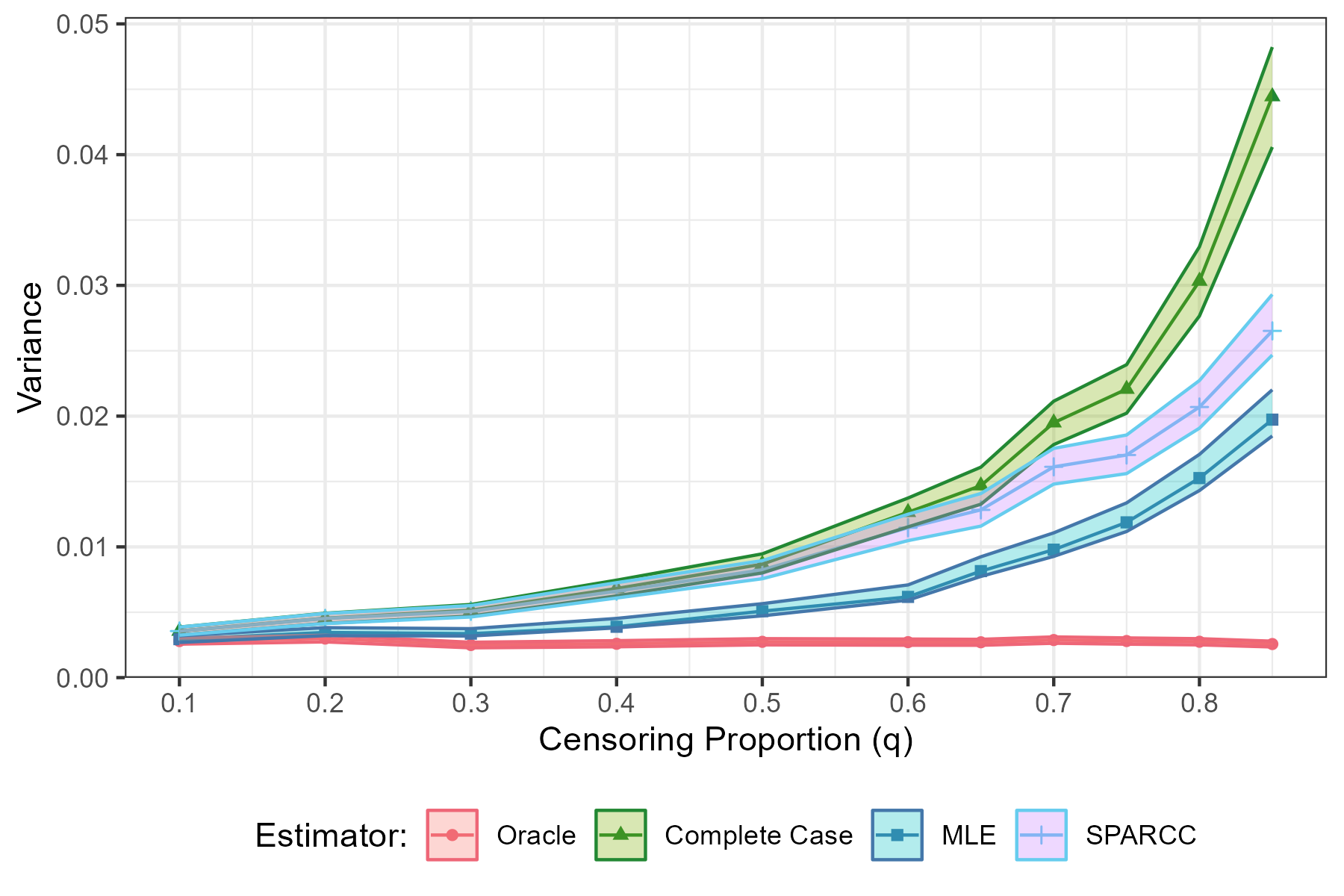}
    \caption{Empirical Variance (and 95\% Confidence Interval) of
      $\wh\beta_1$ by Censoring Proportion, based on 1000 Simulations
      per Setting}
    \label{fig:sim1_2}
\end{figure}

Table \ref{tab:sim1} shows the empirical bias, empirical standard
error, average estimated standard error, and 95\%  confidence interval
coverage probability for each of the four estimators of $\beta_1$, for
censoring proportions $q \in \{0.4, 0.8\}$, and based on $1000$
simulations per setting. Similar results for other components of $\bb$
are in Table \ref{tab:sim1-2}  (Supplementary Material).
\begin{table}[!ht]
    \centering
    \scalebox{0.91}
    {\begin{tabular}{lccrrrrrrrr}
\toprule
&\multicolumn{2}{c}{} & \multicolumn{4}{c}{$\mathbf{q = 0.4}$} & \multicolumn{4}{c}{$\mathbf{q = 0.8}$} \\
\cmidrule(l{3pt}r{3pt}){4-7} \cmidrule(l{3pt}r{3pt}){8-11}
\textbf{Estimator} & \textbf{$\mathbf{X|Z}$} & \textbf{$\mathbf{C|Z}$} & \textbf{Bias} & \textbf{ESE} & \textbf{ASE} & \textbf{Cov} & \textbf{Bias} & \textbf{ESE} & \textbf{ASE} & \textbf{Cov}\\
\midrule
SPARCC & Correct & Correct & 0.08 & 0.81 & 0.81 & 94.6 & 0.15 & 1.44 & 1.46 & 94.9\\
  & Correct & Incorrect & 0.09 & 0.81 & 0.81 & 94.6 & 0.15 & 1.44 & 1.46 & 94.8\\
  & Incorrect & Correct & 0.05 & 0.82 & 0.82 & 94.8 & -0.44 & 1.64 & 1.66 & 93.7\\
  & Incorrect & Incorrect & 0.06 & 0.82 & 0.82 & 94.5 & -0.46 & 1.64 & 1.66 & 93.7\\
  & Nonpar & Nonpar & 0.05 & 0.82 & 0.81 & 94.9 & -0.45 & 1.58 & 1.52 & 92.8\\
\addlinespace
MLE & Correct & - & 0.17 & 0.62 & 0.64 & 94.8 & 0.21 & 1.24 & 1.21 & 93.6\\
  & Incorrect & - & -4.49 & 0.63 & 0.66 & 0.0 & -17.84 & 1.71 & 1.76 & 0.0\\
  & Nonpar & - & -0.40 & 0.84 & 1.06 & 93.2 & -2.21 & 2.79 & 4.14 & 88.8\\
Complete Case & - & - & 0.06 & 0.83 & 0.82 & 94.8 & -0.03 & 1.74 & 1.76 & 95.1\\
Oracle & - & - & 0.03 & 0.51 & 0.52 & 95.2 & 0.00 & 0.52 & 0.52 & 94.6\\
\bottomrule
\end{tabular}}
\caption{Simulation results based on 1000 replicates per
  setting. $X|Z$ and $C|Z$: nuisance distribution model (correct
  parametric, incorrect parametric, Nonpar: nonparametric), $q$:
  censoring proportion; Bias: 10 times the empirical bias of parameter
  estimate; ESE: 10 times the empirical standard error of parameter
  estimate; ASE: 10 times the average (estimated) standard error; Cov:
  empirical percent coverage of 95$\%$ confidence interval. 
}
    \label{tab:sim1}
\end{table}
Figure \ref{fig:sim1_2} shows the empirical variance of the four estimators of $\beta_1$ over a sequence of censoring proportions $q\in[0.1,0.85]$, with the SPARCC estimator and MLE using correct parametric models for nuisance distributions.

When we used correct parametric working models for the nuisance
distributions,  all estimators had minimal bias ($\leq 0.021$ in
absolute value; Table \ref{tab:sim1}), but varying efficiency. The
oracle estimator, which is by
definition unaffected by censoring, had the lowest empirical standard
errors among all estimators regardless of censoring proportion
$q$. The MLE followed with the next lowest empirical standard errors,
then the SPARCC estimator, and finally the complete case estimator
(Figure \ref{fig:sim1_2}). This ordering highlights the SPARCC
estimator's semiparametric efficiency: it is higher than that of the
complete case estimator, which loses efficiency by discarding
incomplete cases, but somewhat lower than the MLE's maximal parametric
efficiency.

When we used an incorrect parametric working model for $\eta_1(x,\z)$,
the MLE had extreme bias and coverage probabilities of
zero. Modeling $\eta_1(x,\z)$ nonparametrically still resulted in
moderate bias of the MLE and coverage probability of the corresponding confidence interval significantly below the 95\% nominal level. Also
in this setting, the variance estimator for the MLE had several large
outliers, leading to an average estimated standard error much larger
than the empirical standard error (for $q=0.8$, the median estimated standard error of 2.64 was closer than the mean estimated standard error of 4.14 to the empirical mean of 2.79). In contrast, the SPARCC estimator
had low bias ($\leq 0.045$ in absolute value), had smaller empirical
standard errors than the complete case estimator, had near-matching
empirical and average estimated standard errors, and its confidence
interval achieved close to the nominal 95\% coverage probability in a
wide range of settings. These settings include when either one of
$\eta_1(x,\z)$ or $\eta_2(c,\z)$ had a correct parametric
specification, or when both $\eta_1(x,\z)$ and $\eta_2(c,\z)$ were
modeled nonparametrically. Interestingly, the SPARCC estimator
performed reasonably well even with incorrect parametric working
models for both $\eta_1(x,\z)$ and $\eta_2(c,\z)$. However, there are
no theoretical guarantees for such performance in general. 

All results suggest that the SPARCC estimator strikes a balance between
robustness and efficiency. It is more efficient than the complete case estimator and
remains consistent across a wider range of modeling scenarios than the
MLE. 

%%%%%%%%%%%%%%%%%%%%%%%%%%%%%%%%%%%%%%%%%%%%%%%%%%

\section{Application to Enroll-HD}
\label{sec:dat}

The SPARCC estimator was applied to data from Enroll-HD, a large
observational study of people living with, or at risk of developing,
Huntington's disease \citep{sathe_enroll-hd_2021}. We modeled symptom
severity as a function of time until diagnosis to learn how symptoms of Huntington's disease progress during the period leading up to a
diagnosis. During this pre-manifest period, symptoms can progress
rapidly, but have not yet significantly impacted patient function,
making this period a promising target for candidate interventions in
future clinical trials \citep{scahill_biological_2020}.  

In this analysis, we considered individuals who have at least 36 CAG repeats in their Huntingtin gene, making them genetically predisposed to Huntington’s disease. We analyzed symptom severity based on three common outcome metrics: (i) total motor score (TMS), (ii) symbol digit
modalities test (SDMT) score, and (iii) the composite Unified
Huntington Disease Rating Scale (cUHDRS) score. TMS is a measure of
motor function based on a battery of assessments, with a higher TMS
indicating more severe motor impairment
\citep{mestre_rating_2018}. The SDMT measures cognitive function based
on a participant's ability to match symbols and digits, with low SDMT
scores indicating high impairment \citep{ryan_normative_2020}. The
cUHDRS is a combined measure of cognitive, motor, and functional
symptoms based on several assessment scores, with a low cUHDRS score
corresponding to high impairment \citep{schobel_motor_2017}. For each
analysis, we adjusted for the genetic component of Huntington's
disease using the CAG-Age-product (CAP) score. This score, defined as
(age at study entry) $\times$ (CAG - 33.66), combines age and CAG
repeat length to measure the cumulative toxicity of mutant Huntingtin protein,
with higher CAP scores corresponding to older age and longer CAG
repeat length \citep{long_tracking_2014}. In this application, we
dichotomized the CAP score at 368---the cutoff for high vs. low-medium
CAP scores in \cite{long_tracking_2014}---to create a binary
uncensored covariate included in both the outcome model and the
nuisance models. Diagnosis of Huntington's disease was defined as the
first visit where a patient received a diagnostic confidence level
(DCL) of 4. DCL is a measure of how confident a clinician is that a
patient's motor symptoms represent Huntington's disease, rated on a
scale of 1-4, with a DCL of 4 corresponding to unequivocal confidence
\citep{hogarth_interrater_2005}. Using this definition, patients who
exit the study before receiving a DCL of 4 have a right-censored time
until diagnosis. 

At annual Enroll-HD visits, DCL, TMS, and SDMT data, as well as
data used to derive the cUHDRS score, are collected from study
participants. Since the SPARCC estimator is for cross-sectional data,
only the baseline visit data were used for model fitting. Note, however,
that subsequent visits were used to determine time of diagnosis or
study exit. Our sample was restricted to individuals with CAG repeat length $\geq 36$ who
were at least 18 years old, had not been diagnosed upon study entry
(i.e., had a DCL less than 4 at baseline), and had at least two annual
study visits. Based on these filters, the analysis data set had a
sample size of $n=4530$; among these, $q=81.9\%$ of participants had a
censored time until diagnosis ($q=63.7\%$ in the low-medium CAP group
and $q=93.0\%$ in the high CAP group). For each analysis, we excluded
participants with missing outcome values (0.7\% for TMS and SDMT, and
2.1\% for cUHDRS). The model used for each of the three outcomes was
\bse
\label{eq:hd-model}
    Y \sim N(\beta_0 + \beta_1 X + \beta_2 Z + \beta_3 XZ, \sigma^2),
\ese
where $Y$ is a given outcome, $X$ is time until diagnosis censored by
time until study exit $C$, and $Z$ is an indicator for a participant
having a high CAP score.

\begin{figure}[!ht]
    \centering
    \includegraphics[width=0.7\textwidth]{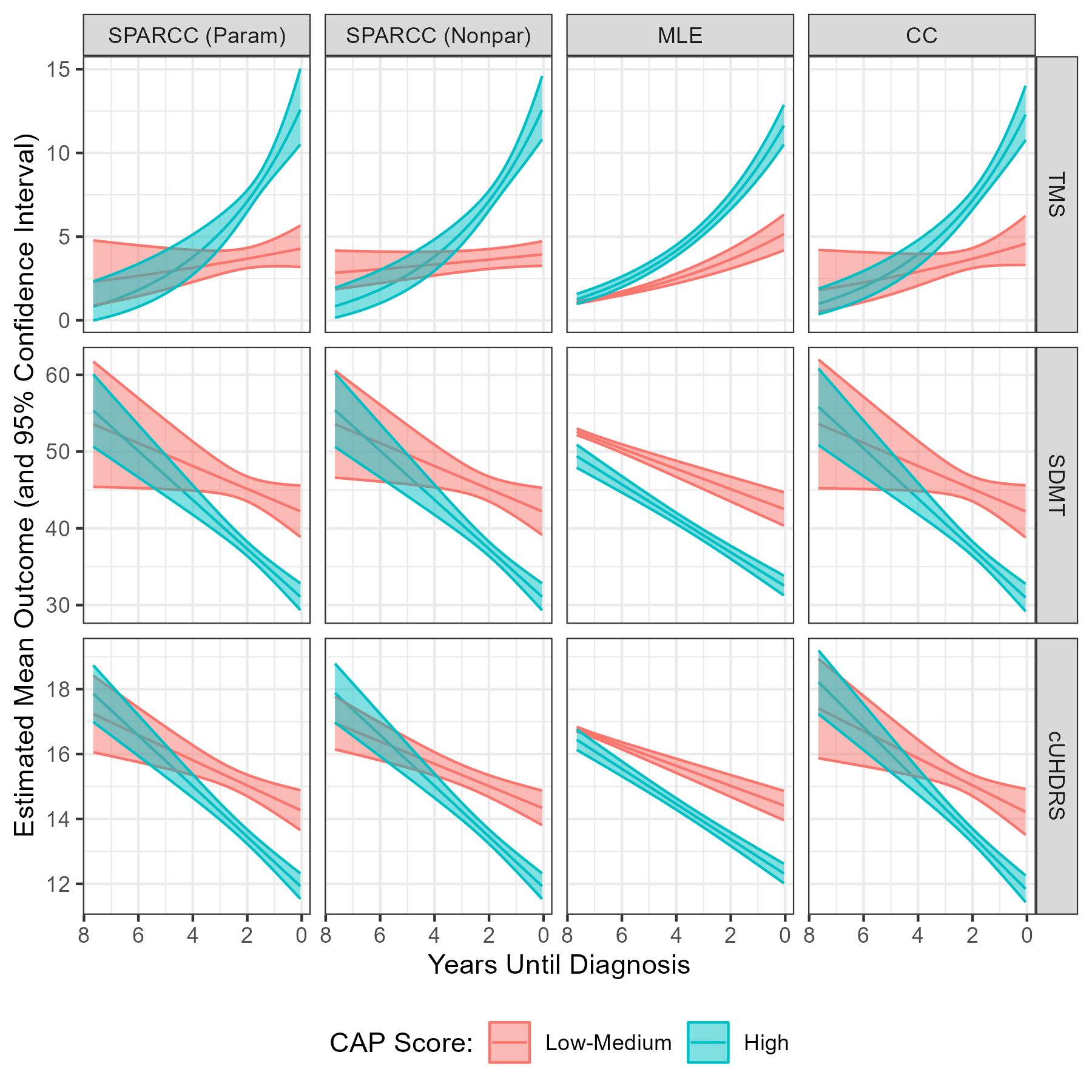}
    \caption{Estimated Mean Outcome vs. Time Until Diagnosis (in Years)
      for Three Outcomes, Stratified by CAP Score Group (Low-Medium
      and High). TMS, SDMT, and cUHDRS as in Table~\ref{tab:dat1}.} 
    \label{fig:dat1}
\end{figure}

We fit the model for each outcome using the SPARCC estimator with
parametric working models and nonparametric models for
$\eta_1(x,\z),\eta_2(c,\z)$. We compared the results to those from the
complete case estimator and the MLE with a parametric working model
for $\eta_1(x,\z)$. For parametric working models, we used the beta
distributions specified in Section~\ref{sec:sim}. For nonparametric
models, we used the B-spline density estimator from
Section~\ref{sec:non_est}. To facilitate nuisance modeling, $X$ and
$C$ were scaled based on the sample maximum of $W=\min(X,C)$ to be
within $(0,1)$. Since the TMS outcome is right-skewed, we transformed
it using $Y \mapsto \log(Y + 1)$; results are presented on the
original TMS scale. Estimated curves of mean outcome over time until
diagnosis and stratified by CAP group are plotted in
Figure~\ref{fig:dat1}. 

Estimates and standard errors for  all parameters are presented in Table~\ref{tab:dat1}.
  \begin{table}[!ht]
    \centering
   \scalebox{0.75}{
      \begin{tabular}{ccccrrrrr}
\toprule
\multicolumn{4}{c}{ } & \multicolumn{5}{c}{\textbf{Estimate (Standard Error)}} \\
\cmidrule(l{3pt}r{3pt}){5-9}
\textbf{Outcome} & \textbf{Estimator} & \textbf{X$|$Z} & \textbf{C$|$Z} & $\pmb{\beta_0}$ & $\pmb{\beta_1}$ & $\pmb{\beta_2}$ & $\pmb{\beta_3}$ & $\pmb{\log\sigma^2}$\\
\midrule
TMS & SPARCC & Param & Param & 1.66(0.12) & -0.47(0.38) & 0.94(0.15) & -1.54(0.54) & -0.30(0.05)\\
TMS & SPARCC & Nonpar & Nonpar & 1.60(0.08) & -0.25(0.18) & 1.01(0.10) & -1.75(0.34) & -0.30(0.05)\\
TMS & MLE & Param & - & 1.82(0.09) & -1.10(0.09) & 0.72(0.10) & -0.62(0.13) & -0.34(0.02)\\
TMS & Complete Case & - & - & 1.72(0.13) & -0.69(0.42) & 0.87(0.15) & -1.22(0.49) & -0.30(0.05)\\
\addlinespace\hline
SDMT & SPARCC & Param & Param & 42.22(1.71) & 11.35(5.58) & -11.13(1.97) & 12.92(6.65) & 4.94(0.05)\\
SDMT & SPARCC & Nonpar & Nonpar & 42.22(1.57) & 11.36(4.80) & -11.14(1.83) & 12.98(5.92) & 4.94(0.05)\\
SDMT & MLE & Param & - & 42.53(1.11) & 10.05(1.14) & -9.98(1.29) & 6.80(1.67) & 4.86(0.03)\\
SDMT & Complete Case & - & - & 42.21(1.74) & 11.40(5.73) & -11.23(1.96) & 13.45(6.60) & 4.94(0.05)\\
\addlinespace\hline
cUHDRS & SPARCC & Param & Param & 14.27(0.31) & 2.96(0.85) & -2.34(0.37) & 2.97(1.05) & 2.02(0.07)\\
cUHDRS & SPARCC & Nonpar & Nonpar & 14.34(0.27) & 2.61(0.60) & -2.41(0.34) & 3.34(0.87) & 2.02(0.07)\\
cUHDRS & MLE & Param & - & 14.41(0.23) & 2.36(0.24) & -2.09(0.28) & 1.77(0.35) & 1.49(0.04)\\
cUHDRS & Complete Case & - & - & 14.22(0.36) & 3.19(1.09) & -2.37(0.42) & 3.18(1.27) & 2.02(0.07)\\
\bottomrule
\end{tabular}
         }
           \caption{Results from the Enroll-HD application. $X|Z$ and $C|Z$: nuisance distribution model (Param: parametric, Nonpar: nonparametric), TMS: Total Motor Score, SDMT: Symbol Digit Modalities Test, cUHDRS: composite Unified Huntington's Disease Rating Scale.
           }
    \label{tab:dat1}
  \end{table}
All estimators showed a clear pattern of worsening symptom severity over
time in the period leading up to diagnosis (accounting for the fact
that increased TMS corresponds to increased impairment;
Figure~\ref{fig:dat1}). This result is consistent with earlier studies (e.g., \citet{long_tracking_2014})
that showed the progressive nature of Huntington's disease, which
based their analyses on the complete case estimator. While estimates the complete case and
SPARCC estimates from our analyses were similar, the SPARCC estimator offered smaller
standard errors (Table \ref{tab:dat1}), and thus more precise
inference. In the context of Huntington's disease, more precise
inference means that studies can make more efficient use of data and
have more power to identify candidate interventions as effective. This
is particularly important given the difficulty in recruiting patients
for long-term observational studies and the burden placed on those
patients.

Other symptom trajectory analyses in the Huntington's disease
literature have relied on an imputation estimator, plugging in an
\emph{estimated} time until diagnosis \citep{scahill_biological_2020,
  zhangetal2021}. Since these imputation approaches rely on a
correctly specified model for $\eta_1(x,\z)$, they are no more robust
than the MLE, and, moreover, they are generally less efficient than the
MLE. For these reasons, in our analysis we included the MLE---and not
an imputation estimator---as a comparator. If the model for
$\eta_1(x,\z)$ was in fact misspecified, then the MLE, with its low
standard error, gives a false sense of confidence of the symptom
trajectories in Figure \ref{fig:dat1}. In Huntington's disease
research, this false sense of confidence could lead to erroneously
concluding that a null intervention is effective, or that an effective
intervention is not effective. Because the SPARCC estimator is either
doubly robust to misspecification of parametric models for nuisance
distributions, or consistent with nonparametric nuisance models, it is
an overall more reliable choice than the MLE.

%%%%%%%%%%%%%%%%%%%%%%%%%%%%%%%%%%%%%%%%%%%%%%%%%%

\section{Conclusion}
\label{sec:con}

We proposed a semiparametric estimator that handles randomly
right-censored covariates robustly and efficiently. This SPARCC
estimator is the root of the efficient score vector $\S\eff$, which is derived using
semiparametric theory and is constructed by substituting models for
the nuisance distributions of the right-censored covariate,
$\eta_1(x,\z)$, and the censoring variable, $\eta_2(c,\z)$. These
nuisance distributions can be estimated with parametric working models
or using nonparametric/machine learning methods. The SPARCC estimator promises consistent
estimation that is doubly robust to misspecification of parametric working
models, and achieves semiparametric efficient estimation as long as
the nuisance distributions are estimated consistently and with a
sufficient convergence rate. The efficiency enables more precise
inference compared to existing semiparametric estimators in the
censored covariate literature. We also rigorously investigated large
sample properties under mild conditions and demonstrated asymptotic
normality, which can be used for inference on parameters of interest. Finally, we showed
how the SPARCC estimator can be applied to Huntington's disease data to
learn the progression patterns of impairments for different patient
groups without needing to accurately model both the time until
diagnosis and censoring distributions. By protecting against model
misspecification, the SPARCC estimator can help researchers better
understand progression patterns in other neurodegenerative diseases,
like sporadic Alzheimer's, where incomplete penetrance 
makes it even more difficult to appropriately model time until
diagnosis and censoring distributions.

Another doubly robust estimator in the censored covariate literature
is that from \citet{ahn2018cox}, yet that estimator is designed for a
different setting. The \cite{ahn2018cox} estimator is for a Cox
proportional hazards  model analyzing a time-to-event outcome, while
the SPARCC estimator is for a cross-sectional regression
model. Moreover, \cite{ahn2018cox} consider a binary covariate that
may change from 0 to 1, with time of change subject to censoring. This
is in contrast to the SPARCC estimator, which is for a continuous
censored covariate. 
Lastly, their estimator is doubly robust to misspecification of
$f_{\Delta|Y,\Z}$ or $f_{X|Y,\Z}$, whereas the SPARCC estimator
is doubly robust to misspecification of $f_{X|\Z}$ or $f_{C|\Z}$. As
there is no one-to-one correspondence between these sets of
distributions, the SPARCC estimator and its double robustness
property are unique.

There are several avenues for future work to build on the SPARCC estimator. One is to
investigate whether the double robustness property is also valid
when both nuisance models are estimated at a slower convergence rate. Similar investigations of double robustness with
nonparametrically estimated nuisance distributions have been done in
different problem contexts \citep{kennedy2017non, bennett2023source}.
A second is to extend the SPARCC estimator to the case where multiple
covariates are right-censored, or where one or more covariates is
interval-censored. A third is extending this cross-sectional regression model
to a longitudinal model. These extensions would improve the versatility of the SPARCC estimator and allow for robust and efficient estimation in a broader range of applications.

%%%%%%%%%%%%%%%%%%%%%%%%%%%%%%%%%%%%%%%%%%%%%%%%%%

\section*{Acknowledgements}

% These acknowledgements are required according to Enroll-HD publication guidelines (https://enroll-hd.org/for-researchers/enroll-hd-publication-policy/)

%This research was supported in part by National Institute of Neurological Disorders and Stroke grant R01NS131225 and National Institute of Environmental Health Sciences grant T32ES007018.

%This research was supported by the National Institute of Neurological Disorders and Stroke (grant R01NS131225) and the National Institute of Environmental Health Sciences (grant T32ES007018).

This research was supported by the National Institutes of Neurological Disorders and Stroke (grant R01NS131225) and of Environmental Health Sciences (grant T32ES007018).

%%%%%%%%%%%%%%%%%%%%%%%%%%%%%%%%%%%%%%%%%%%%%%%%%%

\section*{Data Availability Statement}

Data used in this work were generously provided by the by CHDI Foundation, Inc. The Enroll-HD study would not be possible without the vital contribution of the research participants and their families. These data are available upon request from Enroll-HD.

% \clearpage
\bibliographystyle{agsm}
\bibliography{indepcc}

%%%%%%%%%%%%%%%%%%%%%%%%%%%%%%%%%%%%%%%%%%%%%%%%%%

\clearpage
\pagenumbering{arabic}
{\centering
\section*{Supplementary material}
}
\setcounter{subsection}{0}\renewcommand{\thesubsection}{S.\arabic{subsection}}
\setcounter{subsubsection}{0}\renewcommand{\thesubsubsection}{S.\arabic{subsection}.\arabic{subsubsection}}
\setcounter{equation}{0}\renewcommand{\theequation}{S.\arabic{equation}}
\setcounter{lemma}{0}\renewcommand{\thelemma}{S.\arabic{lemma}}

\setcounter{table}{0}
	\renewcommand{\thetable}{S.\arabic{table}}

\setcounter{figure}{0}
	\renewcommand{\thefigure}{S.\arabic{figure}}

\setcounter{section}{0}
	\renewcommand{\thesection}{S.\arabic{section}}

%%%%%%%%%%%%%%%%%%%%%%%%%%%%%%%%%%%%%%%%%%%%%%%%%%

\subsection{Proof of Lemma \ref{lem:identifiability}}\label{sec:lem-identifiability-proof}

We will prove identifiability by showing that given identical likelihoods under two sets of parameters, the parameters are indeed identical. Let $(\fxz,\fcz,\fz,\bb)$ and $(\wt f_{X|\Z},\wt f_{C|\Z},\wt f_\Z,\wt\bb)$ be the two sets of parameters, and suppose the likelihoods are identical; that is,
\bse
&&\left\{\fyxz(y,w,\z,\bb)\fxz(w,\z)\int_w^\infty\fcz(c,\z)dc\right\}^{\delta}\\
&&\times\left\{\int_w^\infty\fyxz(y,x,\z,\bb)\fxz(x,\z)dx\fcz(w,\z)\right\}^{1-\delta}\fz(\z)\\
&=&\left\{\fyxz(y,w,\z,\wt\bb)\wt f_{X|\Z}(w,\z)\int_w^\infty\wt f_{C|\Z}(c,\z)dc\right\}^{\delta}\\
&&\times\left\{\int_w^\infty\fyxz(y,x,\z,\wt\bb)\wt f_{X|\Z}(x,\z)dx\wt f_{C|\Z}(w,\z)\right\}^{1-\delta}\wt f_\Z(\z)
\ese
for all $(y,w,\delta,\z)$. When $\delta=1$ and $\delta=0$, the above equality is equivalent to
\be
\fyxz(y,x,\z,\bb)\fxz(x,\z)S_{C|\Z}(x,\z)\fz(\z)
&=&\fyxz(y,x,\z,\wt\bb)\wt f_{X|\Z}(x,\z)\wt S_{C|\Z}(x,\z)\wt f_\Z(\z),\label{eq:fdelta1}\\
\int_c^\infty\fyxz(y,x,\z,\bb)\fxz(x,\z)dx\fcz(c,\z)\fz(\z)
&=&\int_c^\infty\fyxz(y,x,\z,\wt\bb)\wt f_{X|\Z}(x,\z)dx\wt f_{C|\Z}(c,\z)\wt f_\Z(\z),\n
\ee
where $S_{C|\Z}(x,\z)\equiv\int_x^\infty\fcz(c,\z)dc$ and $\wt S_{C|\Z}(x,\z)\equiv\int_x^\infty\wt f_{C|\Z}(c,\z)dc$. Integrating these two equalities with respect to $y$ from $-\infty$ to $\infty$ yields
\be
\fxz(x,\z)S_{C|\Z}(x,\z)\fz(\z)
&=&\wt f_{X|\Z}(x,\z)\wt S_{C|\Z}(x,\z)\wt f_\Z(\z),\label{eq:fxsc}\\
S_{X|\Z}(c,\z)\fcz(c,\z)\fz(\z)
&=&\wt S_{X|\Z}(c,\z)\wt f_{C|\Z}(c,\z)\wt f_\Z(\z),\label{eq:sxfc}
\ee
where $S_{X|\Z}(c,\z)\equiv\int_c^\infty\fxz(x,\z)dx$ and $\wt S_{X|\Z}(c,\z)\equiv\int_c^\infty\wt f_{X|\Z}(x,\z)dx$.

Let $t$ be any real scalar. Integrating the terms in \eqref{eq:fxsc} from $t$ to $\infty$ leads to
\bse
\int_t^\infty\fxz(x,\z)S_{C|\Z}(x,\z)dx\fz(\z)
=\int_t^\infty\wt f_{X|\Z}(x,\z)\wt S_{C|\Z}(x,\z)dx\wt f_\Z(\z).
\ese
Applying integration by parts to the above equality results in
\be\label{eq:FSFf}
&&\left\{-F_{X|\Z}(t,\z)S_{C|\Z}(t,\z)+\int_t^\infty F_{X|\Z}(x,\z)\fcz(x,\z)dx\right\}\fz(\z)\n\\
&=&\left\{-\wt F_{X|\Z}(t,\z)\wt S_{C|\Z}(t,\z)+\int_t^\infty\wt F_{X|\Z}(x,\z)\wt f_{C|\Z}(x,\z)dx\right\}\wt f_\Z(\z),
\ee
where $F_{X|\Z}(t,\z)\equiv\int_{-\infty}^t\fxz(x,\z)dx$ and $\wt F_{X|\Z}(t,\z)\equiv\int_{-\infty}^t\wt f_{X|\Z}(x,\z)dx$.
Applying a similar argument to the terms in  \eqref{eq:sxfc}, where we integrate the terms from $t$ to $\infty$, gives
\be\label{eq:Sf}
\int_t^\infty S_{X|\Z}(c,\z)\fcz(c,\z)dc\fz(\z)
=\int_t^\infty\wt S_{X|\Z}(c,\z)\wt f_{C|\Z}(c,\z)dc\wt f_\Z(\z).
\ee
Note that
\be\label{eq:SS}
&&\left\{-F_{X|\Z}(t,\z)S_{C|\Z}(t,\z)+\int_t^\infty F_{X|\Z}(x,\z)\fcz(x,\z)dx\right\}
+\int_t^\infty S_{X|\Z}(c,\z)\fcz(c,\z)dc\n\\
&=&-F_{X|\Z}(t,\z)S_{C|\Z}(t,\z)+\int_t^\infty\fcz(c,\z)dc\n\\
&=&\{1-F_{X|\Z}(t,\z)\}S_{C|\Z}(t,\z)\n\\
&=&S_{X|\Z}(t,\z)S_{C|\Z}(t,\z).
\ee
Combining \eqref{eq:FSFf}, \eqref{eq:Sf}, and \eqref{eq:SS} leads to
\be\label{eq:sxsc}
S_{X|\Z}(t,\z)S_{C|\Z}(t,\z)\fz(\z)=\wt S_{X|\Z}(t,\z)\wt S_{C|\Z}(t,\z)\wt f_\Z(\z).
\ee
Letting $t\to-\infty$ yields $f_\Z(\z)=\wt f_\Z(\z)$. Subsequently, with $f_\Z(\z)=\wt f_\Z(\z)$,   \eqref{eq:fxsc}, \eqref{eq:sxfc}, and \eqref{eq:sxsc} simplify to
\be
\fxz(x,\z)S_{C|\Z}(x,\z)
&=&\wt f_{X|\Z}(x,\z)\wt S_{C|\Z}(x,\z),\label{eq:iden-equalities}\\
S_{X|\Z}(c,\z)\fcz(c,\z)
&=&\wt S_{X|\Z}(c,\z)\wt f_{C|\Z}(c,\z),\nonumber\\
S_{X|\Z}(t,\z)S_{C|\Z}(t,\z)
&=&\wt S_{X|\Z}(t,\z)\wt S_{C|\Z}(t,\z)\nonumber.
\ee
Taking the ratio of the first and last equalities in \eqref{eq:iden-equalities} gives $\fxz(x,\z)/S_{X|\Z}(x,\z)=\wt f_{X|\Z}(x,\z)/\wt S_{X|\Z}(x,\z)$. Then, since 
\bse
1-\exp\left\{-\int_{-\infty}^x\frac{\fxz(t,\z)}{S_{X|\Z}(t,\z)}dt\right\}
&=&1-\exp\left\{\int_{-\infty}^x\frac{\partial\log S_{X|\Z}(t,\z)}{\partial t}dt\right\}\\
&=&1-\exp\{\log S_{X|\Z}(x,\z)\}\\
&=&F_{X|\Z}(x,\z),
\ese
we have $F_{X|\Z}(x,\z)=\wt F_{X|\Z}(x,\z)$, leading to $\fxz(x,\z)=\partial F_{X|\Z}(x,\z)/\partial x=\partial \wt F_{X|\Z}(x,\z)/\partial x=\wt f_{X|\Z}(x,\z)$. Likewise, the ratio of the second and last equalities in \eqref{eq:iden-equalities} leads to $\fcz(c,\z)=\wt f_{C|\Z}(c,\z)$. Up to now, assuming the likelihoods are identical, we have shown that $(\fxz,\fcz,\fz)$ and $(\wt f_{X|\Z},\wt f_{C|\Z},\wt f_{\Z})$ are in fact the same. Because these densities are the same,  \eqref{eq:fdelta1} is then simplified to $\fyxz(y,x,\z,\bb)=\fyxz(y,x,\z,\wt\bb)$. The only way for this equality to hold is $\bb=\wt\bb$. Therefore, all densities $\fxz$, $\fcz$, $\fz$, and $\bb$ are identifiable.
\hfill$\blacksquare$

%%%%%%%%%%%%%%%%%%%%%%%%%%%%%%%%%%%%%%%%%%%%%%%%%%

\subsection{Proof of Proposition \ref{pro:lambda}}\label{sec:pro-lambda-proof}

From \eqref{eq:model}, it is straightforward to derive that the nuisance scores associated with $\eta_1, \eta_2, \eta_3$, denoted respectively as $\S_1,\S_2,\S_3$, are
\bse
\S_1(y,w,\delta,\z)
&=&\delta\a_1(w,\z)+(1-\delta)\frac{\int_w^\infty\a_1(x,\z)\fyxz(y,x,\z,\bb)\eta_1(x,\z)dx}
{\int_w^\infty\fyxz(y,x,\z,\bb)\eta_1(x,\z)dx}\\
&=&\delta\a_1(w,\z)+(1-\delta)\frac{\E\{\I(X>w)\a_1(X,\z)\mid y,\z\}}{\E\{\I(X>w)\mid y,\z\}},\\
\S_2(y,w,\delta,\z)
&=&\delta\frac{\int_w^\infty\a_2(c,\z)\eta_2(c,\z)dc}{\int_w^\infty\eta_2(c,\z)dc}
+(1-\delta)\a_2(w,\z)\\
&=&\delta\frac{\E\{\I(C\ge w)\a_2(C,\z)\mid\z\}}{\E\{\I(C\ge w)\mid\z\}}
+(1-\delta)\a_2(w,\z),\\
\S_3(y,w,\delta,\z)
&=&\a_3(\z),
\ese
where $\a_1(x,\z)$, $\a_2(c,\z)$, $\a_3(\z)$ satisfy $\E\{\a_1(X,\z)\mid\z\}=\0$, $\E\{\a_2(C,\z)\mid\z\}=\0$, $\E\{\a_3(\Z)\}=\0$, respectively. Then it is immediate that the nuisance spaces associated with $\eta_1, \eta_2, \eta_3$ are $\Lambda_1, \Lambda_2, \Lambda_3$, respectively, as defined in Proposition \ref{pro:lambda}. Now, we show $\Lambda_1\perp\Lambda_2$, $\Lambda_1\perp\Lambda_3$, and $\Lambda_2\perp\Lambda_3$. 

First, $\Lambda_1\perp\Lambda_2$ since for $\S_1\in\Lambda_1$ and $\S_2\in\Lambda_2$,
\bse
&&\E\{\S_1(Y,W,\Delta,\Z)\S_2\trans(Y,W,\Delta,\Z)\}\\
&=&\E\left(\left[\Delta\a_1(W,\Z)+(1-\Delta) 
\frac{\E\{\I(X>W)\a_1(X,\Z)\mid Y,W,\Z\}}{\E\{\I(X>W)\mid Y,W,\Z\}}\right]\right.\\
&&\times\left.\left[\Delta\frac{\E\{\I(C\ge W)\a_2(C,\Z)\mid W,\Z\}}{\E\{\I(C\ge W)\mid W,\Z\}}
+(1-\Delta)\a_2(W,\Z)\right]\trans\right)\\
&=&\int\a_1(w,\z)\frac{\E\{\I(C\ge w)\a_2\trans(C,\z)\mid\z\}}{\E\{\I(C\ge w)\mid\z\}}
\left\{\fyxz(y,w,\z,\bb)\eta_1(w,\z)\int_w^\infty\eta_2(c,\z)dc\right\}\eta_3(\z)dydwd\z\\
&&+\int\frac{\E\{\I(X>w)\a_1(X,\z)\mid y,\z\}}{\E\{\I(X>w)\mid y,\z\}}\a_2\trans(w,\z) 
\left\{\int_w^\infty\fyxz(y,x,\z,\bb)\eta_1(x,\z)dx\,\eta_2(w,\z)\right\}\eta_3(\z)dydwd\z\\
&=&\int\I(c\ge w)\a_1(w,\z)\a_2\trans(c,\z)\eta_1(w,\z)\eta_2(c,\z)\eta_3(\z)dwdcd\z\\
&&+\int\I(x>w)\a_1(x,\z)\a_2\trans(w,\z)\eta_1(x,\z)\eta_2(w,\z)\eta_3(\z)dxdwd\z\\
&=&\int\a_1(x,\z)\a_2\trans(c,\z)\eta_1(x,\z)\eta_2(c,\z)\eta_3(\z)dxdcd\z\\
&=&\int\E\{\a_1(X,\z)\mid\z\}\E\{\a_2\trans(C,\z)\mid\z\}\eta_3(\z)d\z\\
&=&\0,
\ese
where the last equality holds because $\E\{\a_1(X,\z)\mid\z\}=\E\{\a_2(C,\z)\mid\z\}=\0$. Second, $\Lambda_1\perp\Lambda_3$ since for $\S_1\in\Lambda_1$ and $\S_3\in\Lambda_3$,
\bse
&&\E\{\S_1(Y,W,\Delta,\z)\mid\z\}\\
&=&\E\left[\Delta\a_1(W,\z)+(1-\Delta)
\frac{\E\{\I(X>W)\a_1(X,\z)\mid Y,W,\z\}}{\E\{\I(X>W)\mid Y,W,\z\}}\mid\z\right]\\
&=&\int\a_1(w,\z)\left\{\fyxz(y,w,\z,\bb)\eta_1(w,\z)\int_w^\infty\eta_2(c,\z)dc\right\}dydw\\
&&+\int\frac{\E\{\I(X>w)\a_1(X,\z)\mid y,\z\}}{\E\{\I(X>w)\mid y,\z\}}
\left\{\int_w^\infty\fyxz(y,x,\z,\bb)\eta_1(x,\z)dx\,\eta_2(w,\z)\right\}dydw\\
&=&\int\a_1(w,\z)\eta_1(w,\z)\int_w^\infty\eta_2(c,\z)dcdw\\
&&+\int\int_w^\infty\a_1(x,\z)\fyxz(y,x,\z,\bb)\eta_1(x,\z)dx\,\eta_2(w,\z)dydw\\
&=&\int\I(c\ge w)\a_1(w,\z)\eta_1(w,\z)\eta_2(c,\z)dwdc
+\int\I(x>w)\a_1(x,\z)\eta_1(x,\z)\eta_2(w,\z)dxdw\\
&=&\int\a_1(x,\z)\eta_1(x,\z)\eta_2(c,\z)dxdc\\
&=&\E\{\a_1(X,\z)\mid\z\}\\
&=&\0,
\ese
hence $\E\{\S_1(Y,W,\Delta,\Z)\S_3\trans(Y,W,\Delta,\Z)\}=\E[\E\{\S_1(Y,W,\Delta,\Z)\mid\Z\}\a_3\trans(\Z)]=\0$. Lastly, $\Lambda_2\perp\Lambda_3$ since for $\S_2\in\Lambda_2$ and $\S_3\in\Lambda_3$,
\bse
&&\E\{\S_2(Y,W,\Delta,\z)\mid\z\}\\
&=&\E\left[\Delta\frac{\E\{\I(C\ge W)\a_2(C,\z)\mid W,\z\}}{\E\{\I(C\ge W)\mid W,\z\}}+(1-\Delta)\a_2(W,\z)\mid\z\right]\\
&=&\int\frac{\E\{\I(C\ge w)\a_2(C,\z)\mid\z\}}{\E\{\I(C\ge w)\mid\z\}}
\left\{\fyxz(y,w,\z,\bb)\eta_1(w,\z)\int_w^\infty\eta_2(c,\z)dc\right\}dydw\\
&&+\int\a_2(w,\z)\left\{\int_w^\infty\fyxz(y,x,\z,\bb)\eta_1(x,\z)dx\,\eta_2(w,\z)\right\}dydw\\
&=&\int\I(c\ge w)\a_2(c,\z)\eta_1(w,\z)\eta_2(c,\z)dwdc
+\int\I(x>w)\a_2(w,\z)\eta_1(x,\z)\eta_2(w,\z)dxdw\\
&=&\int\a_2(c,\z)\eta_1(x,\z)\eta_2(c,\z)dxdc\\
&=&\E\{\a_2(C,\z)\mid\z\}\\
&=&\0,
\ese
hence $\E\{\S_2(Y,W,\Delta,\Z)\S_3\trans(Y,W,\Delta,\Z)\}=\E[\E\{\S_2(Y,W,\Delta,\Z)\mid\Z\}\a_3\trans(\Z)]=\0$. Therefore we have $\Lambda_1\perp\Lambda_2$, $\Lambda_1\perp\Lambda_3$, and $\Lambda_2\perp\Lambda_3$, which lead to the nuisance tangent space of \eqref{eq:model} being $\Lambda\equiv\Lambda_1\oplus\Lambda_2\oplus\Lambda_3$.
\hfill$\blacksquare$

%%%%%%%%%%%%%%%%%%%%%%%%%%%%%%%%%%%%%%%%%%%%%%%%%%

\subsection{Proof of Proposition \ref{pro:lambdaperp}}\label{sec:pro-lambdaperp-proof}
Let $\g(y,w,\delta,\z)\equiv\delta\g_1(y,w,\z)+(1-\delta)\g_0(y,w,\z)\in\H$, and recall $\Lambda_1, \Lambda_2, \Lambda_3$ given in Proposition \ref{pro:lambda}. First, $\g(y,w,\delta,\z)\perp\Lambda_3$ is equivalent to the following: for any $\a_3(\z)$ such that $\E\{\a_3(\Z)\}=\0$,
\be\label{eq:lambda3perp}
\0
&=&\E\{\g(Y,W,\Delta,\Z)\a_3\trans(\Z)\}\n\\
&=&\E[\E\{\I(X\le C)\g_1(Y,X,\Z)+\I(X>C)\g_0(Y,C,\Z)\mid\Z\}\a_3\trans(\Z)]\n\\
&=&\int\left[\int\left\{\I(x\le c)\g_1(y,x,\z)+\I(x>c)\g_0(y,c,\z)\right\}\fyxz(y,x,\z,\bb)\eta_1(x,\z)\eta_2(c,\z)dydxdc\right]\n\\
&&\times\a_3\trans(\z)\eta_3(\z)d\z.
\ee
Also, $\g(y,w,\delta,\z)\perp\Lambda_1$ is equivalent to the following: for any $\a_1(x,\z)$ such that $\E\{\a_1(X,\z)\mid\z\}=\0$,
\be\label{eq:lambda1perp}
&&\0
=\E\left(\g(Y,W,\Delta,\Z)
\left[\Delta\a_1(W,\Z)
+(1-\Delta)\frac{\E\{\I(X>W)\a_1(X,\Z)\mid Y,W,\Z\}}{\E\{\I(X>W)\mid Y,W,\Z\}}\right]\trans\right)\n\\
&=&\int\g_1(y,w,\z)\a_1\trans(w,\z)
\left\{\fyxz(y,w,\z,\bb)\eta_1(w,\z)\int_w^\infty\eta_2(c,\z)dc\right\}\eta_3(\z)dydwd\z\n\\
&&+\int\g_0(y,w,\z)\frac{\E\{\I(X>w)\a_1\trans(X,\z)\mid y,\z\}}{\E\{\I(X>w)\mid y,\z\}}
\left\{\int_w^\infty\fyxz(y,x,\z,\bb)\eta_1(x,\z)dx\,\eta_2(w,\z)\right\}\eta_3(\z)dydwd\z\n\\
&=&\int\{\I(x\le c)\g_1(y,x,\z)+I(x>c)\g_0(y,c,\z)\}\a_1\trans(x,\z)
\fyxz(y,x,\z,\bb)\eta_1(x,\z)\eta_2(c,\z)\eta_3(\z)dydxdcd\z\n\\
&=&\int\left[\int\{\I(x\le c)\g_1(y,x,\z)+I(x>c)\g_0(y,c,\z)\}
\fyxz(y,x,\z,\bb)\eta_2(c,\z)dydc\right]\n\\
&&\times\a_1\trans(x,\z)\eta_1(x,\z)\eta_3(\z)dxd\z.
\ee
Then, combining \eqref{eq:lambda3perp} and \eqref{eq:lambda1perp} and noting that $\a_1$ and $\a_3$ are arbitrary with $\E\{\a_1(X,\z)\mid\z\}=\0$ and $\E\{\a_3(\Z)\}=\0$, we have that $\g(y,w,\delta,\z)\perp(\Lambda_1\oplus\Lambda_3)$ is equivalent to
\be\label{eq:lambda13perp}
\0
&=&\int\{\I(x\le c)\g_1(y,x,\z)+I(x>c)\g_0(y,c,\z)\}\fyxz(y,x,\z,\bb)\eta_2(c,\z)dydc\n\\
&=&\E\{\I(x\le C)\g_1(Y,x,\z)+\I(x>C)\g_0(Y,C,\z)\mid x,\z\}.
\ee
Now, $\g(y,w,\delta,\z)\perp\Lambda_2$ is equivalent to the following: for any $\a_2(c,\z)$ such that $\E\{\a_2(C,\z)\mid\z\}=\0$,
\be\label{eq:lambda2perp}
\0
&=&\E\left(\g(Y,W,\Delta,\Z)
\left[\Delta\frac{\E\{\I(C\ge W)\a_2(C,\Z)\mid W,\Z\}}{\E\{\I(C\ge W)\mid W,\Z\}}
+(1-\Delta)\a_2(W,\Z)\right]\trans\right)\n\\
&=&\int\g_1(y,w,\z)\frac{\E\{\I(C\ge w)\a_2\trans(C,\z)\mid\z\}}{\E\{\I(C\ge w)\mid\z\}}
\left\{\fyxz(y,w,\z,\bb)\eta_1(w,\z)\int_w^\infty\eta_2(c,\z)dc\right\}\eta_3(\z)dydwd\z\n\\
&&+\int\g_0(y,w,\z)\a_2(w,\z)
\left\{\int_w^\infty\fyxz(y,x,\z,\bb)\eta_1(x,\z)dx\,\eta_2(w,\z)\right\}\eta_3(\z)dydwd\z\n\\
&=&\int\{\I(x\le c)\g_1(y,x,\z)+I(x>c)\g_0(y,c,\z)\}
\a_2\trans(c,\z)\fyxz(y,x,\z,\bb)\eta_1(x,\z)\eta_2(c,\z)\eta_3(\z)dydxdcd\z\n\\
&=&\int\left[\int\{\I(x\le c)\g_1(y,x,\z)+I(x>c)\g_0(y,c,\z)\}
\fyxz(y,x,\z,\bb)\eta_1(x,\z)dydx\right]\n\\
&&\times\a_2\trans(c,\z)\eta_2(c,\z)\eta_3(\z)dcd\z.
\ee
Then, combining \eqref{eq:lambda3perp} and \eqref{eq:lambda2perp}, since $\a_2$ and $\a_3$ are arbitrary with $\E\{\a_2(C,\z)\mid\z\}=\0$ and $\E\{\a_3(\Z)\}=\0$, $\g(y,w,\delta,\z)\perp(\Lambda_2\oplus\Lambda_3)$ is equivalent to
\be\label{eq:lambdap23perp}
\0
&=&\int\{\I(x\le c)\g_1(y,x,\z)+I(x>c)\g_0(y,c,\z)\}\fyxz(y,x,\z,\bb)\eta_1(x,\z)dydx\n\\
&=&\E\{\I(X\le c)\g_1(Y,X,\z)+I(X>c)\g_0(Y,c,\z)\mid c,\z\}.
\ee
Noting that the nuisance tangent space $\Lambda=\Lambda_1\oplus\Lambda_2\oplus\Lambda_3$ by Proposition \ref{pro:lambda}, \eqref{eq:lambda13perp} and \eqref{eq:lambdap23perp} lead to the result in Proposition \ref{pro:lambdaperp}.
\hfill$\blacksquare$

%%%%%%%%%%%%%%%%%%%%%%%%%%%%%%%%%%%%%%%%%%%%%%%%%%

\subsection{Proof of Proposition \ref{pro:efficientscore}}\label{sec:pro-efficientscore-proof}
We show that $\S\eff$ in Proposition \ref{pro:efficientscore} is the efficient score vector for $\bb$ by showing that $\S_\bb-\S\eff\in\Lambda$ and $\S\eff\in\Lambda^\perp$. First, we show that $\S_\bb-\S\eff\in\Lambda$. Note that, from the definition of $\a(x,\z,\bb)$ in Proposition \ref{pro:efficientscore},
\bse
\0&=&\E[\E\{\I(X\le C)\mid X,\z\}\a(X,\z,\bb)\mid\z]\\
&&+\E\left(\I(X>C)\frac{\E[\I(X>C)\{\a(X,\z,\bb)-\S_\bb^{\rm F}(Y,X,\z,\bb)\}\mid Y,C,\z]}{\E\{\I(X>C)\mid Y,C,\z\}}\mid\z\right)\\
&=&\E\{\I(X\le C)\a(X,\z,\bb)\mid\z\}\\
&&+\E\left(\E[\I(X>C)\{\a(X,\z,\bb)-\S_\bb^{\rm F}(Y,X,\z,\bb)\}\mid Y,C,\z]\mid\z\right)\\
&=&\E[\I(X\le C)\a(X,\z,\bb)+\I(X>C)\{\a(X,\z,\bb)-\S_\bb^{\rm F}(Y,X,\z,\bb)\}\mid\z]\\
&=&\E\{\a(X,\z,\bb)\mid\z\},
\ese 
where the last equality holds because
\bse
\E\{\I(X>C)\S_\bb^{\rm F}(Y,X,\z,\bb)\mid\z\}
=\E[\I(X>C)\E\{\S_\bb^{\rm F}(Y,X,\z,\bb)\mid X,\z\}\mid\z]
=\0.
\ese
Hence,
\be\label{eq:s1}
\delta\a(w,\z,\bb)+(1-\delta)\frac{\E\{\I(X>w)\a(X,\z,\bb)\mid y,\z\}}{\E\{\I(X>w)\mid y,\z\}}\in\Lambda_1
\ee
by the definition of $\Lambda_1$ in Proposition \ref{pro:lambda}. Then, \eqref{eq:s1} leads to $\S_\bb-\S\eff\in\Lambda_1\subset\Lambda$ by Proposition \ref{pro:lambda}.

Now we show that $\S\eff\in\Lambda^\perp$. First, $\S\eff\perp\Lambda_1$ since, for any $\b_1(x,\z)$ such that $\E\{\b_1(X,\z)\mid\z\}=\0$,
\bse
&&\E\left(\S\eff(Y,W,\Delta,\Z,\bb)
\left[\Delta\b_1(W,\Z)+(1-\Delta)\frac{\E\{\I(X>W)\b_1(X,\Z)\mid Y,W,\Z\}}{\E\{\I(X>W)\mid Y,W,\Z\}}\right]\trans\right)\\
&=&\E\left[\Delta\{\S_\bb^{\rm F}(Y,X,\Z,\bb)-\a(X,\Z,\bb)\}\b_1\trans(X,\Z)\right]\\
&&+\E\left((1-\Delta)\frac{\E[(1-\Delta)\{\S_\bb^{\rm F}(Y,X,\Z,\bb)-\a(X,\Z,\bb)\}\mid Y,C,\Z]}{\E(1-\Delta\mid Y,C,\Z)}
\frac{\E\{(1-\Delta)\b_1\trans(X,\Z)\mid Y,C,\Z\}}{\E(1-\Delta\mid Y,C,\Z)}\right)\\
&=&\E\left[\Delta\{\S_\bb^{\rm F}(Y,X,\Z,\bb)-\a(X,\Z,\bb)\}\b_1\trans(X,\Z)\right]\\
&&+\E\left(\frac{\E[(1-\Delta)\{\S_\bb^{\rm F}(Y,X,\Z,\bb)-\a(X,\Z,\bb)\}\mid Y,C,\Z]}{\E(1-\Delta\mid Y,C,\Z)}
\E\{(1-\Delta)\b_1\trans(X,\Z)\mid Y,C,\Z\}\right)\\
&=&\E\left\{\Delta\S_\bb^{\rm F}(Y,X,\Z,\bb)\b_1\trans(X,\Z)\right\}
-\E\left\{\Delta\a(X,\Z,\bb)\b_1\trans(X,\Z)\right\}\\
&&+\E\left((1-\Delta)\frac{\E[(1-\Delta)\{\S_\bb^{\rm F}(Y,X,\Z,\bb)-\a(X,\Z,\bb)\}\mid Y,C,\Z]}{\E(1-\Delta\mid Y,C,\Z)}\b_1\trans(X,\Z)\right)\\
&=&\E[\I(X\le C)\E\{\S_\bb^{\rm F}(Y,X,\Z,\bb)\mid X,\Z\}\b_1\trans(X,\Z)]-\E\Bigg[\Bigg\{\E\{\I(X\le C)\mid X,\Z\}\a(X,\Z,\bb)\\
&&-\E\left(\I(X>C)\frac{\E[\I(X>C)\{\S_\bb^{\rm F}(Y,X,\Z,\bb)-\a(X,\Z,\bb)\}\mid Y,C,\Z]}
{\E\{\I(X>C)\mid Y,C,\Z\}}\mid X,\Z\right)\Bigg\}\b_1\trans(X,\Z)\Bigg]\\
&=&\0,
\ese
where the last equality holds by $\E\{\S_\bb^{\rm F}(Y,x,\z,\bb)\mid x,\z\}=\0$ and the definition of $\a(x,\z,\bb)$ in Proposition \ref{pro:efficientscore}. Second, $\S\eff\perp\Lambda_2$ since, for any $\b_2(c,\z)$ such that $\E\{\b_2(C,\z)\mid\z\}=\0$,
\bse
&&\E\left(\S\eff(Y,W,\Delta,\Z,\bb)
\left[\Delta\frac{\E\{\I(C\ge W)\b_2(C,\Z)\mid W,\Z\}}{\E\{\I(C\ge W)\mid W,\Z\}}+(1-\Delta)\b_2(W,\Z)\right]\trans\right)\\
&=&\E\left[\Delta\S_\bb^{\rm F}(Y,X,\Z,\bb)\frac{\E\{\Delta\b_2\trans(C,\Z)\mid X,\Z\}}{\E(\Delta\mid X,\Z)}\right]\\
&&+\E\left[(1-\Delta)\frac{\E\{(1-\Delta)\S_\bb^{\rm F}(Y,X,\Z,\bb)\mid Y,C,\Z\}}{\E(1-\Delta\mid Y,C,\Z)}\b_2\trans(C,\Z)\right]\\
&=&\E\left\{\Delta\S_\bb^{\rm F}(Y,X,\Z,\bb)\b_2\trans(C,\Z)
+(1-\Delta)\S_\bb^{\rm F}(Y,X,\Z,\bb)\b_2\trans(C,\Z)\right\}\\
&=&\E[\E\{\S_\bb^{\rm F}(Y,X,\Z,\bb)\mid X,\Z\}\b_2\trans(C,\Z)]\\
&=&\0,
\ese
where the first equality holds by \eqref{eq:s1} and $\Lambda_2\perp\Lambda_1$ from Proposition \ref{pro:lambda}, and the last equality holds by $\E\{\S_\bb^{\rm F}(Y,x,\z,\bb)\mid x,\z\}=\0$. Lastly, we show that $\S\eff\perp\Lambda_3$. Note that $\S_\bb\perp\Lambda_3$ since
\bse
&&\E\{\S_\bb(Y,W,\Delta,\z,\bb)\mid\z\}\\
&=&\E\left[\Delta\S_\bb^{\rm F}(Y,X,\z,\bb)
+(1-\Delta)\frac{\E\{(1-\Delta)\S_\bb^{\rm F}(Y,X,\z,\bb)\mid Y,C,\z\}}{\E(1-\Delta\mid Y,C,\z)}\mid\z\right]\\
&=&\E\left\{\Delta\S_\bb^{\rm F}(Y,X,\z,\bb)+(1-\Delta)\S_\bb^{\rm F}(Y,X,\z,\bb)\mid\z\right\}\\
&=&\E[\E\{\S_\bb^{\rm F}(Y,X,\z,\bb)\mid X,\z\}\mid\z]\\
&=&\0,
\ese
hence
$\E\{\S_\bb(Y,W,\Delta,\Z)\b_3\trans(\Z)\}=\E[\E\{\S_\bb(Y,W,\Delta,\Z)\mid\Z\}\b_3\trans(\Z)]=\0$
for any $\b_3(\z)$ such that $\E\{\b_3(\Z)\}=\0$. Then,
$\S\eff\perp\Lambda_3$ by \eqref{eq:s1} and
$\Lambda_1\perp\Lambda_3$ from Proposition
\ref{pro:lambda}. Therefore, $\S\eff$ is the efficient score vector for
$\bb$. 
\hfill$\blacksquare$

%%%%%%%%%%%%%%%%%%%%%%%%%%%%%%%%%%%%%%%%%%%%%%%%%%

\subsection{Numerical Solution for the Efficient Score Vector}\label{sec:num-soln-Seff}

We show how to numerically solve the efficient score vector
$\S\eff^{*\dagger}(y,w,\delta,\z,\bb)$ given in Proposition
\ref{pro:efficientscore}. The first step of this procedure is to find
the solution $\a(x,z,\bb)$ to the integral equation \eqref{eq:solve-a}. Roughly, this solution is found by discretizing the integrals with respect to $x$, which converts the integral equation into a system of linear equations that can be directly solved. This discretization is performed separately at each unique observed value of $\z$.

More formally, consider a quadrature rule $\{(x_j,r_j):j=1,\dots,m\}$ that assigns point mass $r_j(\z) = \eta_1(x_j,\z) / \sum_{k=1}^m \eta_1(x_k,\z)$ to each of $m$ nodes $x_j$ in the support of $X|\z$. In practice, we can choose the support based on some combination of the observed $W|\z$, the estimated distribution of $X|\z$, and prior knowledge about $X$. Define the $p\times m$ matrix $\A(\z,\bb)=\{\a(x_1,\z,\bb)r_1(\z),\dots,\a(x_m,\z,\bb)r_m(\z)\}$, with columns corresponding to the values of $\a(x_j,\z,\bb)r_j(\z)$ evaluated at each node $x_j$. Then, using this quadrature rule, we can rewrite the left-hand side ($\textrm{LHS}_j$) of \eqref{eq:solve-a} evaluated at $x=x_j$ as follows:
\be
\rm{LHS}_j&=&\E\{\I(x_j\le C)\mid\z\}\a(x_j,\z,\bb) +
\E\left[\I(x_j>C)\frac{\E\{\I(X>C)\a(X,\z,\bb)\mid Y,C,\z\}}
{\E\{\I(X>C)\mid Y,C,\z\}}\mid x_j,\z\right]\nonumber\\
&=&\a(x_j,\z,\bb)\int_{x_j}^{\infty}\eta_2(c,\z)dc
+\int\int\frac
{\sum_{k=1}^{m}\I(x_k>c)\a(x_k,\z,\bb)r_k(\z)f_{Y|X,\Z}(y,x_k,\z,\bb)}
{\sum_{l=1}^{m}\I(x_l>c)r_l(\z)f_{Y|X,\Z}(y,x_l,\z,\bb)}\nonumber\\
&&\times\I(x_j>c)f_{Y|X,\Z}(y,x_j,\z,\bb)\eta_2(c,\z)dydc\nonumber\\
&=&\a(x_j,\z,\bb)r_j(\z)
\underbrace{\frac{1}{r_j(\z)}\int_{x_j}^{\infty}\eta_2(c,\z)dc}_{\equiv d_j(\z)}+\sum_{k=1}^{m}\a(x_k,\z,\bb)r_k(\z)\nonumber\\
&&\times\underbrace{\int\int\frac
{\I(x_j\wedge x_k>c)f_{Y|X,\Z}(y,x_k,\z,\bb)f_{Y|X,\Z}(y,x_j,\z,\bb)}
{\sum_{l=1}^{m}\I(x_l>c)r_l(\z)f_{Y|X,\Z}(y,x_l,\z,\bb)}
\eta_2(c,\z)dydc}_{\equiv M_{jk}(\z,\bb)}\nonumber\\
&=&\a(x_j,\z,\bb)r_j(\z)d_j(\z)+\sum_{k=1}^{m}\a(x_k,\z,\bb)r_k(\z)M_{jk}(\z)\nonumber\\
&=&\a(x_j,\z,\bb)r_j(\z)d_j(\z)+\A(\z,\bb)\pmb{M}_j(\z)\nonumber,
\ee
where $x_j\wedge x_k$ denotes the minimum of $x_j$ and $x_k$ and $\pmb{M}_j(\z)=\{M_{j1}(\z),\dots,M_{jm}(\z)\}\trans$ is an $m \times 1$ vector. Then the right-hand side ($\textrm{RHS}_j$) of \eqref{eq:solve-a} evaluated at $x=x_j$ is
\be
\rm{RHS}_j&=&\E\left[\I(x_j>C)\frac{\E\{\I(X>C)\S_\bb^{\rm F}(Y,X,\z,\bb)\mid Y,C,\z\}}{\E\{\I(X>C)\mid Y,C,\z\}}\mid x_j,\z\right]\nonumber\\
&=&\int\int\frac
{\sum_{k=1}^{m}\I(x_k>c)\S_\bb^{\rm F}(y,x_k,\z,\bb)r_k(\z)f_{Y|X,\Z}(y,x_k,\z,\bb)}
{\sum_{l=1}^{m}\I(x_l>c)r_l(\z)f_{Y|X,\Z}(y,x_l,\z,\bb)}\nonumber\\
&&\times\I(x_j>c)f_{Y|X,\Z}(y,x_j,\z,\bb)\eta_2(c,\z)dydc\nonumber\\
&\equiv&\pmb{b}_j(\z,\bb).\nonumber
\ee
Then, setting $\rm{LHS}_j=\rm{RHS}_j$ for $j=1,\dots,m$ gives
\be
&&\A(\z,\bb)
\left(\begin{bmatrix}
    d_1(\z) & & \\
    & \ddots & \\
    & & d_m(\z)
\end{bmatrix}+
\begin{bmatrix}
    \vert & & \vert \\
    \pmb{M}_1(\z,\bb) & \dots & \pmb{M}_m(\z,\bb)   \\
    \vert & & \vert
\end{bmatrix}\right)\nonumber
=\begin{bmatrix}
    \vert & & \vert \\
    \pmb{b}_1(\z,\bb) & \dots & \pmb{b}_m(\z,\bb)   \\
    \vert & & \vert
\end{bmatrix}\nonumber\\
&&\iff\A(\z,\bb)\left\{\pmb{D}(\z)+\pmb{M}(\z,\bb)\right\}=\pmb{B}(\z,\bb),\nonumber\\
&&\iff\A(\z,\bb) = \pmb{B}(\z,\bb)\left\{\pmb{D}(\z)+\pmb{M}(\z,\bb)\right\}^{-1},\nonumber
\ee
where $\pmb{D}(\z)$ is a diagonal matrix with diagonal entries $d_j(\z)$, $\pmb{M}(\z,\bb)=\{\pmb{M}_1(\z,\bb) \dots \pmb{M}_m(\z,\bb)\}$ is a symmetric matrix, and $\pmb{B}(\z,\bb)=\{\pmb{b}_1(\z,\bb) \dots \pmb{b}_m(\z,\bb)\}$. Since $D(\z)$ and $\pmb{M}(\z,\bb)$ are both symmetric, their sum is also symmetric and can thus be inverted stably and efficiently using the Cholesky decomposition.

The solution $\A(\z,\bb)$ provides approximations for $\a(x_j,\z,\bb)r_j(\z)$ evaluated at the nodes $x_j$ ($j=1,\dots,m$), from which an approximation for $\a(x_j,\z,\bb)$ can be recovered by dividing by $r_j(\z)$. Recall, however, that $\S\eff(y_i,w_i,\delta_i,\z_i,\bb)$ involves $\a(x_i,\z,\bb)$ evaluated at each $x_i$ with $\delta_i=1$. For such $x_i$ that are not nodes, we can interpolate $\a(x_i,\z,\bb)$ using the closest nodes $x_j, x_k$ on either side of $x_i$. In particular,
\bse
\frac{\a(x_i,\z,\bb)\eta_1(x_i,\z)}{\sum_{k=1}^m \eta_1(x_k,\z)}  &\approx&
\frac{\a(x_j,\z,\bb)r_j(\z)(x_k-x_i) + \a(x_k,\z,\bb)r_k(\z)(x_j-x_i)}{x_k-x_j} \\
\implies \a(x_i,\z,\bb) &\approx&
\frac{\sum_{k=1}^m \eta_1(x_k,\z)}{\eta_1(x_i,\z)}
\left\{\frac{\a(x_j,\z,\bb)r_j(\z)(x_k-x_i) + \a(x_k,\z,\bb)r_k(\z)(x_j-x_i)}{x_k-x_j}\right\}.
\ese
Finally, note that the components of $\pmb{B}(\z,\bb), \pmb{D}(\z)$, and $\pmb{M}(\z,\bb)$ involve integrals with respect to $c$ and $y$. These integrals can generally be approximated using Simpson's rule. If $C|\z$ or $Y|x,\z$ follow particular distributions (e.g., normal), then these integrals can be computed more efficiently using other quadrature rules (e.g., Guass-Hermite).

%%%%%%%%%%%%%%%%%%%%%%%%%%%%%%%%%%%%%%%%%%%%%%%%%%

\subsection{Proof of Proposition \ref{pro:consistency}}\label{sec:pro-consistency-proof}
We show the consistency of $\wh\bb$ following Theorem 2.6 of
\cite{newey1994large}. The existence of
$\partial\S\eff^{*\dagger}(y,w,\delta,\z,\bb)/\partial\bb\trans$
automatically implies that $\S\eff^{*\dagger}(y,w,\delta,\z,\bb)$ is
continuous with respect to $\bb$. In addition, the invertibility of
$\E\{\partial\S\eff^{*\dagger}(Y,W,\Delta,\Z,\bb_0)/\partial\bb\trans\}$
implies that, in the neighborhood of $\bb_0$, the solution to
$\E\{\S\eff^{*\dagger}(Y,W,\Delta,\Z,\bb)\}=\0$ is unique if it
exists. Hence it suffices to show that
$\E\{\S\eff^{*\dagger}(Y,W,\Delta,\Z,\bb_0)\}=\0$ given either
$\eta_1^*(x,\z)=\eta_1(x,\z)$ or
$\eta_2^\dagger(c,\z)=\eta_2(c,\z)$. Since
$\E\{\S\eff(Y,W,\Delta,\Z,\bb_0)\}=\0$ by the definition of $\S\eff$,
we only need to show $\E\{\S\eff^*(Y,W,\Delta,\Z,\bb_0)\}=\0$ and
$\E\{\S\eff^\dagger(Y,W,\Delta,\Z,\bb_0)\}=\0$. 

First, we will show $\E\{\S\eff^*(Y,W,\Delta,\z,\bb_0)\mid\z\}=\0$. This is because
\bse
&&\E\{\S\eff^*(Y,W,\Delta,\z,\bb_0)\mid\z\}\\
&=&\E\{\I(X\le C)\S_\bb^{\rm F}(Y,X,\z,\bb_0)\mid\z\}
-\E\{\I(X\le C)\a^*(X,\z,\bb_0)\mid\z\}\\
&&+\E\left[\I(X>C)\frac{\E^*\{\I(X>C)\S_\bb^{\rm F}(Y,X,\z,\bb_0)\mid Y,C,\z\}}
{\E^*\{\I(X>C)\mid Y,C,\z\}}\mid\z\right]\\
&&-\E\left[\I(X>C)\frac{\E^*\{\I(X>C)\a^*(X,\z,\bb_0)\mid Y,C,\z\}}
{\E^*\{\I(X>C)\mid Y,C,\z\}}\mid\z\right]\\
&=&\E[\I(X\le C)\E\{\S_\bb^{\rm F}(Y,X,\z,\bb_0)\mid X,\z\}\mid\z]\\
&=&\0,
\ese
where the second equality holds by
\eqref{eq:solve-a} with parametric working models under $\eta_2^\dagger(c,\z)=\eta_2(c,\z)$, and the last
equality holds by $\E\{\S_\bb^{\rm F}(Y,x,\z,\bb_0)\mid x,\z\}=\0$.

Now, we will show $\E\{\S\eff^\dagger(Y,W,\Delta,\z,\bb_0)\mid\z\}=\0$. Note that
\bse
&&\E\{\S\eff^\dagger(Y,W,\Delta,\z,\bb_0)\mid\z\}\\
&=&\E\left[\Delta\{\S_\bb^{\rm F}(Y,X,\z,\bb_0)-\a^\dagger(X,\z,\bb_0)\}\mid\z\right]\\
&&+\E\left((1-\Delta)\frac{\E[(1-\Delta)\{\S_\bb^{\rm F}(Y,X,\z,\bb_0)-\a^\dagger(X,\z,\bb_0)\}\mid Y,C,\z]}
{\E(1-\Delta\mid Y,C,\z)}\mid\z\right)\\
&=&\E\left[\Delta\{\S_\bb^{\rm F}(Y,X,\z,\bb_0)-\a^\dagger(X,\z,\bb_0)\}
+(1-\Delta)\{\S_\bb^{\rm F}(Y,X,\z,\bb_0)-\a^\dagger(X,\z,\bb_0)\}\mid\z\right]\\
&=&\E\left\{\S_\bb^{\rm F}(Y,X,\z,\bb_0)-\a^\dagger(X,\z,\bb_0)\mid\z\right\}\\
&=&\E\left\{\a^\dagger(X,\z,\bb_0)\mid\z\right\},
\ese
where the last equality holds by $\E\{\S_\bb^{\rm F}(Y,x,\z,\bb_0)\mid x,\z\}=\0$. Also, \eqref{eq:solve-a} with parametric working models under $\eta_1^*(x,\z)=\eta_1(x,\z)$ leads to
\bse
\0
&=&\E[\E\{\I(X\le C)\mid X,\z\}\a^\dagger(X,\z,\bb_0)\mid\z]\\
&&+\E\left(\I(X>C)\frac{\E[\I(X>C)\{\a^\dagger(X,\z,\bb_0)-\S_\bb^{\rm F}(Y,X,\z,\bb_0)\}\mid Y,C,\z]}{\E\{\I(X>C)\mid Y,C,\z\}}\mid\z\right)\\
&=&\E\left[\I(X\le C)\a^\dagger(X,\z,\bb_0)
+\I(X>C)\{\a^\dagger(X,\z,\bb_0)-\S_\bb^{\rm F}(Y,X,\z,\bb_0)\}\mid\z\right]\\
&=&\E\left[\a^\dagger(X,\z,\bb_0)
+\I(X>C)\E\{\S_\bb^{\rm F}(Y,X,\z,\bb_0)\mid X,\z\}\mid\z\right]\\
&=&\E\{\a^\dagger(X,\z,\bb_0)\mid\z\}.
\ese
Thus, we have $\E\{\S\eff^\dagger(Y,W,\Delta,\z,\bb_0)\mid\z\}=\E\{\a^\dagger(X,\z,\bb_0)\mid\z\}=\0$.
\hfill$\blacksquare$

%%%%%%%%%%%%%%%%%%%%%%%%%%%%%%%%%%%%%%%%%%%%%%%%%%

\subsection{Regularity Conditions for Theorem~\ref{th:parprofile}}\label{sec:reg-parprofile}

Define
\bse
\A_1(\bb)&\equiv&\E[\partial\S\eff^{*\dagger}\{Y,W,\Delta,\Z,\ba_1^*(\bb),\ba_2^\dagger,\bb\}/\partial\ba_1\trans],\\
\A_2(\bb)&\equiv&\E[\partial\S\eff^{*\dagger}\{Y,W,\Delta,\Z,\ba_1^*(\bb),\ba_2^\dagger,\bb\}/\partial\ba_2\trans],\\
\B&\equiv&\E[d\S\eff^{*\dagger}\{Y,W,\Delta,\Z,\ba_1^*(\bb_0),\ba_2^\dagger,\bb_0\}/d\bb\trans],
\ese
where
\bse
&&\frac{d\S\eff^{*\dagger}\{y,w,\delta,\z,\ba_1(\bb),\ba_2,\bb\}}{d\bb\trans}\\
&\equiv&\frac{\partial\S\eff^{*\dagger}\{y,w,\delta,\z,\ba_1(\bb),\ba_2,\bb\}}{\partial\bb\trans}
+\frac{\partial\S\eff^{*\dagger}\{y,w,\delta,\z,\ba_1(\bb),\ba_2,\bb\}}{\partial\ba_1\trans}
\frac{\partial\ba_1(\bb)}{\partial\bb\trans}.
\ese
The following regularity conditions are used for the asymptotic results  in Theorem~\ref{th:parprofile}.
\begin{enumerate}[label=(A\arabic*),ref=(A\arabic*)]
    \item\label{con:bb}
    $\bb_0\in\Omega$, and $\Omega$ is compact.
    \item\label{con:seffub'}
    $\E[\sup_{\bb\in\Omega}\|\S\eff^{*\dagger}\{Y,W,\Delta,\Z,\ba_1^*(\bb),\ba_2^\dagger,\bb\}\|_2]<\infty$.
    \item\label{con:A'}
    $\A_1(\bb), \A_2(\bb)$ have bounded eigenvalues for any $\bb\in\Omega$.
    \item\label{con:B'} $\B$ has bounded eigenvalues and is invertible.
\end{enumerate}
Conditions \ref{con:bb} and \ref{con:seffub'} are used to show consistency of the estimator and are standard assumptions in large sample theory for the method of estimating equations \citep{newey1994large}. Condition \ref{con:A'} regulates the impact of nuisance estimation on the resulting estimating equation by assuming finite eigenvalues of derivative matrices, which is a reasonable assumption since $\ba_1,\ba_2$ have finite dimensions. Finally, Condition \ref{con:B'} guarantees that the estimating equation has a unique solution and a well-behaved derivative, which are also standard conditions for asymptotic theory.

%%%%%%%%%%%%%%%%%%%%%%%%%%%%%%%%%%%%%%%%%%%%%%%%%%

\subsection{Proof of Theorem \ref{th:parprofile}}\label{sec:th-parprofile-proof}
For notational brevity, we define
\bse
&&\b^{*\dagger}(y,w,\delta,\z,\ba_1,\ba_2,\bb)\\
&\equiv&\delta\a^{*\dagger}(w,\z,\ba_1,\ba_2,\bb)\\
&&+(1-\delta)\frac{\E^*[\I(X>w)\{\a^{*\dagger}(X,\z,\ba_1,\ba_2,\bb)
-\S_\bb^{\rm F}(y,X,\z,\bb)\}\mid y,\z,\ba_1,\bb]}{\E^*\{\I(X>w)\mid y,\z,\ba_1,\bb\}}.
\ese
Then the efficient score vector in Proposition \ref{pro:efficientscore} with parametric working models is equivalent to
\bse
\S\eff^{*\dagger}(y,w,\delta,\z,\ba_1,\ba_2,\bb)
=\delta\S_\bb^{\rm F}(y,w,\z,\bb)-\b^{*\dagger}(y,w,\delta,\z,\ba_1,\ba_2,\bb),
\ese
and \eqref{eq:solve-a} with parametric working models is equivalent to
\be\label{eq:b}
\E^\dagger\{\b^{*\dagger}(Y,W,\Delta,\z,\ba_1,\ba_2,\bb)\mid x,\z,\ba_2,\bb\}=\0.
\ee

We first derive some basic properties of $\S\eff^{*\dagger}$. \eqref{eq:b} leads to
\be\label{eq:beta1}
\0&=&\E^*[\E^\dagger\{\b^{*\dagger}(Y,W,\Delta,\z,\ba_1,\ba_2,\bb_0)\mid X,\z,\ba_2\}\mid\z,\ba_1]\n\\
&=&\E^\dagger[\E^*\{\b^{*\dagger}(Y,W,\Delta,\z,\ba_1,\ba_2,\bb_0)\mid C,\z,\ba_1\}\mid\z,\ba_2]\n\\
&=&\E^\dagger\left\{\E^*\left(\Delta\a^{*\dagger}(X,\z,\ba_1,\ba_2,\bb_0)\right.\right.\n\\
&&\left.\left.+(1-\Delta)\frac{\E^*[(1-\Delta)\{\a^{*\dagger}(X,\z,\ba_1,\ba_2,\bb_0)
-\S_\bb^{\rm F}(Y,X,\z,\bb_0)\}\mid Y,C,\z,\ba_1]}{\E^*(1-\Delta\mid Y,C,\z,\ba_1)}\mid C,\z,\ba_1\right)\mid\z,\ba_2\right\}\n\\
&=&\E^\dagger\{\E^*(\Delta\a^{*\dagger}(X,\z,\ba_1,\ba_2,\bb_0)\n\\
&&+\E^*[(1-\Delta)\{\a^{*\dagger}(X,\z,\ba_1,\ba_2,\bb_0)
-\S_\bb^{\rm F}(Y,X,\z,\bb_0)\}\mid Y,C,\z,\ba_1]\mid C,\z,\ba_1)\mid\z,\ba_2\}\n\\
&=&\E^\dagger[\E^*\{\a^{*\dagger}(X,\z,\ba_1,\ba_2,\bb_0)
-(1-\Delta)\S_\bb^{\rm F}(Y,X,\z,\bb_0)\mid C,\z,\ba_1\}\mid\z,\ba_2]\n\\
&=&\E^*[\a^{*\dagger}(X,\z,\ba_1,\ba_2,\bb_0)
-\E^\dagger(1-\Delta\mid X,\z,\ba_2)\E\{\S_\bb^{\rm F}(Y,X,\z,\bb_0)\mid X,\z\}\mid\z,\ba_1]\n\\
&=&\E^*\{\a^{*\dagger}(X,\z,\ba_1,\ba_2,\bb_0)\mid\z,\ba_1\},
\ee
where the last equality holds by $\E\{\S_\bb^{\rm F}(Y,x,\z,\bb_0)\mid x,\z\}=\0$.
Then we have
\bse
&&\E^*\{\S\eff^{*\dagger}(Y,W,\Delta,\z,\ba_1,\ba_2,\bb_0)\mid c,\z,\ba_1\}\n\\
&=&\E^*\{\Delta\S_\bb^{\rm F}(Y,W,\z,\bb_0)-\b^{*\dagger}(Y,W,\Delta,\z,\ba_1,\ba_2,\bb_0)\mid c,\z,\ba_1\}\n\\
&=&\E^*\left(\Delta\S_\bb^{\rm F}(Y,X,\z,\bb_0)-\Delta\a^{*\dagger}(X,\z,\ba_1,\ba_2,\bb_0)\right.\n\\
&&\left.-(1-\Delta)\frac{\E^*[(1-\Delta)\{\a^{*\dagger}(X,\z,\ba_1,\ba_2,\bb_0)
-\S_\bb^{\rm F}(Y,X,\z,\bb_0)\}\mid Y,c,\z,\ba_1]}{\E^*(1-\Delta\mid Y,c,\z,\ba_1)}\mid c,\z,\ba_1\right)\n\\
&=&\E^*\left(\Delta\S_\bb^{\rm F}(Y,X,\z,\bb_0)-\Delta\a^{*\dagger}(X,\z,\ba_1,\ba_2,\bb_0)\right.\n\\
&&\left.-\E^*[(1-\Delta)\{\a^{*\dagger}(X,\z,\ba_1,\ba_2,\bb_0)
-\S_\bb^{\rm F}(Y,X,\z,\bb_0)\}\mid Y,c,\z,\ba_1]\mid c,\z,\ba_1\right)\n\\
&=&\E^*\{\S_\bb^{\rm F}(Y,X,\z,\bb_0)-\a^{*\dagger}(X,\z,\ba_1,\ba_2,\bb_0)\mid\z,\ba_1\}\n\\
&=&\0,
\ese
where the last equality holds by $\E\{\S_\bb^{\rm F}(Y,x,\z,\bb_0)\mid
x,\z\}=\0$ and \eqref{eq:beta1}. Subsequently, when $\eta_1^*\{x,\z,\ba_1^*(\bb_0)\}=\eta_1(x,\z)$,
\be
&&\E[\S\eff^\dagger\{Y,W,\Delta,\Z,\ba_1^*(\bb_0),\ba_2,\bb_0\}]\n\\
&=&\E(\E^*[\S\eff^\dagger\{Y,W,\Delta,\Z,\ba_1^*(\bb_0),\ba_2,\bb_0\}\mid C,\Z,\ba_1^*(\bb_0)])\n\\
&=&\0,\label{eq:seffeta1profile}\\
&&\E\left[\frac{\partial\S\eff^\dagger\{Y,W,\Delta,\Z,\ba_1^*(\bb_0),\ba_2,\bb_0\}}{\partial\ba_2\trans}\right]\n\\
&=&\E\left(\frac{\partial\E^*[\S\eff^\dagger\{Y,W,\Delta,\Z,\ba_1^*(\bb_0),\ba_2,\bb_0\}\mid C,\Z,\ba_1^*(\bb_0)]}{\partial\ba_2\trans}\right)\n\\
&=&\0.\label{eq:A2eta1profile}
\ee
In addition, we have
\bse
&&\E^\dagger\{\S\eff^{*\dagger}(Y,W,\Delta,\z,\ba_1,\ba_2,\bb_0)\mid x,\z,\ba_2\}\\
&=&\E^\dagger\{\Delta\S_\bb^{\rm F}(Y,W,\z,\bb_0)-\b^{*\dagger}(Y,W,\Delta,\z,\ba_1,\ba_2,\bb_0)\mid x,\z,\ba_2\}\\
&=&\E^\dagger(\Delta\mid x,\z,\ba_2)\E\{\S_\bb^{\rm F}(Y,X,\z,\bb_0)\mid x,\z\}\\
&=&\0,
\ese
where the second equality holds by \eqref{eq:b} and the last equality holds by $\E\{\S_\bb^{\rm F}(Y,x,\z,\bb_0)\mid x,\z\}=\0$.
Subsequently, when $\eta_2^\dagger(c,\z,\ba_2^\dagger)=\eta_2(c,\z)$,
\be
\E\{\S\eff^*(Y,W,\Delta,\Z,\ba_1,\ba_2^\dagger,\bb_0)\}
&=&\E[\E^\dagger\{\S\eff^*(Y,W,\Delta,\Z,\ba_1,\ba_2^\dagger,\bb_0)\mid X,\Z,\ba_2^\dagger\}]\n\\
&=&\0,\label{eq:seffeta2}\\
\E\left\{\frac{\partial\S\eff^*(Y,W,\Delta,\Z,\ba_1,\ba_2^\dagger,\bb_0)}{\partial\ba_1\trans}\right\}
&=&\E\left[\frac{\partial\E^\dagger\{\S\eff^*(Y,W,\Delta,\Z,\ba_1,\ba_2^\dagger,\bb_0)\mid X,\Z,\ba_2^\dagger\}}{\partial\ba_1\trans}\right]\n\\
&=&\0.\label{eq:A1eta2}
\ee
On the other hand, $\E[\bphi_1\{Y,W,\Delta,\Z,\ba_1^*(\bb),\bb\}]=\0$ for any $\bb$, and this leads to
\bse
\0&=&\E\left[\frac{d\bphi_1\{Y,W,\Delta,\Z,\ba_1^*(\bb),\bb\}}{d\bb\trans}\right]\\
&=&\E\left[\frac{\partial\bphi_1\{Y,W,\Delta,\Z,\ba_1^*(\bb),\bb\}}{\partial\bb\trans}\right]
+\E\left[\frac{\partial\bphi_1\{Y,W,\Delta,\Z,\ba_1^*(\bb),\bb\}}{\partial\ba_1\trans}\right]
\frac{\partial\ba_1^*(\bb)}{\partial\bb\trans},
\ese
and further,
\be\label{eq:babbeta1}
\frac{\partial\ba_1^*(\bb_0)}{\partial\bb\trans}
&=&-\left(\E\left[\frac{\partial\bphi_1\{Y,W,\Delta,\Z,\ba_1^*(\bb_0),\bb_0\}}{\partial\ba_1\trans}\right]\right)^{-1}
\E\left[\frac{\partial\bphi_1\{Y,W,\Delta,\Z,\ba_1^*(\bb_0),\bb_0\}}{\partial\bb\trans}\right]\n\\
&=&\V_1\E\left[\frac{\partial\bphi_1\{Y,W,\Delta,\Z,\ba_1^*(\bb_0),\bb_0\}}{\partial\bb\trans}\right].
\ee

We now show the consistency of $\wh\bb$ following Theorem 2.1 of \cite{newey1994large}. We can view solving $\E[\S\eff^{*\dagger}\{Y,W,\Delta,\Z,\ba_1^*(\bb),\ba_2^\dagger,\bb\}]=\0$ as maximizing
\bse
Q_0(\bb)\equiv-\|\E[\S\eff^{*\dagger}\{Y,W,\Delta,\Z,\ba_1^*(\bb),\ba_2^\dagger,\bb\}]\|_2^2.
\ese
Then (i) $Q_0(\bb)$ is uniquely maximized at $\bb_0$ since
$\E[\S\eff^{*\dagger}\{Y,W,\Delta,\Z,\ba_1^*(\bb_0),\ba_2^\dagger,\bb_0\}]=\0$ by
\eqref{eq:seffeta1profile} and \eqref{eq:seffeta2} given that either
$\eta_1^*\{x,\z,\ba_1^*(\bb_0)\}=\eta_1(x,\z)$ or
$\eta_2^\dagger(c,\z,\ba_2^\dagger)=\eta_2(c,\z)$, and the uniqueness is
by \ref{con:B'}. Also, (ii) $\Omega$ is compact by \ref{con:bb}, and
(iii) $Q_0(\bb)$ is continuous since it is differentiable by
\ref{con:B'}. Hence, it suffices to show (iv) 
\bse
\wh Q_n(\bb)\equiv-\left\|n^{-1}\sumi\S\eff^{*\dagger}\{y_i,w_i,\delta_i,\z_i,\wh\ba_1(\bb),\wh\ba_2,\bb\}\right\|_2^2
\ese
converges uniformly in probability to $Q_0(\bb)$. This holds because
\bse
&&n^{-1}\sumi\S\eff^{*\dagger}\{y_i,w_i,\delta_i,\z_i,\wh\ba_1(\bb),\wh\ba_2,\bb\}\\
&=&n^{-1}\sumi\S\eff^{*\dagger}\{y_i,w_i,\delta_i,\z_i,\ba_1^*(\bb),\ba_2^\dagger,\bb\}
+\E\left[\frac{\partial\S\eff^{*\dagger}\{Y,W,\Delta,\Z,\ba_1^*(\bb),\ba_2^\dagger,\bb\}}{\partial\ba_1\trans}\right]\{\wh\ba_1(\bb)-\ba_1^*(\bb)\}\\
&&+\E\left[\frac{\partial\S\eff^{*\dagger}\{Y,W,\Delta,\Z,\ba_1^*(\bb),\ba_2^\dagger,\bb\}}{\partial\ba_2\trans}\right](\wh\ba_2-\ba_2^\dagger)
+o_p(n^{-1/2})\\
&=&n^{-1}\sumi\S\eff^{*\dagger}\{y_i,w_i,\delta_i,\z_i,\ba_1^*(\bb),\ba_2^\dagger,\bb\}+O_p(n^{-1/2})\\
&=&\E[\S\eff^{*\dagger}\{Y,W,\Delta,\Z,\ba_1^*(\bb),\ba_2^\dagger,\bb\}]+o_p(1),
\ese
where the first equality holds by
$\|\wh\ba_1(\bb)-\ba_1^*(\bb)\|_2=O_p(n^{-1/2})$ and
$\|\wh\ba_2-\ba_2^\dagger\|_2=O_p(n^{-1/2})$, the second equality holds
by \ref{con:A'}, and the last equality holds by \ref{con:seffub'}.
Also, this holds uniformly with respect to $\bb$ by \ref{con:bb}.
Therefore, $\wh\bb$ is consistent for $\bb_0$.

We now derive the asymptotic distribution of $\wh\bb$. By the definition of $\wh\bb$,
\bse
\0&=&n^{-1/2}\sumi\S\eff^{*\dagger}\{y_i,w_i,\delta_i,\z_i,\wh\ba_1(\wh\bb),\wh\ba_2,\wh\bb\}\\
&=&n^{-1/2}\sumi\S\eff^{*\dagger}\{y_i,w_i,\delta_i,\z_i,\wh\ba_1(\bb_0),\wh\ba_2,\bb_0\}+\wh\B\sqrt{n}(\wh\bb-\bb_0)\\
&=&n^{-1/2}\sumi\S\eff^{*\dagger}\{y_i,w_i,\delta_i,\z_i,\ba_1^*(\bb_0),\ba_2^\dagger,\bb_0\}\\
&&+\E\left[\frac{\partial\S\eff^{*\dagger}\{Y,W,\Delta,\Z,\ba_1^*(\bb_0),\ba_2^\dagger,\bb_0\}}{\partial\ba_1\trans}\right]
\sqrt{n}\{\wh\ba_1(\bb_0)-\ba_1^*(\bb_0)\}\\
&&+\E\left[\frac{\partial\S\eff^{*\dagger}\{Y,W,\Delta,\Z,\ba_1^*(\bb_0),\ba_2^\dagger,\bb_0\}}{\partial\ba_2\trans}\right]
\sqrt{n}(\wh\ba_2-\ba_2^\dagger)
+o_p(1)+\wh\B\sqrt{n}(\wh\bb-\bb_0)\\
&=&n^{-1/2}\sumi\left[\S\eff^{*\dagger}\{y_i,w_i,\delta_i,\z_i,\ba_1^*(\bb_0),\ba_2^\dagger,\bb_0\}
+\A_1(\bb_0)\bphi_1\{y_i,w_i,\delta_i,\z_i,\ba_1^*(\bb_0),\bb_0\}\right.\\
&&\left.+\A_2(\bb_0)\bphi_2(w_i,\delta_i,\z_i)\right]
+\wh\B\sqrt{n}(\wh\bb-\bb_0)+o_p(1),
\ese
where $\wh\B\equiv n^{-1}\sumi d\S\eff^{*\dagger}\{y_i,w_i,\delta_i,\z_i,\wh\ba_1(\wt\bb),\wh\ba_2,\wt\bb\}/d\bb\trans$ and $\wt\bb\equiv t\wh\bb+(1-t)\bb_0$ for some $t\in(0,1)$, the third equality holds by $\|\wh\ba_1(\bb)-\ba_1^*(\bb)\|_2=O_p(n^{-1/2})$ and $\|\wh\ba_2-\ba_2^\dagger\|_2=O_p(n^{-1/2})$, and the last equality holds by the asymptotic expressions of $\wh\ba_1(\bb)-\ba_1^*(\bb)$ and $\wh\ba_2-\ba_2^\dagger$ given as the assumption.
Then,
\bse
&&\sqrt{n}(\wh\bb-\bb_0)\\
&=&-\wh\B^{-1}\left(n^{-1/2}\sumi\left[\S\eff^{*\dagger}\{y_i,w_i,\delta_i,\z_i,\ba_1^*(\bb_0),\ba_2^\dagger,\bb_0\}\right.\right.\\
&&\left.\left.+\A_1(\bb_0)\bphi_1\{y_i,w_i,\delta_i,\z_i,\ba_1^*(\bb_0),\bb_0\}
+\A_2(\bb_0)\bphi_2(w_i,\delta_i,\z_i)\right]+o_p(1)\right)\\
&=&-\{\B^{-1}+o_p(1)\}\left(n^{-1/2}\sumi\left[\S\eff^{*\dagger}\{y_i,w_i,\delta_i,\z_i,\ba_1^*(\bb_0),\ba_2^\dagger,\bb_0\}\right.\right.\\
&&\left.\left.+\A_1(\bb_0)\bphi_1\{y_i,w_i,\delta_i,\z_i,\ba_1^*(\bb_0),\bb_0\}
+\A_2(\bb_0)\bphi_2(w_i,\delta_i,\z_i)\right]+o_p(1)\right)\\
&=&-\B^{-1}n^{-1/2}\sumi\left[\S\eff^{*\dagger}\{y_i,w_i,\delta_i,\z_i,\ba_1^*(\bb_0),\ba_2^\dagger,\bb_0\}\right.\\
&&\left.+\A_1(\bb_0)\bphi_1\{y_i,w_i,\delta_i,\z_i,\ba_1^*(\bb_0),\bb_0\}
+\A_2(\bb_0)\bphi_2(w_i,\delta_i,\z_i)\right]+o_p(1),
\ese
where the second equality holds by \ref{con:B'}, $\|\wh\ba_1(\bb)-\ba_1^*(\bb)\|_2=O_p(n^{-1/2})$, $\|\wh\ba_2-\ba_2^\dagger\|_2=O_p(n^{-1/2})$, and $\|\wt\bb-\bb_0\|_2\leq\|\wh\bb-\bb_0\|_2=o_p(1)$ by the consistency of $\wh\bb$. Also, the third equality holds because of \ref{con:A'}, \ref{con:B'}, and
\bse
n^{-1/2}\sumi\S\eff^{*\dagger}\{y_i,w_i,\delta_i,\z_i,\ba_1^*(\bb_0),\ba_2^\dagger,\bb_0\}&=&O_p(1),\\
n^{-1/2}\sumi\bphi_1\{y_i,w_i,\delta_i,\z_i,\ba_1^*(\bb_0),\bb_0\}&=&O_p(1),\\
n^{-1/2}\sumi\bphi_2(w_i,\delta_i,\z_i)&=&O_p(1),
\ese
since
$\E[\S\eff^{*\dagger}\{Y,W,\Delta,\Z,\ba_1^*(\bb_0),\ba_2^\dagger,\bb_0\}]=\0$ by
\eqref{eq:seffeta1profile} and \eqref{eq:seffeta2} given that either
$\eta_1^*\{x,\z,\ba_1^*(\bb_0)\}=\eta_1(x,\z)$ or
$\eta_2^\dagger(c,\z,\ba_2^\dagger)=\eta_2(c,\z)$,
$\E[\bphi_1\{Y,W,\Delta,\Z,\ba_1^*(\bb_0),\bb_0\}]=\0$, and
$\E\{\bphi_2(W,\Delta,\Z)\}=\0$.
Since
$\eta_1^*\{x,\z,\ba_1^*(\bb_0)\}=\eta_1(x,\z)$ gives $\A_2(\bb_0)=\0$ by
\eqref{eq:A2eta1profile} and
\be\label{eq:Beta1}
\B
&=&\E\left[\frac{\partial\S\eff^\dagger\{Y,W,\Delta,\Z,\ba_1^*(\bb_0),\ba_2^\dagger,\bb_0\}}{\partial\bb\trans}\right]
+\A_1(\bb_0)\frac{\partial\ba_1^*(\bb_0)}{\partial\bb\trans}\\
&=&\E\left[\frac{\partial\S\eff^\dagger\{Y,W,\Delta,\Z,\ba_1^*(\bb_0),\ba_2^\dagger,\bb_0\}}{\partial\bb\trans}\right]
+\A_1(\bb_0)\V_1\E\left[\frac{\partial\bphi_1\{Y,W,\Delta,\Z,\ba_1^*(\bb_0),\bb_0\}}{\partial\bb\trans}\right]\n
\ee
by \eqref{eq:babbeta1}, we get the result (i). 
Also, $\eta_2^\dagger(c,\z,\ba_2^\dagger)=\eta_2(c,\z)$
gives $\A_1(\bb_0)=\0$ by \eqref{eq:A1eta2} and
\bse
\B
&=&\E\left[\frac{\partial\S\eff^*\{Y,W,\Delta,\Z,\ba_1^*(\bb_0),\ba_2^\dagger,\bb_0\}}{\partial\bb\trans}\right]
+\A_1(\bb_0)\frac{\partial\ba_1^*(\bb_0)}{\partial\bb\trans}\\
&=&\E\left[\frac{\partial\S\eff^*\{Y,W,\Delta,\Z,\ba_1^*(\bb_0),\ba_2^\dagger,\bb_0\}}{\partial\bb\trans}\right],
\ese
then we get the result (ii). Furthermore, when $\eta_1^*\{x,\z,\ba_1^*(\bb_0)\}=\eta_1(x,\z)$ and
$\eta_2^\dagger(c,\z,\ba_2^\dagger)=\eta_2(c,\z)$, we have
$\A_1(\bb_0)=\0$, $\A_2(\bb_0)=\0$, and
\bse
\B&=&\E\left\{\partial\S\eff(Y,W,\Delta,\Z,\bb_0)/\partial\bb\trans\right\}\\
&=&-\E\left\{\S\eff(Y,W,\Delta,\Z,\bb_0)\S_\bb\trans(Y,W,\Delta,\Z,\bb_0)\right\}\\
&=&-\E\left\{\S\eff^{\otimes2}(Y,W,\Delta,\Z,\bb_0)\right\},
\ese
where the second equality holds because $\E_\bb\{\S\eff(Y,W,\Delta,\Z,\bb)\}=\0$ for any $\bb$, and the last equality holds by $\S_\bb-\S\eff\in\Lambda$ and $\S\eff\in\Lambda^\perp$.
These lead to the result (iii).
\hfill$\blacksquare$

%%%%%%%%%%%%%%%%%%%%%%%%%%%%%%%%%%%%%%%%%%%%%%%%%%

\subsection{Proof of Corollary \ref{cor:par}}\label{sec:cor-par-proof}
In Corollary \ref{cor:par}, the results (ii) and (iii) are the same as in Theorem \ref{th:parprofile} with $\ba_1^*(\bb_0)$ replaced by $\ba_1^*$. Furthermore, $\bSig^\dagger$ in the result (i) of Corollary \ref{cor:par} is identical to $\bSig^\dagger$ in the result (i) of Theorem \ref{th:parprofile} with $\ba_1^*(\bb_0), \bphi_1\{Y,W,\Delta,\Z,\ba_1^*(\bb_0),\bb_0\}$ replaced by $\ba_1^*, \bphi_1(W,\Delta,\Z)$. Hence it suffices to show that $\B$ in the result (i) of Theorem \ref{th:parprofile} reduces to $\B$ in the result (i) of Corollary \ref{cor:par}. When $\eta_1^*(x,\z,\ba_1^*)=\eta_1(x,\z)$, by \eqref{eq:Beta1},
\bse
\B
&=&\E\left\{\frac{\partial\S\eff^\dagger(Y,W,\Delta,\Z,\ba_1^*,\ba_2^\dagger,\bb_0)}{\partial\bb\trans}\right\}
+\A_1(\bb_0)\frac{\partial\ba_1^*}{\partial\bb\trans}\\
&=&\E\left\{\frac{\partial\S\eff^\dagger(Y,W,\Delta,\Z,\ba_1^*,\ba_2^\dagger,\bb_0)}{\partial\bb\trans}\right\}
\ese
because $\partial\ba_1^*/\partial\bb\trans=\0$ as $\ba_1^*$ does not depend on $\bb$. Therefore, we get the result (i).
\hfill$\blacksquare$

%%%%%%%%%%%%%%%%%%%%%%%%%%%%%%%%%%%%%%%%%%%%%%%%%%

\subsection{Regularity Conditions for Theorem~\ref{th:non}}\label{sec:reg-non}

Below are regularity conditions used to derive the asymptotic distribution of $\wt\bb$.
\begin{enumerate}[label=(B\arabic*),ref=(B\arabic*),start=1]
    \item\label{con:bb-non}
    $\bb_0\in\Omega$, and $\Omega$ is compact. The support of $(Y,X,C,\Z)$ is compact.
    \item\label{con:seffub-non}
    $\E\{\sup_{\bb\in\Omega}\|\S\eff(Y,W,\Delta,\Z,\bb)\|_2\}<\infty$,
    $\|\S_\bb^{\rm F}(y,x,z,\bb)\|_1<\infty$, and
    $\|\a(x,z,\bb)\|_1<\infty$ on the support of $(Y,X,C,\Z)$ and
    $\Omega$, where $\|\cdot\|_1$ is the vector $L_1$ norm.%$\|\cdot\|_2, \|\cdot\|_1$ are the vector $L_2, L_1$ norms.
    \item\label{con:A-non}
    $\fyxz(y,x,\z,\bb),\eta_1(x,\z),\eta_2(c,\z)$ are bounded and
    bounded away from zero on the support of $(Y,X,C,\Z)$ and $\Omega$.
    \item\label{con:B-non}
    $\B\equiv\E\{\partial\S\eff(Y,W,\Delta,\Z,\bb)/\partial\bb\trans\}$ has bounded eigenvalues and is invertible.
\end{enumerate}
Conditions \ref{con:bb-non}, \ref{con:seffub-non}, and \ref{con:B-non} are similar to conditions \ref{con:bb}, \ref{con:seffub'}, and \ref{con:B'} for Theorem~\ref{th:parprofile}, respectively, with the added assumptions that the support of the data $(Y,X,C,\Z)$ is compact, and the functions $\S_\bb^{\rm F}$ and $\a$ are bounded. The compact support assumption is included to guarantee that the nonparametric estimators converge uniformly over the support, hence can be relaxed to another set of assumptions that ensures the uniform convergence. In addition, Condition \ref{con:A-non} is necessary to guarantee that the conditional expectation estimators $\wh\E_1,\wh\E_2$ are consistent at the same rate as $\wh\eta_1,\wh\eta_2$.

%%%%%%%%%%%%%%%%%%%%%%%%%%%%%%%%%%%%%%%%%%%%%%%%%%

\subsection{Lemmas}
We present lemmas used for the proof of Theorem \ref{th:non} in Section \ref{sec:th-non-proof}.
We first show the convergence rate of $\wh\E_1, \wh\E_2$. For clarity,
we denote the true conditional expectations under $\eta_1,\eta_2$ as
$\E_{10},\E_{20}$, respectively.
\begin{lemma}\label{lem:E}
Assume the conditions \ref{con:bb-non} and \ref{con:A-non},
$\sup_{x,\z}|(\wh\eta_1-\eta_1)(x,\z)|=o_p(n^{-1/4})$, and
$\sup_{c,\z}|(\wh\eta_2-\eta_2)(c,\z)|=o_p(n^{-1/4})$. Let
$\g(y,x,c,\z,\bb)$ be such that
$\|\g(y,x,c,\z,\bb)\|_1<\infty$ on the support of $(Y,X,C,\Z)$ and $\Omega$,
where $\|\cdot\|_1$ is the vector $L_1$ norm.
Then,
\bse
\|(\wh\E_1-\E_{10})\{\g(y,X,c,\z,\bb)\mid y,\z,\bb\}\|_1&=&o_p(n^{-1/4}),\\
\|(\wh\E_2-\E_{20})\{\g(Y,x,C,\z,\bb)\mid x,\z,\bb\}\|_1&=&o_p(n^{-1/4}),
\ese
uniformly on the support of $(Y,X,C,\Z)$ and $\Omega$.
\end{lemma}
Proof.
For a given function $\g(y,x,c,\z,\bb)$,
\bse
&&\|(\wh\E_1-\E_{10})\{\g(y,X,c,\z,\bb)\mid y,\z,\bb\}\|_1\n\\
&=&\sum_{j=1}^p\left|\frac{\int g_j(y,x,c,\z,\bb)\fyxz(y,x,\z,\bb)\wh\eta_1(x,\z)dx}{\int\fyxz(y,x,\z,\bb)\wh\eta_1(x,\z)dx}
-\frac{\int g_j(y,x,c,\z,\bb)\fyxz(y,x,\z,\bb)\eta_1(x,\z)dx}{\int\fyxz(y,x,\z,\bb)\eta_1(x,\z)dx}\right|\n\\
&=&\sum_{j=1}^p\left|\frac{\int g_j(y,x,c,\z,\bb)\fyxz(y,x,\z,\bb)(\wh\eta_1-\eta_1)(x,\z)dx}{\int\fyxz(y,x,\z,\bb)\eta_1(x,\z)dx}\right.\n\\
&&\left.-\frac{\int g_j(y,x,c,\z,\bb)\fyxz(y,x,\z,\bb)\wh\eta_1(x,\z)dx\int\fyxz(y,x,\z,\bb)(\wh\eta_1-\eta_1)(x,\z)dx}
{\int\fyxz(y,x,\z,\bb)\wh\eta_1(x,\z)dx\int\fyxz(y,x,\z,\bb)\eta_1(x,\z)dx}\right|\n\\
&\leq&\frac{\int\sum_{j=1}^p|g_j(y,x,c,\z,\bb)\fyxz(y,x,\z,\bb)(\wh\eta_1-\eta_1)(x,\z)|dx}{\int\fyxz(y,x,\z,\bb)\eta_1(x,\z)dx}\n\\
&&+\|\wh\E_1\{\g(y,X,c,\z,\bb)\mid y,\z,\bb\}\|_1
\frac{\int|\fyxz(y,x,\z,\bb)(\wh\eta_1-\eta_1)(x,\z)|dx}{\int\fyxz(y,x,\z,\bb)\eta_1(x,\z)dx}\n\\
&\leq&\frac{\int\sum_{j=1}^p|g_j(y,x,c,\z,\bb)|\fyxz(y,x,\z,\bb)\eta_1(x,\z)dx}{\int\fyxz(y,x,\z,\bb)\eta_1(x,\z)dx}
\sup_{x,\z}\frac{|(\wh\eta_1-\eta_1)(x,\z)|}{\eta_1(x,\z)}\n\\
&&+\wh\E_1\{\|\g(y,X,c,\z,\bb)\|_1\mid y,\z,\bb\}\sup_{x,\z}\frac{|(\wh\eta_1-\eta_1)(x,\z)|}{\eta_1(x,\z)}\n\\
&=&o_p(n^{-1/4})(\E_{10}+\wh\E_1)\{\|\g(y,X,c,\z,\bb)\|_1\mid y,\z,\bb\}\n\\
&=&o_p(n^{-1/4}),
\ese
where the third equality holds because
$\sup_{x,\z}|(\wh\eta_1-\eta_1)(x,\z)|=o_p(n^{-1/4})$ and $\eta_1$ is
bounded away from zero by \ref{con:A-non}, and the last
equality holds because $\|\g(y,x,c,\z,\bb)\|_1$ is bounded.
This holds uniformly with respect to $(y,\z,\bb)$ by \ref{con:bb-non}. On the other hand,
\bse
&&\|(\wh\E_2-\E_{20})\{\g(Y,x,C,\z,\bb)\mid x,\z,\bb\}\|_1\n\\
&=&\sum_{j=1}^p\left|\int g_j(y,x,c,\z,\bb)\fyxz(y,x,\z,\bb)(\wh\eta_2-\eta_2)(c,\z)dydc\right|\n\\
&\leq&\int\sum_{j=1}^p|g_j(y,x,c,\z,\bb)\fyxz(y,x,\z,\bb)(\wh\eta_2-\eta_2)(c,\z)|dydc\n\\
&\leq&\int\sum_{j=1}^p|g_j(y,x,c,\z,\bb)|\fyxz(y,x,\z,\bb)\eta_2(c,\z)dydc\sup_{c,\z}\frac{|(\wh\eta_2-\eta_2)(c,\z)|}{\eta_2(c,\z)}\n\\
&=&o_p(n^{-1/4})\E_{20}\{\|\g(Y,x,C,\z,\bb)\|_1\mid x,\z,\bb\}\n\\
&=&o_p(n^{-1/4}),
\ese
where the second equality holds because
$\sup_{c,\z}|(\wh\eta_2-\eta_2)(c,\z)|=o_p(n^{-1/4})$ and $\eta_2$ is
bounded away from zero by \ref{con:A-non}, and the last equality holds
because $\|\g(y,x,c,\z,\bb)\|_1$ is bounded. This holds uniformly with
respect to $(x,\z,\bb)$ by \ref{con:bb-non}. 
\hfill$\blacksquare$

%--------------------------------------------------

Now, let $d$ be the dimension of $(x,\z\trans)\trans$.
For $\a(x,\z,\bb)=\{a_1(x,\z,\bb),\dots,a_p(x,\z,\bb)\}\trans$, 
let $\|\a\|_*\equiv\sup_{x,\z,\bb}\sum_{j=1}^p|a_j(x,\z,\bb)|$ and
$L^*(\mR^{d+p})\equiv\{\a(x,\z,\bb):\|\a\|_*<\infty\}$. 
We also define a linear operator $\calL:L^*(\mR^{d+p})\to L^*(\mR^{d+p})$ as 
\bse
&&\calL(\a)(x,\z,\bb)\\
&\equiv&\E\{\I(x\le C)\mid\z\}\a(x,\z,\bb)
+\E\left[\I(x>C)\frac{\E\{\I(X>C)\a(X,\z,\bb)\mid Y,C,\z,\bb\}}
{\E\{\I(X>C)\mid Y,C,\z,\bb\}}\mid x,\z,\bb\right].
\ese
In addition, for clarity, we denote $\a(x,\z,\bb)$ defined in Proposition \ref{pro:efficientscore} as $\a_0(x,\z,\bb)$ and
\bse
\u_0(x,\z,\bb)
&\equiv&\E\left[\I(x>C)\frac{\E\{\I(X>C)\S_\bb^{\rm F}(Y,X,\z,\bb)\mid Y,C,\z,\bb\}}
{\E\{\I(X>C)\mid Y,C,\z,\bb\}}\mid x,\z,\bb\right].
\ese
Then we have $\calL(\a_0)(x,\z,\bb)=\u_0(x,\z,\bb)$ by the definition of $\a_0$.
Below, we show that $\calL$ is invertible and well-behaved.
\begin{lemma}\label{lem:L}
Under the conditions \ref{con:bb-non} and \ref{con:A-non},
\begin{enumerate}[label=(\roman*)]
    \item $\calL$ is invertible,
    \item there exist constants $c_1, c_2$ such that $0<c_1,c_2<\infty$,
    $\|\calL(\a)\|_*\leq c_1\|\a\|_*$ for $\a\in L^*(\mR^{d+p})$, and
    $\|\calL^{-1}(\u)\|_*\leq c_2\|\u\|_*$ for $\u\in L^*(\mR^{d+p})$.
\end{enumerate}
\end{lemma}
Proof.
We first show that $\calL$ is invertible at $\u_0$ by contradiction.
Suppose there are $\a_1(x,\z,\bb)$ and $\a_2(x,\z,\bb)$ such that
$\a_1\neq \a_2$ and $\calL(\a_1)(x,\z,\bb)=\calL(\a_2)(x,\z,\bb)=\u_0(x,\z,\bb)$.
Note that the efficient score vector given in Proposition \ref{pro:efficientscore} is
\bse
\S\eff(y,w,\delta,\z,\bb)
&\equiv&\delta\S_\bb^{\rm F}(y,w,\z,\bb)+(1-\delta)\frac{\E\{\I(X>w)\S_\bb^{\rm F}(y,X,\z,\bb)\mid y,\z,\bb\}}{\E\{\I(X>w)\mid y,\z,\bb\}}\\
&&-\left[\delta\a(w,\z,\bb)+(1-\delta)\frac{\E\{\I(X>w)\a(X,\z,\bb)\mid y,\z,\bb\}}{\E\{\I(X>w)\mid y,\z,\bb\}}\right],
\ese
where $\a(x,\z,\bb)$ satisfies $\calL(\a)(x,\z,\bb)=\u_0(x,\z,\bb)$.
Then letting $\a=\a_1$ and $\a=\a_2$ gives two distinct efficient score vectors,
and this contradicts the uniqueness of the efficient score vector.
Hence, $\a_0$ is a unique solution to $\calL(\a)(x,\z,\bb)=\u_0(x,\z,\bb)$.
Therefore, $\calL$ is invertible at $\u_0$.

We now show $\calL$ is invertible at any $\u$ in the range
  of $\calL$ by contradiction.
Suppose there are $\a_1(x,\z,\bb)$ and $\a_2(x,\z,\bb)$ such that
$\a_1\neq \a_2$ and $\calL(\a_1)(x,\z,\bb)=\calL(\a_2)(x,\z,\bb)=\u(x,\z,\bb)$.
This yields $\calL(\a_0+\a_1-\a_2)=\calL(\a_0)+\calL(\a_1)-\calL(\a_2)=\u_0$.
Then letting $\a=\a_0$ and $\a=\a_0+\a_1-\a_2$ gives two distinct efficient score vectors,
and this again contradicts the uniqueness of the efficient score vector.
Hence, there is a unique solution to $\calL(\a)(x,\z,\bb)=\u(x,\z,\bb)$.
Therefore, $\calL$ is invertible. 

We now show the existence of the constants $c_1,c_2$.
It is immediate that there exists a constant $c_1$
such that $0<c_1<\infty$ and $\|\calL(\a)\|_*\leq c_1\|\a\|_*$
by the definition of $\calL(\a)$ and \ref{con:bb-non}. 
In addition, we have $\|\calL^{-1}(\u)\|_*\leq c_2\|\u\|_*$
for some constant $c_2$ such that $0<c_2<\infty$ by the bounded inverse theorem,
since $\calL$ is bounded and invertible.
\hfill$\blacksquare$

%--------------------------------------------------

We now analyze the derivatives of $\S\eff$ with respect to $\E_1,\E_2$.
We define the $k$th Gateaux derivative of $\g(\mu,\cdot)$
with respect to $\mu$ at $\mu_0$ in the direction $d$ as 
\bse
\frac{\partial^k\g(\mu_0,\cdot)}{\partial\mu^k}(d)
\equiv\left.\frac{\partial^k\g(\mu_0+hd,\cdot)}{\partial h^k}\right|_{h=0}.
\ese
In addition, we define  
\bse
&&\b(y,w,\delta,\z,\E_1,\E_2,\bb)\\
&\equiv&\delta\a(w,\z,\E_1,\E_2,\bb)+(1-\delta)\frac{\E_1[\I(X>w)\{\a(X,\z,\E_1,\E_2,\bb)-\S_\bb^{\rm F}(y,X,\z,\bb)\}\mid y,\z,\bb]}{\E_1\{\I(X>w)\mid y,\z,\bb\}},
\ese
where $\a(x,\z,\E_1,\E_2,\bb)$ satisfies
\be\label{eq:a-non2}
&&\E_2\{\I(x\le C)\mid\z\}\a(x,\z,\E_1,\E_2,\bb)\n\\
&&+\E_2\left[\I(x>C)\frac{\E_1\{\I(X>C)\a(X,\z,\E_1,\E_2,\bb)\mid Y,C,\z,\bb\}}{\E_1\{\I(X>C)\mid Y,C,\z,\bb\}}\mid x,\z,\bb\right]\n\\
&=&\E_2\left[\I(x>C)\frac{\E_1\{\I(X>C)\S_\bb^{\rm F}(Y,X,\z,\bb)\mid Y,C,\z,\bb\}}{\E_1\{\I(X>C)\mid Y,C,\z,\bb\}}\mid x,\z,\bb\right].
\ee
Then Proposition \ref{pro:efficientscore} leads to
\bse
\S\eff(y,w,\delta,\z,\E_{10},\E_{20},\bb)&=&\delta\S_\bb^{\rm F}(y,w,\z,\bb)-\b(y,w,\delta,\z,\E_{10},\E_{20},\bb),\\
\E_{20}\{\b(Y,W,\Delta,\z,\E_{10},\E_{20},\bb)\mid x,\z,\bb\}&=&\0,
\ese
and 
% \eqref{eq:seff-non} and \eqref{eq:a-non} are equivalent to
the efficient score vector with nonparametric models $\S\eff(y,w,\delta,\z,\wh\E_1,\wh\E_2,\bb)$ and the corresponding integral equation for $\wh\a(x,\z,\bb)$ are equivalent to
\bse
\S\eff(y,w,\delta,\z,\wh\E_1,\wh\E_2,\bb)&=&\delta\S_\bb^{\rm F}(y,w,\z,\bb)-\b(y,w,\delta,\z,\wh\E_1,\wh\E_2,\bb),\n\\
\wh\E_2\{\b(Y,W,\Delta,\z,\wh\E_1,\wh\E_2,\bb)\mid x,\z,\bb\}&=&\0.
\ese
We can write the two results generally as
\be
\S\eff(y,w,\delta,\z,\E_1,\E_2,\bb)&=&\delta\S_\bb^{\rm F}(y,w,\z,\bb)-\b(y,w,\delta,\z,\E_1,\E_2,\bb),\n\\
\E_2\{\b(Y,W,\Delta,\z,\E_1,\E_2,\bb)\mid x,\z,\bb\}&=&\0.\label{eq:b-non}
\ee

\begin{lemma}\label{lem:dseff}
Assume conditions \ref{con:bb-non}-\ref{con:seffub-non},
$\sup_{x,\z}|(\wh\eta_1-\eta_1)(x,\z)|=o_p(n^{-1/4})$, and $\sup_{c,\z}|(\wh\eta_2-\eta_2)(c,\z)|=o_p(n^{-1/4})$.
Then
\bse
\left\|\frac{\partial\S\eff(y,w,\delta,\z,\E_{10},\E_{20},\bb)}{\partial\E_1}(\wh\E_1-\E_{10})\right\|_1&=&o_p(n^{-1/4}),\\
\left\|\frac{\partial^2\S\eff(y,w,\delta,\z,\E_{10},\E_{20},\bb)}{\partial{\E_1}^2}(\wh\E_1-\E_{10})\right\|_1&=&o_p(n^{-1/2}),\\
\left\|\frac{\partial\S\eff(y,w,\delta,\z,\E_{10},\E_{20},\bb)}{\partial\E_2}(\wh\E_2-\E_{20})\right\|_1&=&o_p(n^{-1/4}),\\
\left\|\frac{\partial^2\S\eff(y,w,\delta,\z,\E_{10},\E_{20},\bb)}{\partial{\E_2}^2}(\wh\E_2-\E_{20})\right\|_1&=&o_p(n^{-1/2}),\\
\left\|\frac{\partial^2\S\eff(y,w,\delta,\z,\E_{10},\E_{20},\bb)}{\partial\E_1\partial\E_2}(\wh\E_1-\E_{10},\wh\E_2-\E_{20})\right\|_1&=&o_p(n^{-1/2}),
\ese
uniformly on the support of $(Y,W,\Delta,\Z)$ and $\Omega$.
\end{lemma}
Proof. First of all, the results below hold uniformly on the support of $(Y,W,\Delta,\Z)$ and $\Omega$ by \ref{con:bb-non}.
First,
\be\label{eq:dseffdE1}
&&\frac{\partial\S\eff(y,w,\delta,\z,\E_1,\E_{20},\bb)}{\partial\E_1}(\wh\E_1-\E_{10})\\
&=&-\delta\frac{\partial\a(w,\z,\E_1,\E_{20},\bb)}{\partial\E_1}(\wh\E_1-\E_{10})
-(1-\delta)\frac{\E_1\{\I(X>w)\frac{\partial\a(X,\z,\E_1,\E_{20},\bb)}{\partial\E_1}(\wh\E_1-\E_{10})\mid y,\z,\bb\}}{\E_1\{\I(X>w)\mid y,\z,\bb\}}\n\\
&&-(1-\delta)\frac{(\wh\E_1-\E_{10})[\I(X>w)\{\a(X,\z,\E_1,\E_{20},\bb)-\S_\bb^{\rm F}(y,X,\z,\bb)\}\mid y,\z,\bb]}{\E_1\{\I(X>w)\mid y,\z,\bb\}}\n\\
&&+(1-\delta)\E_1[\I(X>w)\{\a(X,\z,\E_1,\E_{20},\bb)-\S_\bb^{\rm F}(y,X,\z,\bb)\}\mid y,\z,\bb]\n\\
&&\times\frac{(\wh\E_1-\E_{10})\{\I(X>w)\mid y,\z,\bb\}}{[\E_1\{\I(X>w)\mid y,\z,\bb\}]^2},\n
\ee
and 
\be\label{eq:d2seffdE1}
&&\frac{\partial^2\S\eff(y,w,\delta,\z,\E_1,\E_{20},\bb)}{\partial{\E_1}^2}(\wh\E_1-\E_{10})\\
&=&-\delta\frac{\partial^2\a(w,\z,\E_1,\E_{20},\bb)}{\partial{\E_1}^2}(\wh\E_1-\E_{10})
-(1-\delta)\frac{\E_1\{\I(X>w)\frac{\partial^2\a(X,\z,\E_1,\E_{20},\bb)}{\partial{\E_1}^2}(\wh\E_1-\E_{10})\mid y,\z,\bb\}}{\E_1\{\I(X>w)\mid y,\z,\bb\}}\n\\
&&-2(1-\delta)\frac{(\wh\E_1-\E_{10})\{\I(X>w)\frac{\partial\a(X,\z,\E_1,\E_{20},\bb)}{\partial\E_1}(\wh\E_1-\E_{10})\mid y,\z,\bb\}}{\E_1\{\I(X>w)\mid y,\z,\bb\}}\n\\
&&+2(1-\delta)\E_1\left\{\I(X>w)\frac{\partial\a(X,\z,\E_1,\E_{20},\bb)}{\partial\E_1}(\wh\E_1-\E_{10})\mid y,\z,\bb\right\}
\frac{(\wh\E_1-\E_{10})\{\I(X>w)\mid y,\z,\bb\}}{[\E_1\{\I(X>w)\mid y,\z,\bb\}]^2}\n\\
&&+(1-\delta)\frac{2(\wh\E_1-\E_{10})\{\I(X>w)\mid y,\z,\bb\}}{[\E_1\{\I(X>w)\mid y,\z,\bb\}]^3}\n\\
&&\times\left((\wh\E_1-\E_{10})[\I(X>w)\{\a(X,\z,\E_1,\E_{20},\bb)-\S_\bb^{\rm F}(y,X,\z,\bb)\}\mid y,\z,\bb]\E_1\{\I(X>w)\mid y,\z,\bb\}\right.\n\\
&&\left.-\E_1[\I(X>w)\{\a(X,\z,\E_1,\E_{20},\bb)-\S_\bb^{\rm F}(y,X,\z,\bb)\}\mid y,\z,\bb]
(\wh\E_1-\E_{10})\{\I(X>w)\mid y,\z,\bb\}\right).\n
\ee 
On the other hand, differentiating \eqref{eq:a-non2} with respect to $\E_1$ yields
\bse
&&\calL\left\{\frac{\partial\a(x,\z,\E_{10},\E_{20},\bb)}{\partial\E_1}(\wh\E_1-\E_{10})\right\}\\
&=&\E_{20}\left(-\I(x>C)\frac{(\wh\E_1-\E_{10})[\I(X>C)\{\a(X,\z,\E_{10},\E_{20},\bb)-\S_\bb^{\rm F}(Y,X,\z,\bb)\}\mid Y,C,\z,\bb]}{\E_{10}\{\I(X>C)\mid Y,C,\z,\bb\}}\right.\\
&&+\I(x>C)\E_{10}[\I(X>C)\{\a(X,\z,\E_{10},\E_{20},\bb)-\S_\bb^{\rm F}(Y,X,\z,\bb)\}\mid Y,C,\z,\bb]\\
&&\left.\times\frac{(\wh\E_1-\E_{10})\{\I(X>C)\mid Y,C,\z,\bb\}}{[\E_{10}\{\I(X>C)\mid Y,C,\z,\bb\}]^2}\mid x,\z,\bb\right)\\
&=&o_p(n^{-1/4})
\ese
in terms of its vector $L_1$ norm by Lemma \ref{lem:E} and \ref{con:seffub-non},
and this holds uniformly with respect to $(x,\z,\bb)$ by
\ref{con:bb-non}. 
Also, differentiating \eqref{eq:a-non2} twice with respect to $\E_1$ yields
\bse
&&\calL\left\{\frac{\partial^2\a(x,\z,\E_{10},\E_{20},\bb)}{\partial{\E_1}^2}(\wh\E_1-\E_{10})\right\}\\
&=&\E_{20}\left\{-2\I(x>C)\frac{(\wh\E_1-\E_{10})\{\I(X>C)\frac{\partial\a(X,\z,\E_{10},\E_{20},\bb)}{\partial\E_1}(\wh\E_1-\E_{10})\mid Y,C,\z,\bb\}}{\E_{10}\{\I(X>C)\mid Y,C,\z,\bb\}}\right.\n\\
&&+2\I(x>C)\E_{10}\left\{\I(X>C)\frac{\partial\a(X,\z,\E_{10},\E_{20},\bb)}{\partial\E_1}(\wh\E_1-\E_{10})\mid Y,C,\z,\bb\right\}\\
&&\times\frac{(\wh\E_1-\E_{10})\{\I(X>C)\mid Y,C,\z,\bb\}}{[\E_{10}\{\I(X>C)\mid Y,C,\z,\bb\}]^2}\n\\
&&+\I(x>C)\frac{2(\wh\E_1-\E_{10})\{\I(X>C)\mid Y,C,\z,\bb\}}{[\E_{10}\{\I(X>C)\mid Y,C,\z,\bb\}]^3}\n\\
&&\times\left((\wh\E_1-\E_{10})[\I(X>C)\{\a(X,\z,\E_{10},\E_{20},\bb)-\S_\bb^{\rm F}(Y,X,\z,\bb)\}\mid Y,C,\z,\bb]\right.\\
&&\times\E_{10}\{\I(X>C)\mid Y,C,\z,\bb\}\n\\
&&-\E_{10}[\I(X>C)\{\a(X,\z,\E_{10},\E_{20},\bb)-\S_\bb^{\rm F}(Y,X,\z,\bb)\}\mid Y,C,\z,\bb]\\
&&\left.\left.\times(\wh\E_1-\E_{10})\{\I(X>C)\mid Y,C,\z,\bb\}\right)\mid x,\z,\bb\right\}.
\ese 
These two lead to
\be
\sup_{x,\z,\bb}\left\|\frac{\partial\a(x,\z,\E_{10},\E_{20},\bb)}{\partial\E_1}(\wh\E_1-\E_{10})\right\|_1&=&o_p(n^{-1/4}),\label{eq:dadE1}\\
\sup_{x,\z,\bb}\left\|\frac{\partial^2\a(x,\z,\E_{10},\E_{20},\bb)}{\partial{\E_1}^2}(\wh\E_1-\E_{10})\right\|_1&=&o_p(n^{-1/2}),\n
\ee
where the first equality holds by Lemma \ref{lem:L},
and the second equality holds by Lemmas \ref{lem:L} and \ref{lem:E},
the first equality, and \ref{con:seffub-non}. 
These further lead to
\bse
\left\|\frac{\partial\S\eff(y,w,\delta,\z,\E_{10},\E_{20},\bb)}{\partial\E_1}(\wh\E_1-\E_{10})\right\|_1&=&o_p(n^{-1/4}),\\
\left\|\frac{\partial^2\S\eff(y,w,\delta,\z,\E_{10},\E_{20},\bb)}{\partial{\E_1}^2}(\wh\E_1-\E_{10})\right\|_1&=&o_p(n^{-1/2}),
\ese
where the first equality holds by \eqref{eq:dseffdE1}, Lemma
\ref{lem:E}, and \ref{con:seffub-non}, 
and the second equality holds by \eqref{eq:d2seffdE1}, Lemma
\ref{lem:E}, and \ref{con:seffub-non}. 

Now,
\be\label{eq:dseffdE2}
&&\frac{\partial^k\S\eff(y,w,\delta,\z,\E_{10},\E_2,\bb)}{\partial{\E_2}^k}(\wh\E_2-\E_{20})\\
&=&-\delta\frac{\partial^k\a(w,\z,\E_{10},\E_2,\bb)}{\partial{\E_2}^k}(\wh\E_2-\E_{20})
-(1-\delta)\frac{\E_{10}\{\I(X>w)\frac{\partial^k\a(X,\z,\E_{10},\E_2,\bb)}{\partial{\E_2}^k}(\wh\E_2-\E_{20})\mid y,\z,\bb\}}{\E_{10}\{\I(X>w)\mid y,\z,\bb\}}.\n
\ee 
On the other hand, differentiating \eqref{eq:a-non2} with respect to $\E_2$ yields
\bse
\calL\left\{\frac{\partial\a(x,\z,\E_{10},\E_{20},\bb)}{\partial\E_2}(\wh\E_2-\E_{20})\right\}
&=&-(\wh\E_2-\E_{20})\{\b(Y,W,\Delta,\z,\E_{10},\E_{20},\bb)\mid x,\z,\bb\}\\
&=&o_p(n^{-1/4})
\ese
in terms of its vector $L_1$ norm by Lemma \ref{lem:E} and \ref{con:seffub-non},
and this holds uniformly with respect to $(x,\z,\bb)$ by
\ref{con:bb-non}. 
Also, differentiating \eqref{eq:a-non2} twice with respect to $\E_2$ yields
\bse
&&\calL\left\{\frac{\partial^2\a(x,\z,\E_{10},\E_{20},\bb)}{\partial{\E_2}^2}(\wh\E_2-\E_{20})\right\}\\
&=&-2(\wh\E_2-\E_{20})\{\I(x\le C)\mid\z\}\frac{\partial\a(x,\z,\E_{10},\E_{20},\bb)}{\partial\E_2}(\wh\E_2-\E_{20})\n\\
&&-2(\wh\E_2-\E_{20})\left[\I(x>C)\frac{\E_1\{\I(X>C)\frac{\partial\a(x,\z,\E_{10},\E_{20},\bb)}{\partial\E_2}(\wh\E_2-\E_{20})\mid Y,C,\z,\bb\}}{\E_1\{\I(X>C)\mid Y,C,\z,\bb\}}\mid x,\z,\bb\right].
\ese
These two lead to
\be
\sup_{x,\z,\bb}\left\|\frac{\partial\a(x,\z,\E_{10},\E_{20},\bb)}{\partial\E_2}(\wh\E_2-\E_{20})\right\|_1&=&o_p(n^{-1/4}),\label{eq:dadE2}\\
\sup_{x,\z,\bb}\left\|\frac{\partial^2\a(x,\z,\E_{10},\E_{20},\bb)}{\partial{\E_2}^2}(\wh\E_2-\E_{20})\right\|_1&=&o_p(n^{-1/2}),\n
\ee
where the first equality holds by Lemma \ref{lem:L},
and the second equality holds by Lemmas \ref{lem:L}, \ref{lem:E}, and
the first equality. These with \eqref{eq:dseffdE2} further lead to
\bse
\left\|\frac{\partial\S\eff(y,w,\delta,\z,\E_{10},\E_{20},\bb)}{\partial\E_2}(\wh\E_2-\E_{20})\right\|_1=o_p(n^{-1/4}),\\
\left\|\frac{\partial^2\S\eff(y,w,\delta,\z,\E_{10},\E_{20},\bb)}{\partial{\E_2}^2}(\wh\E_2-\E_{20})\right\|_1=o_p(n^{-1/2}).
\ese  

Finally,
\be\label{eq:d2seffdE1dE2}
&&\frac{\partial^2\S\eff(y,w,\delta,\z,\E_1,\E_2,\bb)}{\partial\E_1\partial\E_2}(\wh\E_1-\E_{10},\wh\E_2-\E_{20})\\
&=&-\delta\frac{\partial^2\a(w,\z,\E_1,\E_2,\bb)}{\partial\E_1\partial\E_2}(\wh\E_1-\E_{10},\wh\E_2-\E_{20})\n\\
&&-(1-\delta)\frac{\E_1\{\I(X>w)\frac{\partial^2\a(X,\z,\E_1,\E_2,\bb)}{\partial\E_1\partial\E_2}(\wh\E_1-\E_{10},\wh\E_2-\E_{20})\mid y,\z,\bb\}}{\E_1\{\I(X>w)\mid y,\z,\bb\}}\n\\
&&-(1-\delta)\frac{(\wh\E_1-\E_{10})\{\I(X>w)\frac{\partial\a(x,\z,\E_1,\E_2,\bb)}{\partial\E_2}(\wh\E_2-\E_{20})\mid y,\z,\bb\}}{\E_1\{\I(X>w)\mid y,\z,\bb\}}\n\\
&&+(1-\delta)\E_1\left\{\I(X>w)\frac{\partial\a(x,\z,\E_1,\E_2,\bb)}{\partial\E_2}(\wh\E_2-\E_{20})\mid y,\z,\bb\right\}\frac{(\wh\E_1-\E_{10})\{\I(X>w)\mid y,\z,\bb\}}{[\E_1\{\I(X>w)\mid y,\z,\bb\}]^2}.\n
\ee
On the other hand, differentiating \eqref{eq:a-non2} with respect to $\E_1$ and $\E_2$ yields
\bse
&&\calL\left\{\frac{\partial^2\a(x,\z,\E_{10},\E_{20},\bb)}{\partial\E_1\partial\E_2}(\wh\E_1-\E_{10},\wh\E_2-\E_{20})\right\}\\
&=&-(\wh\E_2-\E_{20})\{\I(x\le C)\mid\z\}\frac{\partial\a(x,\z,\E_{10},\E_{20},\bb)}{\partial\E_1}(\wh\E_1-\E_{10})\\
&&+(\wh\E_2-\E_{20})\left(-\I(x>C)\frac{\E_{10}\{\I(X>C)\frac{\partial\a(x,\z,\E_{10},\E_{20},\bb)}{\partial\E_1}(\wh\E_1-\E_{10})\mid Y,C,\z,\bb\}}{\E_{10}\{\I(X>C)\mid Y,C,\z,\bb\}}\right.\\
&&-\I(x>C)\frac{(\wh\E_1-\E_{10})[\I(X>C)\{\a(X,\z,\E_{10},\E_{20},\bb)-\S_\bb^{\rm F}(Y,X,\z,\bb)\}\mid Y,C,\z,\bb]}{\E_{10}\{\I(X>C)\mid Y,C,\z,\bb\}}\\
&&+\I(x>C)\E_{10}[\I(X>C)\{\a(X,\z,\E_{10},\E_{20},\bb)-\S_\bb^{\rm F}(Y,X,\z,\bb)\}\mid Y,C,\z,\bb]\\
&&\left.\times\frac{(\wh\E_1-\E_{10})\{\I(X>C)\mid Y,C,\z,\bb\}}{[\E_{10}\{\I(X>C)\mid Y,C,\z,\bb\}]^2}\mid x,\z,\bb\right)\\
&&+\E_{20}\left(-\I(x>C)\frac{(\wh\E_1-\E_{10})\{\I(X>C)\frac{\partial\a(x,\z,\E_{10},\E_{20},\bb)}{\partial\E_2}(\wh\E_2-\E_{20})\mid Y,C,\z,\bb\}}{\E_{10}\{\I(X>C)\mid Y,C,\z,\bb\}}\right.\\
&&+\I(x>C)\E_{10}\left\{\I(X>C)\frac{\partial\a(x,\z,\E_{10},\E_{20},\bb)}{\partial\E_2}(\wh\E_2-\E_{20})\mid Y,C,\z,\bb\right\}\\
&&\left.\times\frac{(\wh\E_1-\E_{10})\{\I(X>C)\mid Y,C,\z,\bb\}}{[\E_{10}\{\I(X>C)\mid Y,C,\z,\bb\}]^2}\mid x,\z,\bb\right)\\
&=&o_p(n^{-1/2})
\ese
in terms of its vector $L_1$ norm by Lemma \ref{lem:E}, \eqref{eq:dadE1}, \ref{con:seffub-non}, and \eqref{eq:dadE2},
and this holds uniformly with respect to $(x,\z,\bb)$ by
\ref{con:bb-non}.
This with Lemma \ref{lem:L} leads to
\bse
\sup_{x,\z,\bb}\left\|\frac{\partial^2\a(x,\z,\E_{10},\E_{20},\bb)}{\partial\E_1\partial\E_2}(\wh\E_1-\E_{10},\wh\E_2-\E_{20})\right\|_1&=&o_p(n^{-1/2}).
\ese
This with \eqref{eq:d2seffdE1dE2}, Lemma \ref{lem:E}, and \eqref{eq:dadE2} further leads to
\bse
\left\|\frac{\partial^2\S\eff(y,w,\delta,\z,\E_{10},\E_{20},\bb)}{\partial\E_1\partial\E_2}(\wh\E_1-\E_{10},\wh\E_2-\E_{20})\right\|_1&=&o_p(n^{-1/2}).
\ese
\hfill$\blacksquare$ 

%%%%%%%%%%%%%%%%%%%%%%%%%%%%%%%%%%%%%%%%%%%%%%%%%%

\subsection{Proof of Theorem \ref{th:non}}\label{sec:th-non-proof}
We first derive some basic properties of $\S\eff$. \eqref{eq:b-non} yields
\bse
&&\E_2\{\S\eff(Y,W,\Delta,\z,\E_1,\E_2,\bb)\mid x,\z,\bb\}\\
&=&\E_2\{\Delta\S_\bb^{\rm F}(Y,W,\z,\bb)-\b(Y,W,\Delta,\z,\E_1,\E_2,\bb)\mid x,\z,\bb\}\\
&=&\E_2(\Delta\mid x,\z)\E\{\S_\bb^{\rm F}(Y,x,\z,\bb)\mid x,\z,\bb\}\\
&=&\0,
\ese
where the last equality holds by $\E\{\S_\bb^{\rm F}(Y,x,\z,\bb)\mid x,\z,\bb\}=\0$.
Subsequently, for any $\E_1$ and $d_1$,
\be
\E_{20}\left\{\frac{\partial\S\eff(Y,W,\Delta,\z,\E_1,\E_{20},\bb_0)}{\partial\E_1}(d_1)\mid x,\z,\bb_0\right\}
=\0.\label{eq:A1-non}
\ee
In addition, \eqref{eq:b-non} yields
\bse
\0&=&\E_1[\E_2\{\b(Y,W,\Delta,\z,\E_1,\E_2,\bb)\mid X,\z,\bb\}\mid\z]\n\\
&=&\E_2[\E_1\{\b(Y,W,\Delta,\z,\E_1,\E_2,\bb)\mid C,\z,\bb\}\mid\z]\n\\
&=&\E_2\left\{\E_1\left(\Delta\a(X,\z,\E_1,\E_2,\bb)\right.\right.\n\\
&&\left.\left.+(1-\Delta)\frac{\E_1[(1-\Delta)\{\a(X,\z,\E_1,\E_2,\bb)-\S_\bb^{\rm F}(Y,X,\z,\bb)\}\mid Y,C,\z,\bb]}
{\E_1(1-\Delta\mid Y,C,\z,\bb)}\mid C,\z,\bb\right)\mid\z\right\}\n\\
&=&\E_2\{\E_1(\Delta\a(X,\z,\E_1,\E_2,\bb)\n\\
&&+\E_1[(1-\Delta)\{\a(X,\z,\E_1,\E_2,\bb)-\S_\bb^{\rm F}(Y,X,\z,\bb)\}\mid Y,C,\z,\bb]\mid C,\z,\bb)\mid\z\}\n\\
&=&\E_2[\E_1\{\a(X,\z,\E_1,\E_2,\bb)-(1-\Delta)\S_\bb^{\rm F}(Y,X,\z,\bb)\mid C,\z,\bb\}\mid\z]\n\\
&=&\E_1[\a(X,\z,\E_1,\E_2,\bb)-\E_2(1-\Delta\mid X,\z)\E\{\S_\bb^{\rm F}(Y,X,\z,\bb)\mid X,\z,\bb\}\mid\z]\n\\
&=&\E_1\{\a(X,\z,\E_1,\E_2,\bb)\mid\z\},
\ese
where the last equality holds by $\E\{\S_\bb^{\rm F}(Y,x,\z,\bb)\mid x,\z,\bb\}=\0$.
This leads to
\bse
&&\E_1\{\S\eff(Y,W,\Delta,\z,\E_1,\E_2,\bb)\mid c,\z,\bb\}\n\\
&=&\E_1\{\Delta\S_\bb^{\rm F}(Y,W,\z,\bb)-\b(Y,W,\Delta,\z,\E_1,\E_2,\bb)\mid c,\z,\bb\}\n\\
&=&\E_1\left(\Delta\S_\bb^{\rm F}(Y,X,\z,\bb)-\Delta\a(X,\z,\E_1,\E_2,\bb)\right.\n\\
&&\left.-(1-\Delta)\frac{\E_1[(1-\Delta)\{\a(X,\z,\E_1,\E_2,\bb)
-\S_\bb^{\rm F}(Y,X,\z,\bb)\}\mid Y,c,\z,\bb]}{\E_1(1-\Delta\mid Y,c,\z,\bb)}\mid c,\z,\bb\right)\n\\
&=&\E_1\left(\Delta\S_\bb^{\rm F}(Y,X,\z,\bb)-\Delta\a(X,\z,\E_1,\E_2,\bb)\right.\n\\
&&\left.-\E_1[(1-\Delta)\{\a(X,\z,\E_1,\E_2,\bb)-\S_\bb^{\rm F}(Y,X,\z,\bb)\}\mid Y,c,\z,\bb]\mid c,\z,\bb\right)\n\\
&=&\E_1\{\S_\bb^{\rm F}(Y,X,\z,\bb)-\a(X,\z,\E_1,\E_2,\bb)\mid\z,\bb\}\n\\
&=&\0,
\ese
where the last equality holds by $\E\{\S_\bb^{\rm F}(Y,x,\z,\bb)\mid
x,\z,\bb\}=\0$.  
Subsequently, for any $\E_2$ and $d_2$,
\be
\E_{10}\left\{\frac{\partial\S\eff(Y,W,\Delta,\z,\E_{10},\E_2,\bb_0)}{\partial\E_2}(d_2)\mid c,\z,\bb_0\right\}=\0.\label{eq:A2-non}
\ee

We now show the consistency of $\wt\bb$ following Theorem 2.1 of \cite{newey1994large}.
We can view solving $\E\{\S\eff(Y,W,\Delta,\Z,\bb)\}=\0$ as maximizing
\bse
Q_0(\bb)\equiv-\|\E\{\S\eff(Y,W,\Delta,\Z,\bb)\}\|_2^2.
\ese
Then (i) $Q_0(\bb)$ is uniquely maximized at $\bb_0$ in a
  neighborhood of $\bb_0$ because
$\E\{\S\eff(Y,W,\Delta,\Z,\bb_0)\}=\0$ and the uniqueness is by \ref{con:B-non}.
Also, (ii) $\Omega$ is compact by \ref{con:bb-non}, and
(iii) $Q_0(\bb)$ is continuous since it is differentiable by \ref{con:B-non}.
Hence it suffices to show (iv)
\bse
\wh Q_n(\bb)\equiv-\left\|n^{-1}\sumi\S\eff(y_i,w_i,\delta_i,\z_i,\wh\E_1,\wh\E_2,\bb)\right\|_2^2
\ese
converges uniformly in probability to $Q_0(\bb)$. This holds because
\bse
&&n^{-1}\sumi\S\eff(y_i,w_i,\delta_i,\z_i,\wh\E_1,\wh\E_2,\bb)\\
&=&n^{-1}\sumi\S\eff(y_i,w_i,\delta_i,\z_i,\E_{10},\E_{20},\bb)
+n^{-1}\sumi\frac{\partial\S\eff(y_i,w_i,\delta_i,\z_i,\E_{10},\E_{20},\bb)}{\partial\E_1}(\wh\E_1-\E_{10})\\
&&+n^{-1}\sumi\frac{\partial\S\eff(y_i,w_i,\delta_i,\z_i,\E_{10},\E_{20},\bb)}{\partial\E_2}(\wh\E_2-\E_{20})
+o_p(n^{-1/2})\\
&=&n^{-1}\sumi\S\eff(y_i,w_i,\delta_i,\z_i,\bb)+o_p(n^{-1/4})\\
&=&\E\{\S\eff(Y,W,\Delta,\Z,\bb)\}+o_p(1),
\ese
where the first equality holds by Lemma \ref{lem:dseff},
the second equality holds by Lemma \ref{lem:dseff},
and the last equality holds by \ref{con:seffub-non}.
Also, this holds uniformly with respect to $\bb$ by \ref{con:bb-non}.
Therefore, $\wt\bb$ is consistent for $\bb_0$. 

We now derive the asymptotic distribution of $\wt\bb$. By the definition of $\wt\bb$,
\bse
\0&=&n^{-1/2}\sumi\S\eff(y_i,w_i,\delta_i,\z_i,\wh\E_1,\wh\E_2,\wt\bb)\\
&=&n^{-1/2}\sumi\S\eff(y_i,w_i,\delta_i,\z_i,\wh\E_1,\wh\E_2,\bb_0)+\wh\B\sqrt{n}(\wt\bb-\bb_0)\\
&=&n^{-1/2}\sumi\S\eff(y_i,w_i,\delta_i,\z_i,\E_{10},\E_{20},\bb_0)
+n^{-1/2}\sumi\frac{\partial\S\eff(y_i,w_i,\delta_i,\z_i,\E_{10},\E_{20},\bb_0)}{\partial\E_1}(\wh\E_1-\E_{10})\\
&&+n^{-1/2}\sumi\frac{\partial\S\eff(y_i,w_i,\delta_i,\z_i,\E_{10},\E_{20},\bb_0)}{\partial\E_2}(\wh\E_2-\E_{20})
+o_p(1)+\wh\B\sqrt{n}(\wt\bb-\bb_0)\\
&=&n^{-1/2}\sumi\S\eff(y_i,w_i,\delta_i,\z_i,\bb_0)
+\wh\B\sqrt{n}(\wt\bb-\bb_0)+o_p(1),
\ese
where $\wh\B\equiv n^{-1}\sumi\partial\S\eff(y_i,w_i,\delta_i,\z_i,\wh\E_1,\wh\E_2,\wc\bb)/\partial\bb\trans$ and $\wc\bb\equiv t\wt\bb+(1-t)\bb_0$ for some $t\in(0,1)$, and the third equality holds by Lemma \ref{lem:dseff}. Finally, the last equality holds because
\bse
&&n^{-1/2}\sumi\frac{\partial\S\eff(y_i,w_i,\delta_i,\z_i,\E_{10},\E_{20},\bb_0)}{\partial\E_1}(\wh\E_1-\E_{10})\\
&=&n^{1/4}\left[n^{-1}\sumi n^{1/4}\frac{\partial\S\eff(y_i,w_i,\delta_i,\z_i,\E_{10},\E_{20},\bb_0)}{\partial\E_1}(\wh\E_1-\E_{10})\right.\\
&&\left.-\int\left\{n^{1/4}\frac{\partial\S\eff(y,w,\delta,\z,\E_{10},\E_{20},\bb_0)}{\partial\E_1}(\wh\E_1-\E_{10})\right\}f_{Y,W,\Delta,\Z}(y,w,\delta,\z,\bb_0)dydwd\delta d\z\right]\\
&=&n^{1/4}o_p(n^{-1/2})=o_p(n^{-1/4})
\ese
where
the first equality holds by \eqref{eq:A1-non} and the second equality holds by Lemma \ref{lem:dseff}, and
\bse
&&n^{-1/2}\sumi\frac{\partial\S\eff(y_i,w_i,\delta_i,\z_i,\E_{10},\E_{20},\bb_0)}{\partial\E_2}(\wh\E_2-\E_{20})\\
&=&n^{1/4}\left[n^{-1}\sumi n^{1/4}\frac{\partial\S\eff(y_i,w_i,\delta_i,\z_i,\E_{10},\E_{20},\bb_0)}{\partial\E_2}(\wh\E_2-\E_{20})\right.\\
&&\left.-\int\left\{n^{1/4}\frac{\partial\S\eff(y,w,\delta,\z,\E_{10},\E_{20},\bb_0)}{\partial\E_2}(\wh\E_2-\E_{20})\right\}f_{Y,W,\Delta,\Z}(y,w,\delta,\z,\bb_0)dydwd\delta d\z\right]\\
&=&n^{1/4}o_p(n^{-1/2})=o_p(n^{-1/4}),
\ese
where the first equality holds by \eqref{eq:A2-non} and the second equality holds by Lemma \ref{lem:dseff}.
Then,
\bse
\sqrt{n}(\wt\bb-\bb_0)
&=&-\wh\B^{-1}\left\{n^{-1/2}\sumi\S\eff(y_i,w_i,\delta_i,\z_i,\bb_0)+o_p(1)\right\}\\
&=&-\{\B^{-1}+o_p(1)\}\left\{n^{-1/2}\sumi\S\eff(y_i,w_i,\delta_i,\z_i,\bb_0)+o_p(1)\right\}\\
&=&-\B^{-1}n^{-1/2}\sumi\S\eff(y_i,w_i,\delta_i,\z_i,\bb_0)+o_p(1),
\ese
where the second equality holds by \ref{con:B-non}, Lemma \ref{lem:E}, and $\|\wc\bb-\bb_0\|_2\leq\|\wt\bb-\bb_0\|_2=o_p(1)$ by the consistency of $\wt\bb$. Also, the third equality holds by \ref{con:B-non} and $n^{-1/2}\sumi\S\eff(y_i,w_i,\delta_i,\z_i,\bb_0)=O_p(1)$
since $\E\{\S\eff(Y,W,\Delta,\Z,\bb_0)\}=\0$.
Furthermore,
\bse
\B&=&\E\left\{\partial\S\eff(Y,W,\Delta,\Z,\bb_0)/\partial\bb\trans\right\}\\
&=&-\E\left\{\S\eff(Y,W,\Delta,\Z,\bb_0)\S_\bb\trans(Y,W,\Delta,\Z,\bb_0)\right\}\\
&=&-\E\left\{\S\eff^{\otimes2}(Y,W,\Delta,\Z,\bb_0)\right\},
\ese
where the second equality holds because $\E_\bb\{\S\eff(Y,W,\Delta,\Z,\bb)\}=\0$ for any $\bb$, and the last equality holds by $\S_\bb-\S\eff\in\Lambda$ and $\S\eff\in\Lambda^\perp$.
Thus, we obtain the result of Theorem \ref{th:non}.
\hfill$\blacksquare$

%%%%%%%%%%%%%%%%%%%%%%%%%%%%%%%%%%%%%%%%%%%%%%%%%%

\subsection{Additional Simulation Results}\label{sec:add-sims}
\spacingset{1.0}
\begin{table}[!ht]
    \centering
    \scalebox{0.87}{\begin{tabular}{ccccrrrrrrrr}
\toprule
\multicolumn{4}{c}{ } & \multicolumn{4}{c}{$\mathbf{q = 0.4}$} & \multicolumn{4}{c}{$\mathbf{q = 0.8}$} \\
\cmidrule(l{3pt}r{3pt}){5-8} \cmidrule(l{3pt}r{3pt}){9-12}
\textbf{Param} & \textbf{Estimator} & $\mathbf{X|Z}$ & $\mathbf{C|Z}$ & \textbf{Bias} & \textbf{ESE} & \textbf{ASE} & \textbf{Cov} & \textbf{Bias} & \textbf{ESE} & \textbf{ASE} & \textbf{Cov}\\
\midrule
$\beta_1$ & SPARCC & Correct & Correct & -0.04 & 0.47 & 0.47 & 93.9 & -0.05 & 0.68 & 0.69 & 95.4\\
  &   & Correct & Incorrect & -0.04 & 0.47 & 0.47 & 94.0 & -0.05 & 0.68 & 0.69 & 95.3\\
  &   & Incorrect & Correct & -0.03 & 0.47 & 0.47 & 94.0 & 0.07 & 0.73 & 0.74 & 95.0\\
  &   & Incorrect & Incorrect & -0.03 & 0.47 & 0.47 & 94.1 & 0.07 & 0.73 & 0.74 & 95.0\\
  &   & Nonpar & Nonpar & -0.03 & 0.47 & 0.47 & 93.9 & 0.13 & 0.71 & 0.70 & 94.0\\
\addlinespace
  & MLE & Correct & - & -0.06 & 0.40 & 0.41 & 95.8 & -0.05 & 0.68 & 0.65 & 94.1\\
  &   & Incorrect & - & 2.80 & 0.40 & 0.42 & 0.0 & 9.75 & 0.76 & 0.78 & 0.0\\
  &   & Nonpar & - & 0.18 & 0.49 & 0.64 & 95.0 & 0.94 & 1.41 & 2.14 & 89.1\\
\addlinespace
  & Complete Case & - & - & -0.03 & 0.47 & 0.47 & 94.0 & 0.00 & 0.75 & 0.75 & 95.1\\
\addlinespace
  & Oracle & - & - & -0.02 & 0.35 & 0.36 & 95.4 & 0.01 & 0.36 & 0.36 & 95.0\\
\midrule
$\beta_3$ & SPARCC & Correct & Correct & 0.03 & 0.36 & 0.36 & 95.2 & 0.04 & 0.61 & 0.60 & 94.4\\
  &   & Correct & Incorrect & 0.03 & 0.36 & 0.36 & 95.3 & 0.04 & 0.61 & 0.61 & 94.5\\
  &   & Incorrect & Correct & 0.02 & 0.36 & 0.36 & 95.2 & 0.04 & 0.63 & 0.63 & 94.7\\
  &   & Incorrect & Incorrect & 0.02 & 0.36 & 0.36 & 95.1 & 0.04 & 0.63 & 0.63 & 94.8\\
  &   & Nonpar & Nonpar & 0.02 & 0.36 & 0.36 & 95.1 & -0.04 & 0.62 & 0.61 & 94.2\\
\addlinespace
  & MLE & Correct & - & 0.01 & 0.32 & 0.32 & 95.6 & -0.01 & 0.70 & 0.67 & 94.4\\
  &   & Incorrect & - & -2.64 & 0.31 & 0.32 & 0.0 & -10.88 & 0.60 & 0.61 & 0.0\\
  &   & Nonpar & - & -0.09 & 0.37 & 0.47 & 95.6 & -0.54 & 1.22 & 1.84 & 89.1\\
\addlinespace
  & Complete Case & - & - & 0.02 & 0.36 & 0.36 & 95.4 & 0.02 & 0.63 & 0.62 & 94.6\\
\addlinespace
  & Oracle & - & - & 0.01 & 0.27 & 0.26 & 94.7 & 0.00 & 0.25 & 0.26 & 95.5\\
\midrule
$\log\sigma^2$ & SPARCC & Correct & Correct & -0.02 & 0.20 & 0.20 & 95.5 & -0.03 & 0.37 & 0.35 & 94.3\\
  &   & Correct & Incorrect & -0.02 & 0.20 & 0.20 & 95.5 & -0.03 & 0.37 & 0.35 & 94.3\\
  &   & Incorrect & Correct & -0.02 & 0.20 & 0.20 & 95.5 & -0.03 & 0.37 & 0.35 & 94.3\\
  &   & Incorrect & Incorrect & -0.02 & 0.20 & 0.20 & 95.5 & -0.03 & 0.37 & 0.35 & 94.3\\
  &   & Nonpar & Nonpar & -0.02 & 0.20 & 0.20 & 95.5 & -0.03 & 0.37 & 0.35 & 94.3\\
\addlinespace
  & MLE & Correct & - & 0.01 & 0.18 & 0.19 & 95.7 & 0.05 & 0.30 & 0.29 & 94.3\\
  &   & Incorrect & - & 0.10 & 0.18 & 0.24 & 93.3 & 2.21 & 0.37 & 0.38 & 0.0\\
  &   & Nonpar & - & -0.03 & 0.19 & 0.21 & 94.6 & 0.01 & 0.34 & 0.46 & 95.7\\
\addlinespace
  & Complete Case & - & - & -0.02 & 0.20 & 0.20 & 95.5 & -0.03 & 0.37 & 0.35 & 94.3\\
\addlinespace
  & Oracle & - & - & -0.01 & 0.15 & 0.16 & 95.3 & -0.01 & 0.16 & 0.16 & 95.4\\
\bottomrule
\end{tabular}}
\caption{Results from the simulation study based on 1000 simulated replicates per setting and for additional parameters. Param: parameter estimated. $X|Z$, $C|Z$, q, Bias, ESE, ASE, Cov as in Table~\ref{tab:sim1}.}
    \label{tab:sim1-2}
\end{table}
%%%%%%%%%%%%%%%%%%%%%%%%%%%%%%%%%%%%%%%%%%%%%%%%%%

\end{document}